\documentclass[11pt]{article}
\usepackage{amsmath,amsthm,amsfonts,amssymb, graphicx,mathabx,epsfig}
\usepackage{bbm}
\usepackage{authblk}
\usepackage{bm}
\usepackage{graphicx,color}
\usepackage{footmisc}
\usepackage{epstopdf}
\usepackage{dsfont}
\usepackage{pgfplots}
\usepackage{subfig}
\usetikzlibrary{fit}
\usepackage{slashed}
\usepackage[colorlinks]{hyperref}
\hypersetup{
    colorlinks=true,
    linkcolor=blue,
    citecolor=red,
    filecolor=magenta,      
    urlcolor=cyan,
}
\usepackage{enumitem}

\numberwithin{equation}{section}

\newtheorem{thm}{Theorem}[section]

\newtheorem{lemma}[thm]{Lemma}
\newtheorem{prop}[thm]{Proposition}

\def\ps{\slashed{\pp}}
\def\qs{\slashed{\qq}}
\def\ks{\slashed{\kk}}

\def\kk{{\bf k}} 
\def\qq{{\bf q}} \def\pp{{\bf p}}
 \def\xx{{\bf x}} \def\yy{ {\bf y}} \def\zz{{\bf z}}
\def\kk{{\bf k}}

\def\\{\hfill\break}
\let\0=\noindent

\def\otto{\,{\kern-1.truept\leftarrow\kern-5.truept\to\kern-1.truept}\,}
\def\lis{\overline}

\def \be{\begin{equation}}
\def \ee{\end{equation}}
\def \bea{\begin{eqnarray}}
\def \eea{\end{eqnarray}}

\def\l{\lambda}
\def\r{\rho}
\def\s{\sigma}
\def\a{\alpha}
\def\b{\beta}
\def\c{\chi}
\def\d{\delta}

\def\m{\mu}
\def\g{\gamma}
\def\G{\Gamma}
\def\e{\varepsilon}

\def\p{\pi}

\let\n=\nu
\let\o=\omega
\def\Tr{{\rm{Tr}}}
\def\tr{{\rm{Tr}}}

\begin{document}

\title{Anomaly non-renormalization in interacting Weyl semimetals}

\author[1,2]{Alessandro Giuliani}
\affil[1]{Universit\`a degli Studi Roma Tre, Department of Mathematics and Physics, L.go S. L. Murialdo 1, 00146 Roma, Italy}
\affil[2]{Centro Linceo Interdisciplinare {\it Beniamino Segre}, Accademia Nazionale dei Lincei, Palazzo Corsini, Via della Lungara 10,
00165 Roma, Italy}
\author[3]{Vieri Mastropietro}
\affil[3]{Universit\`a di Milano, Department of Mathematics ``F. Enriquez'', Via C. Saldini 50, 20133 Milano, Italy}
\author[4]{Marcello Porta}
%\affil[4]{University of T\"ubingen, Department of Mathematics, Auf der Morgenstelle 10, 72076 T\"ubingen, Germany}
\affil[4]{Scuola Internazionale Superiore di Studi Avanzati (SISSA), Mathematics Area, Via Bonomea 265, 34136 Trieste, Italy}
\maketitle

\begin{abstract} Weyl semimetals are 3D condensed matter systems characterized by a degenerate Fermi surface, consisting of a pair of `Weyl nodes'. 
Correspondingly, in the infrared limit, these systems behave effectively as Weyl fermions in $3+1$ dimensions. 
We consider a class of interacting 3D lattice models for Weyl semimetals and prove that the quadratic response of the quasi-particle flow between the Weyl nodes
is universal, that is, independent of the interaction strength and form. 
Universality is the counterpart
of the Adler-Bardeen non-renormalization property of the chiral anomaly for the infrared emergent description, which is proved here 
in the presence of a lattice and at a non-perturbative level.
Our proof relies on constructive bounds for the Euclidean ground state correlations combined with lattice Ward Identities, and it is valid arbitrarily close to the critical point where the 
Weyl points merge and the relativistic description breaks down.
\end{abstract}

\maketitle

\section{Introduction}

The behaviour of relativistic fermionic particles is described by the Dirac equation and by its 
interacting extensions, 
such as Quantum ElectroDynamics (QED) or other standard Quantum Field Theory (QFT) models, 
whose spectacular and often counter-intuitive 
predictions have been tested for decades in high energy experiments. 
On the other hand, the electrons present in ordinary matter, like the conduction electrons in metals, which live at much lower energy scales, 
are described by the many-body Schr\"odinger equation, which breaks Lorentz invariance explicitly. 

By scale separation, the Dirac high energy physics is not expected to play directly any role in the behaviour of conduction electrons. Nevertheless, 
there are cases in which the interaction of the conduction electrons with the underlying lattice can produce an effective description 
in terms of Dirac particles, with a velocity of propagation much smaller than the speed of light. Such emergent QFT
description is valid in great generality for one dimensional metals, see \cite{L,MaL,To} (see also \cite{MM} for a recent review), but is also possible in
higher dimensional systems: notable examples were proposed in \cite{A8,A7} and in \cite{2a}. 
The discovery of graphene \cite{Ka,No} and of topological insulators \cite{A0} provides 
a condensed matter realization of systems of 2D Dirac fermions. 
In connection with their emergent QFT description, the transport coefficients of these systems are characterized 
by remarkable universality properties: examples include the optical conductivity of graphene
\cite{A3} and the Hall conductance \cite{A2,TKNdN}. 
A mathematical proof of universality in the presence of interactions
has been established in \cite{AMP,BBDRF1,BBDRF2,GJMP,GMP,GMPbis,GMPhall, GMPtris, HM1,MPspin}.
More recently, 3D Dirac systems have been experimentally realized \cite{Bori,A11, LiuJ, A22}, following the theoretical predictions of
\cite{2a} and \cite{FKM07,MB07,R09}, see \cite{AMV,A0} for reviews;
these semimetals are dubbed `Dirac' or `Weyl' semimetals, depending on whether the Fermi points coincide or are at distinct locations in the Brillouin zone. 
Such experimental discoveries open the way to the observation of the counterpart
of one of the most interesting phenomena characterizing 3D massless Dirac fermions: 
the Adler-Bell-Jackiw (ABJ) axial anomaly \cite{A4,ABJ}, which corresponds to the breaking of the classical chiral symmetry by quantum mechanical radiative
corrections and results in the non-conservation of the axial current.  
Nielsen and Ninomiya \cite{2a} predicted that this quantum anomaly should have a realization, or `simulation', in 
the quadratic response of the quasi-particle flow between the Weyl nodes to an external electromagnetic field, 
see also \cite{2c1,2c2}; in particular, this quasi-particle flow is expected to be proportional to $E\cdot B$, with coefficient given by the value of the axial anomaly in QED$_4$. 
The experimental observation of the `chiral anomaly' in Weyl semimetals has been reported in \cite{o2,Huang,LiKh,o1,Zhang}, see also 
\cite{Betal97} for an observation in the context of superfluid $^3$He.
Other applications of the chiral anomaly in condensed matter physics have been discussed in \cite{FST}, in the context of a gauge theory of phases of matter, see 
also \cite{FP} for new proposed applications and \cite{Fro} for a recent review.

The theoretical analysis in \cite{2a} neglects the effects of many-body interaction.
What happens to the quadratic response of the quasi-particle flow in the presence of electron interactions, which are unavoidable in real systems? One of the distinctive features of the axial anomaly in QFT is the Adler-Bardeen (AB) {\it non-renormalization property} \cite{A6}, which says that all the radiative corrections cancel out at all orders in the fine structure constant, 
so that the axial anomaly equals the value of the chiral triangle graph. However, 
the AB argument cannot be applied to Weyl semimetals: the analysis in \cite{A6}
requires cancellations between classes of Feynman graphs based on Lorentz and chiral symmetries, which are broken explicitly in any lattice model of interacting electrons. 
Strangely enough, the issue of universality has not been addressed theoretically until now, despite the huge literature on the effects of interactions\footnote{An even more extensive and rapidly growing literature is available on the effects of disorder, which we cannot fully account for here, see \cite{AB15,2c3} for a couple of representative results, see also \cite{AFP,Ma_qp} for recent rigorous results. Even in that context, the issue of universality of the `chiral anomaly' for Weyl semi-metals has not been satisfactorily discussed yet.}
\cite{AFP1,DHM13,Gos,GPDS15,MN14,M1,WKA14}. In this paper, we consider a paradigmatic class of lattice models for 3D Weyl semimetals 
with short-range interactions, generalizing a widely used `standard' model for these materials, see \cite{AMV}. For this class of models,

\medskip

{\it we prove that the quadratic response of the quasi-particle flow between the Weyl points, which is the counterpart of the chiral anomaly in the emergent QFT description
of the system, is universal, that is, independent of the interaction strength and form.}

\medskip

Therefore, this quantity can be added to the limited list of universal transport coefficients in condensed matter, like the Hall conductivity and graphene's optical conductivity, and is the 
only known example of universal quantum transport coefficient in three dimensions. 
This quadratic response has been measured \cite{o2,Huang,LiKh,o1,Zhang} in 
different Weyl semimetal, such as Na$_3$Bi, TaAs, GdPtBi, ZrTe$_5$, and the proportionality between
the quasi-particle flow between the Weyl points and $E \cdot B$ has been experimentally verified. The interaction does not appear to qualitatively modify 
such a response, in agreement with our result, even though a precise measurement of the proportionality coefficient remains to be performed. 
Remarkably, {\it our result 
is valid arbitrarily close to the critical point where the Weyl nodes merge} and the natural parameter measuring the strength of the interaction, that is,
the coupling strength divided by the quasi-particle velocity, diverges: this is important for applications to real 
3D Weyl semimetals, where typically the distance between the Weyl nodes is very small, as compared to the size of the Brillouin zone; 
see Section \ref{sec:1.2} for further discussion of this point.

\medskip

In this paper, we focus our attention on the quadratic response coefficient of the `chiral current' to an external electromagnetic field directly, without discussing the 
linear response coefficient, which is not expected to have any remarkable universality feature. 
We stress that in the experiments the `chiral anomaly' is detected by measuring the peak of quasi-particle flow between the Weyl nodes 
as the angle between the external electric and magnetic fields $E$ and $B$ is varied, see e.g. \cite{o1}: such a procedure has the effect of automatically subtracting the linear response, 
which is uninteresting for the effect under investigation. 

\medskip

The paper is organized as follows: in Section \ref{sec:1.2}, we introduce a class of lattice Weyl semimetals, we define the notions of lattice current and lattice `chiral current', and state our main results, summarized in Theorem \ref{thm:main}. In Section \ref{sec:3} we present the proof: first we describe the general strategy and the core argument (Section \ref{sec:3.0}); next we discuss the technical details: the Grassmann representation of the generating function of correlations and its multiscale Renormalization Group (RG) computation are presented in Sections \ref{sec:3.1} and \ref{sec:RG}; the RG flow of the effective couplings is studied in Section \ref{sec:flow}; the regularity properties of the  Euclidean correlations is discussed in Section \ref{sec:sing}, and the relativistic contribution to the quadratic response of the chiral current to the electromagnetic field is computed in Section \ref{sec:3.6}. A few additional technical aspects of the proof are deferred to the appendices: in Appendix \ref{app:symm} we prove a number of key symmetry properties of the kernels of the effective potentials produced by the RG iterations; in Appendix \ref{app.dimbounds} we discuss a few technical aspects of the tree expansion for the effective potentials; 
in Appendix \ref{app:2nd} we compute the  quadratic response coefficient via time-dependent perturbation theory and prove a Wick rotation theorem that allows us to express it in terms of Euclidean correlations. 

\section{Weyl semimetals and response functions}\label{sec:1.2}

\subsection{A basic non-interacting model for Weyl semimetals}

We start by describing a paradigmatic non-interacting lattice model for Weyl semimetal, see \cite[Section II.B.1]{AMV}, 
with broken time-reversal symmetry, for which inversion symmetry is preserved. This example motivates the class of models which our analysis applies to. 
It describes an electron system on a simple cubic lattice subject to an external magnetic field and with the following features: 
the system is magnetically ordered, so that the bands have no spin degeneracy; moreover, every lattice site comes 
with two internal degrees of freedom, $r=1,2$, playing the role of pseudo-spin labels, corresponding to, {\it e.g.}, orbital degrees of freedom;
finally, these orbitals (say, $s,p$ orbitals) have opposite parity. i.e., they behave differently under inversion. The model is defined in terms of 
the following Bloch Hamiltonian 
(see next section for details):
\begin{equation}\label{hop}
\hat H^{0}(k) = \begin{pmatrix} \alpha(k) & \beta(k) \\ \beta^*(k) & -\alpha(k)\end{pmatrix}\;,
\end{equation}
where
\begin{equation}
\alpha(k)=2 + \zeta-\cos k_1-\cos k_2-\cos k_3\;,\qquad \beta(k)=t_1 \sin k_1-it_2\sin k_2
\end{equation}
and $\zeta, t_1,t_2$ are three real parameters. 
We assume that $t_1,t_2>0$ and $-1<\zeta<1$, so that $\hat H^0(k)$ is singular only at two Weyl nodes, or Fermi points, 
$k=p_F^\pm$, with $p_F^\pm=(0,0,\pm\arccos\zeta)$. We shall think of $t_1,t_2$ as being fixed once and for all, while $\zeta$ 
is tunable: it controls the distance among the Weyl nodes, which tend to 
merge in the limit as $|\zeta|\to 1^-$. Note that the relative distance between the two nodes vanishes like $\sqrt{1-|\zeta|}$ as 
$|\zeta|\to 1^-$. 

\medskip

{\bf Remark.} Having a parameter tuning the distance between Weyl nodes is extremely natural: in real systems, 
magnetic fields in orthogonal directions are used to induce and control a `topological' transitions from, e.g., a Dirac semimetal to a magnetic semiconductor, or 
to a Weyl semimetal; once a Weyl semimetal phase is entered, the intensity of the magnetic fields is used to tune the 
location of the Weyl nodes. See \cite[Fig.2]{AMV} and the corresponding discussion; see also \cite{DHM13}. In the limit as the Weyl nodes tend to merge, 
the valence and conduction bands generically tend to touch quadratically at the doubly-degenerate node. 
Experimentally, the Weyl nodes may be quite close to each other, compared to the size of the Brillouin zone.
The fact that the Weyl points can be arbitrarily close is a peculiarity of Weyl semimetals, which distinguish them from other Dirac materials, like graphene.

\medskip

In the vicinity of the Weyl nodes, $k\simeq p_F^\omega$, the Bloch Hamiltonian can be linearized as: 
\begin{equation}\label{eq:linear}
\hat H^{0}(k) = t_1\sigma_1 k_1+t_2 \sigma_2 k_2 +\omega \sqrt{1-\zeta^2} \sigma_3 (k_3-p_{F,3}^\omega) 
+ O(|k-p_F^\o|^2)\;,
\end{equation}
with $\sigma_{i}$ the Pauli matrices:
\begin{equation}
\sigma_{1} = \begin{pmatrix} 0 & 1 \\ 1 & 0 \end{pmatrix}\;,\qquad \sigma_{2} = \begin{pmatrix} 0 & -i \\ i & 0 \end{pmatrix}\;,\qquad \sigma_{3} = \begin{pmatrix} 1 & 0 \\ 0 & -1 \end{pmatrix}\;.
\end{equation}
The parameters $t_1,t_2, \omega\sqrt{1-\zeta^2}$ in front of the three terms in the right side of \eqref{eq:linear} play the role of `Fermi velocities' in the three coordinate directions. 
Note that, as $|\zeta|\to 1$, the third component of the Fermi velocity (the one along which the Weyl nodes are located), 
$v_{3,\omega}^{0} :=\omega\sqrt{1-\zeta^2}$, 
tends to zero at the {\it same speed} as $p_F^+-p_F^-$. This is no accident, it is a generic feature in the limit as 
two Weyl nodes merge. 

In order to provide an asymptotic description of the system that holds unifromly in the limit as the Weyl nodes merge, it 
is convenient to explicitly extract the quadratic part of the Bloch Hamiltonian in the third coordinate direction. In our case: 
\begin{equation}\label{eq:linear2}
\hat H^{0}(k) =  t_1\sigma_1 k_1+t_2 \sigma_2 k_2 +\sigma_3\big[v_{3,\omega}^0 (k_3-p_{F,3}^\omega) + \tfrac{\zeta}2 (k_3-p_{F,3}^\omega)^2\big]+
O(k_1^2,k_2^2, |k_3-p_{F,3}^\omega|^3)\;.\end{equation}
Note that the prefactor of the quadratic term that we extracted, equal to $\zeta/2$, tends to a non-zero constant in the limit as the Weyl nodes tend to merge, 
$|\zeta|\to 1^-$. This is again no accident, it is another generic feature in the limit as two Weyl nodes merge. 

\subsection{A class of non-interacting models for Weyl semimetals}\label{sec:2.1}

Motivated by the model introduced in the previous section, we now consider a class of non-interacting lattice models, including (\ref{hop}) as a special case. 
We will use adapted coordinates, such that, as in the model above, the Weyl nodes are located along the third coordinate direction.

Let $\Lambda_{L}$ be a cubic lattice of side $L$, with periodic boundary conditions; that is $\Lambda_L=\mathbb Z_L^3$, 
with $\mathbb Z_L=\mathbb Z / L\mathbb Z$ the integers modulo $L$.  
The coordinates of the particle on the lattice are specified by $(x,r)$, with $x\in \Lambda_{L}$ the position and $r=1,2$ the internal degree of freedom,
called `color', or `orbital'. 
Thus, the single-particle Hilbert space is $\frak{h}_{L} = \mathbb{C}^{L^{3}} \otimes \mathbb{C}^{2}$. We denote by $H^{0}$ the single-particle Hamiltonian on 
$\frak{h}_{L}$, which we assume to be Hermitian. We consider finite-ranged, translation invariant systems, such that $H^{0}(x,y) \equiv H^{0}(y-x)$. 
We always assume that $L$ is larger than the range of $H^0$. We also assume that, after identifying $y-x$ with its image in $\mathbb Z^3$, 
$H^0(y-x)$ is independent of $L$ (i.e., the kernel of $H^0$ is independent of $L$, up to the periodicity condition). 
Due to periodicity and translation invariance, we can represent the single-particle Hamiltonian as:
\begin{equation}\label{eq:bloch}
H^{0}(x,y) = \frac{1}{L^{3}}\sum_{k\in \mathcal{B}_{L}} e^{-ik\cdot (x-y)} \hat H^{0}(k)\;,
\end{equation}
where $\mathcal{B}_{L}$ is the finite-volume Brillouin zone, defined as:
\begin{equation}\label{eq:BL}
\mathcal{B}_{L} = \frac{2\pi}{L} \mathbb{Z}_L^{3}, 
\end{equation}
and $\hat H^{0}(k)$ is the Bloch Hamiltonian, a two-by-two matrix parametrized by $k\in \mathcal{B}_{L}$. 
Under the above assumptions, the Bloch Hamiltonian is an $L$-independent function, Hermitian and 
analytic in $k\in \mathcal{B}_{\infty} \equiv \mathbb{T}^{3}$. We also allow for $\hat H^0$ to depend smoothly 
on an additional paremeter $\zeta\in [\zeta_0,\zeta_1)\equiv I$, which can be used to tune 
the location of the Weyl points, see below. 

We assume the following {\bf symmetry properties}, valid for all $\zeta\in I$, which are all satisfied by the model in Eq.\eqref{hop}. The first two symmetries are the most physically significant: they characterize the class of Weyl semimetals 
under consideration. The other two are not crucial, but simplify the structure of the linearized Hamiltonian around the Fermi points. 
\begin{enumerate}
\item {\it Broken time-reversal symmetry.} $\hat H^0(k)\neq\big[\hat H^0(-k)]^*$, where $*$ indicates complex conjugation\footnote{We 
denote by $z^*$, rather than by $\bar z$, the complex conjugate of $z\in\mathbb C$.  Given a matrix $A\in \mathbb C^{n\times n}$, we denote by $A^\dagger$ its Hermitian conjugate, and by $A^*$ its complex conjugate, that is, the matrix whose elements satisfy $(A^*)_{ij}=(A_{ij})^*$.}.
\item {\it Inversion symmetry\footnote{As mentioned in the previous section, we 
are assuming that the two orbitals $r=1$ and $r=2$ transform differently under inversion, that is, one is even and one is odd under $x\to -x$, which explains the presence of the third Pauli matrix $\sigma_3$.}.} $\hat H^0(k)=\sigma_3 \hat H^0(-k)\sigma_3$.
\item {\it Reflection about a horizontal plane.} $\hat H^0(k)=\hat H^0\big((k_1,k_2,-k_3)\big)$
\item {\it Reflection about a vertical plane and color exchange.} $\hat H^0(k)=-\sigma_1 \hat H^0\big((-k_1,k_2,k_3)\big)\sigma_1$. 
\end{enumerate}
In addition to these symmetries, we assume the following properties on the energy bands of the Bloch Hamiltonian. 

\medskip

{\bf Weyl points.} We write the Bloch Hamiltonian in the form 
\begin{equation}\label{eq:genrepH0}
\hat H^{0}(k) = \sigma_1 a(k)+ \sigma_2 b(k) + \sigma_3 c(k) + d(k). \end{equation}
Using the symmetries (ii), (iii), (iv), we see that: $a$ is odd in $k_1$, even in $k_2$ and in $k_3$; $b$ is odd in $k_2$, even in $k_1$ and in $k_3$; $c$ is even in 
$k_1$, in $k_2$ and in $k_3$; $d$ is odd in $k_1$ and in $k_2$, even in $k_3$. We let $\varepsilon_\pm(k)$ be the $k$-dependent eigenvalues of the Bloch 
Hamiltonian, also known as the `energy bands':
\begin{equation}\label{eq:detgenrepH0}
\varepsilon_\pm(k)=d(k)\pm\sqrt{a^2(k)+ b^2(k) + c^2(k)}. \end{equation}
For any $\zeta\in I$, we assume that $\varepsilon_+(k)\ge 0\ge \varepsilon_-(k)$, with equalities only at {\it two} points, called the Weyl nodes, 
denoted by $p_F^\pm$, at which we assume $\varepsilon_\pm(k)$ to vanish {\it linearly}. 
Thanks to the symmetries (ii), (iii), (iv), the {\it pair} of Weyl nodes is invariant under 
reflections about any of the coordinate axes. Therefore, in order for the nodes to be exactly two, they need to be located along one of the three axes. 
Moreover, the requirement that $\varepsilon_\pm(k)$ vanish linearly at $p_F^\pm$, combined with the fact that $a,b,c,d$ are all even functions of $k_3$, 
implies that the Weyl nodes are located along the third axis: 
\begin{equation}\label{pF3}p_{F}^{\omega} = (0,0,p^{\omega}_{F,3}),\end{equation}
where $p^{+}_{F,3}=-p^-_{F,3}\neq 0$. 
Of course, $p_F^\pm$ depend 
smoothly on $\zeta$. We allow for the possibility that the Weyl nodes merge in the limit $\zeta\to \zeta_1^-$. For definiteness, we assume once and for all that 
\begin{equation}\label{pF+pF-}\lim_{\zeta\to\zeta_1^-}(p_F^+-p_F^-)=0.\end{equation} 
If the limit were different from zero, the model could be treated in the same way as if $\zeta\in [\zeta_0,\zeta_1']\subset [\zeta_0,\zeta_1)$, which is a sub-case of the 
problem studied in this paper\footnote{In the `non-singular' case, that 
is, in the case that the two Weyl points are well separated uniformly in $\zeta$, the infrared Renormalization Group analysis of the interacting model simplifies, as we 
shall see below. In particular, the regime $h>h_*$, discussed in Section \ref{sec:1stregime}, disappears, see also the remark after \eqref{eq:cross}.}. 

Let us now discuss the structure of the Bloch Hamiltonian in the vicinity of the Weyl nodes. We have  $a(p_F^\pm)=b(p_F^\pm)=c(p_F^\pm)=d(p_F^\pm)=0$.
If we Taylor expand the Bloch Hamiltonian around $p_F^\omega$, with $\omega\in\{\pm\}$, at first order in $k_1,k_2$ and at second order 
in $k_3-p_{F,3}^\omega$, recalling the parity properties of $a,b,c,d$, we obtain the analogue of \eqref{eq:linear2}, namely
\begin{equation}\label{eq:linear2gen}
\hat H^{0}(k) = \sigma_1(v_1^0 k_1+a_R(k))+\sigma_2(v_2^0 k_2+b_R(k)) +\sigma_3\big[\omega v_{3}^0 (k_3-p_{F,3}^\omega) + \tfrac{b^0}2 (k_3-p_{F,3}^\omega)^2+c_{R,\omega}(k)\big]+d_R(k), \end{equation}
where $v_1^0, v_2^0, v_3^0, b^0$ are real variables, smoothly depending on $\zeta$, and $a_R, b_R, c_{R,\omega}, d_R$ are the remainders of the Taylor expansion (note that $d_R=d$). 
We assume that there exists a constant $c_0\ge 1$ such that
\be \min\{|v^0_1|, |v^0_2|, |b^0|\}\ge c_0^{-1}, \qquad c_0^{-1} |v^{0}_{3}| \leq |p_{F}^{+} - p_{F}^{-}| \leq c_0|v^{0}_{3}|\;,\label{c0boundsbelow}
\ee
uniformly in $\zeta$, for $\zeta\in I$. Thanks to their parity properties of $a,b,c,d$, the remainder terms satisfy the following bounds in the vicinity of $p_F^\omega$:
\begin{eqnarray} && |a_R(k)|\le C |k_1|(k_1^2+k_2^2+|k_3-p_{F,3}^\omega|), \quad 
 |b_R(k)|\le C |k_2|(k_1^2+k_2^2+|k_3-p_{F,3}^\omega|), \nonumber \\
 &&  |c_{R,\omega}(k)|\le C (k_1^2+k_2^2+|k_3-p_{F,3}^\omega|^3), \quad  |d_R(k)|\le C |k_1|\,|k_2|, \label{number.1} \end{eqnarray}
for a constant $C$ that we assume to be independent of $\zeta$. By the `vicinity of $p_F^\omega$', we mean $|k-p_F^\omega|\le 2|p_F^+|$. 
In the complementary region, $\min_\omega|k-p_F^\omega|>2|p_F^+|$, we have: 
\begin{eqnarray} && |a_R(k)|\le C |k_1|\, |k|^2, \quad |b_R(k)|\le C |k_2|\, |k|^2, \label{number.2}\\
 &&  |c(k)-c(0)-\tfrac{b^0}{2}k_3^2|\le  C (k_1^2+k_2^2+k_3^4), \quad  |d_R(k)|\le C |k_1|\,|k_2|, \nonumber\end{eqnarray}
for a constant $C$ that, without loss of generality, can be taken to be the same as in \eqref{number.1}. Note that, in view of the bounds \eqref{c0boundsbelow}, 
\eqref{number.1}, \eqref{number.2}, it is possible to find $c_1>0$ such that
\begin{equation} \min_{\omega}|k-p_F^\omega|\le c_1 \quad \Rightarrow \quad |d(k)|\le \tfrac12\sqrt{a^2(k)+b^2(k)+c^2(k)},\label{number.3}\end{equation}
uniformly in $\zeta\in I$. Moreover, by the smoothness of $\hat H^0(k)$ and the fact that $\det \hat H^0(k)$ vanishes only at $\p_F^\pm$, 
there exists $c_2>0$ such that 
\begin{equation} \min_{\omega}|k-p_F^\omega|\ge c_1 \quad \Rightarrow \quad -\det \hat H^0(k)\ge c_2, \label{number.4}\end{equation}
uniformly in $\zeta\in I$.
Under these assumptions, we will be able to construct and analyze the ground state of a many-body interacting version of $H^0$, uniformly in $\zeta$, for $\zeta\in I$. 
In particular, our analysis will be valid in the limit $\zeta\to \zeta_1^-$ as the Weyl nodes merge. 

For other examples of models in this class, in addition to the one discussed in the previous section, see \cite{D1,Ho}.

\subsection{The interacting model}

Let us now include the many-body interaction. We describe the system in the grand-canonical setting, in second quantization. The fermionic Fock space is defined as $\mathcal{F}_{L} = \bigoplus_{N\geq 0} \frak{h}_{L}^{\ \wedge N}$, with $\wedge$ the antisymmetric tensor product. For all $(x, r) \in \Lambda_{L} \times \{1,2\}$, we consider fermionic creation operators $a^{+}_{x,r}:  \frak{h}_{L}^{\wedge N}\to   \frak{h}_{L}^{\wedge N+1}$ and annihilation operators $a^{-}_{x,r}:  \frak{h}_{L}^{\wedge N}\to   \frak{h}_{L}^{\wedge N-1}$
verifying the canonical anticommutation relations:  $\{ a^{+}_{x, r}\,, a^{-}_{ y, r'} \} = \delta_{x,y} \d_{r,r'}$, $ \{ a^{+}_{x, r}\, , a^{+}_{y, r'} \} = \{ a^{-}_{x, r}\, , a^{-}_{y, r'} \} = 0$. The fermionic operators are compatible with the periodic boundary conditions of the model. Their Fourier transforms are defined, for $k\in \mathcal{B}_{L}$, as:
\begin{equation}
\hat a^{\pm}_{k,r} = \sum_{x\in \Lambda_{L}} e^{\mp ik\cdot x} a^{\pm}_{x,r} \iff a^{\pm}_{x,r} = \frac{1}{L^{3}} \sum_{k\in \mathcal{B}_{L}} e^{\pm ik\cdot x} \hat a^{\pm}_{k,r}\;.
\end{equation}
The many-body Hamiltonian is:
\be\label{eq:H}
\mathcal{H}_{L} = \sum_{x, y\in \Lambda_{L}}\sum_{r,r' \in \{1,2\}} a^{+}_{x,r} H^{0}_{r,r'}(x,y)a^{-}_{y, r'} + \lambda \sum_{x, y \in \Lambda_{L}}\sum_{r, r' \in \{1,2\}} (\rho_{x,r}-\tfrac12)  w_{r,r'}(x,y) (\rho_{y,r'}-\tfrac12)  - \nu \mathcal{N}_{L}
\ee
where $\rho_{x,r} = a^{+}_{x, r} a^{-}_{x, r}$ is the density operator and $\mathcal{N}_{L} = \sum_{x\in \Lambda_{L}} (\rho_{x,1} - \rho_{x,2})$ is the staggered number operator. 
The first term in the right side of \eqref{eq:H} is the hopping term, defined in terms of an $H^0$ satisfying the properties listed in the previous section. 
The second term is the many-body interaction, defined in terms of a $w_{r,r'}(x,y)$, which we assume to be 
even, short-ranged, translational invariant, $w_{r,r'}(x,y) = w_{r,r'}(y-x)$, and 
periodic over $\Lambda_L$. As for $H^0(x,y)$, we assume that $w(x,y)$ is $L$-independent, up to the periodicity condition. 
Also, we suppose that the interaction is invariant under the reflections {\it (iv)} and {\it (v)} above: that is, we require that $w_{r,r'}(x)$ is invariant under $x\to (x_1,x_2,-x_3)$ and under\footnote{Since we assumed that $w_{rr'}$ is even, these two reflection symmetries also imply the invariance of $w_{r,r'}$ under $x\to (x_1,-x_2,x_3)$} $x\to (-x_1,x_2,x_3)$.
The interaction strength $\lambda$ will be assumed to be small, compared with the bandwidth $\max_k\|H^0(k)\|$, uniformly in the system size and in the choice of 
the parameter $\zeta$. The $-1/2$ appearing in the factors $(\rho_{x,r}-\tfrac12)$ and $(\rho_{y,r'}-\tfrac12)$ correspond to a specific, convenient, choice of the 
chemical potential. Finally, the third term in the right side of \eqref{eq:H} is a staggered chemical potential,
with $\nu \equiv \nu(\lambda)$ being a free parameter, such that $\nu(0) = 0$, to be chosen in such a way that the Fermi points of the interacting theory do not move as the interaction varies: they will coincide with those of the noninteracting theory, $p_{F}^{\pm}$, in a sense to be made precise later. 

The Gibbs state of the model is:
\be\label{defGibbs}
\langle \cdot \rangle_{\beta,L} = \tr_{\mathcal{F}_{L}} \cdot \rho_{\beta,L}\;,\qquad \rho_{\beta,L} = \frac{e^{-\beta \mathcal{H}_{L}}}{\mathcal{Z}_{\beta,L}}\;,\qquad \mathcal{Z}_{\beta,L} = \Tr_{\mathcal{F}_{L}}e^{-\beta \mathcal{H}_{L}}\;.
\ee
Given a local observable $\mathcal{O}_{x}$, even in the fermionic fields, we denote by $\mathcal{O}_{\xx} := e^{x_{0} \mathcal{H}_{L}} \mathcal{O}_{x} e^{-x_{0}\mathcal{H}_{L}}$ its imaginary-time evolution, for $\xx = (x_{0}, x)$ and $x_{0} \in [0,\beta)$. We denote by $\hat O_{\pp}$ its space-time Fourier transform,
\begin{equation}
\hat{\mathcal{O}}_{\pp} = \int_{0}^{\beta} dx_{0}\, \sum_{x\in \Lambda_{L}} e^{-i\kk\cdot \xx} O_{\xx}\;,
\end{equation}
with $\pp = (p_{0}, p)\in \mathcal{M}_{\beta}^{\text{b}} \times \mathcal{B}_{L}$ and where $\mathcal{M}_{\beta}^{\text{b}} = \frac{2\pi}{\beta} \mathbb{Z}$ is the set of bosonic Matsubara frequencies. If, instead, $\mathcal{O}_{x}$ is an operator which is odd in the fermionic fields, such as $a^{-}_{x}$, 
similar definitions and formulas hold; however, in the definition of Fourier transform, the set of bosonic Matsubara frequencies is replaced by the set of fermionic Matsubara frequencies, $\mathcal{M}_{\beta}^{\text{f}} = \frac{2\pi}{\beta} (\mathbb{Z} + \frac{1}{2})$. Given $\kk\in
\mathcal{M}_{\beta}^{\text{f}} \times \mathcal{B}_{L}$, we denote the space-time Fourier transforms of the fermionic operators as:
\begin{equation}
\hat a^{\pm}_{\kk, r} = \int_{0}^{\beta} dx_{0} \sum_{x\in \Lambda_{L}} e^{\mp i\kk \cdot \xx} \hat a^{\pm}_{\xx, r}\;.
\end{equation}
Later, we shall be interested in the Schwinger correlation functions of the model, defined, for $x_{0,i} \in [0,\beta)$, as:
\begin{equation}\label{eq:Sc}
\langle {\bf T} a^{\varepsilon_{1}}_{\xx_{1}, r_{1}}\cdots a^{\varepsilon_{n}}_{\xx_{n}, r_{n}} \rangle_{\beta, L}\;,
\end{equation}
with ${\bf T}$ the fermionic time-ordering, ordering the operators in decreasing imaginary-time order, see, e.g., \cite[Eq.(3.3)]{GMPhall}. We extend (\ref{eq:Sc}) anti-periodically to all imaginary times, $x_{0,i}\in \mathbb{R}$.  Of particular interest is the two-point Schwinger function,
$$S_{2; r_{1}, r_{2}}^{\beta, L}(\xx, \yy) = \langle {\bf T} a^{-}_{\xx, r_{1}} a^{+}_{\yy, r_{2}} \rangle_{\beta, L},$$
together with its thermodynamic and zero temperature limit\footnote{\label{betaLlim}Whenever we write or refer to the limit $\beta,L\to\infty$, 
we mean $L\to \infty$ first, then $\beta\to \infty$.}
\begin{equation} S_{2;r_1,r_2}(\xx, \yy) = \lim_{\beta,L\to\infty}\langle {\bf T} a^{-}_{\xx, r_{1}} a^{+}_{\yy, r_{2}} \rangle_{\beta, L},\label{intpropx}\end{equation}
which is translationally invariant, whenever it exists. We also denote by $\hat S_{2;r_1,r_2}(\pp)$ its Fourier transform. 

In the absence of interaction, $\lambda = \nu=0$, the two-point function (thought of as a $2\times2$ matrix of elements $S_{2;r_1,r_2}$) 
can be written, for $x_{0} - y_{0} \notin \beta\mathbb{Z}$, as:
\begin{equation}\label{eq:2pt}
S^{\beta, L}_{2}(\xx;\yy)|_{\lambda = 0}  = \frac{1}{\beta L^{3}} \sum_{k_{0} \in \mathcal{M}^{\text{f}}_{\beta}} \sum_{k \in \mathcal{B}_{L}} \frac{e^{i\kk\cdot (\xx - \yy)}}{-ik_{0} + \hat H^{0}(k)} \equiv g_{\beta, L}(\xx,\yy)\;.
\end{equation}
We refer to $g_{\beta,L}$ as the free propagator of the model, we set $g(\cdot) = \lim_{\beta, L\to \infty} g_{\beta, L}(\cdot)$, and we denote 
by $\hat g$ its Fourier transform. Let $\pp_{F}^{\omega} := (0, p_{F}^{\omega})$. 
In the absence of interactions, as $\beta, L\to \infty$, the Fourier transform of the free propagator reads, 
for $\kk = \kk' + \pp_{F}^{\omega}$ and $\kk'$ small,
\begin{equation}\label{pp}
\hat g(\kk'+\pp_{F}^{\omega}) = \begin{pmatrix} -ik_{0} + \omega v^{0}_{3} k'_{3} &  v^{0}_{1} k'_{1} -i v^{0}_{2} k'_{2} \\  v^{0}_{1} k'_{1} +i v^{0}_{2} k'_{2} & -ik_{0} - \omega v^{0}_{3} k'_{3}   \end{pmatrix}^{-1} (1 + R_{\omega}^{0}(\kk'))\;,
\end{equation}
with $\| R^{0}_{\omega}(\kk') \| \leq C |\kk'|$. We see that $g_{\omega}(\kk') := g(\kk' + \pp_{F}^{\omega})$ agrees at leading order with the propagator of chiral 
relativistic fermions, with chirality $\omega = \pm$, and anisotropic velocities. We will prove that, for $\lambda$ small, by fixing $\nu=\nu(\lambda)$ appropriately, 
the singularities of the Fourier transform of the interacting propagator are located at the same points, $\pp_F^\pm$, as the non-interacting one; moreover, at those points, 
it has the same singularity structure: 
\begin{equation}\label{pp_int}
\hat S_2(\kk'+\pp_{F}^{\omega}) =\frac1{Z} \begin{pmatrix} -ik_{0} + \omega v_{3} k'_{3} &v_{1} k'_{1}  -i v_{2} k'_{2} \\ v_{1} k'_{1} +i v_{2} k'_{2}  & -ik_{0} - \omega v_{3} k'_{3}   \end{pmatrix}^{-1} (1 + R_{\omega}(\kk'))\;,
\end{equation}
with $v_j=v^0_j(1+O(\lambda))$, $j=1,2,3$, the interacting Fermi velocities, $Z=1+O(\lambda)$ the wave function renormalization and, for any $\theta\in(0,1)$, the remainder satisfies $\| R_{\omega}(\kk') \| = O(|\kk'|^\theta)$, 
non-uniformly in $\theta$ as $\theta\to 1^-$ (and, possibly, non-uniformly in the distance between the Weyl nodes). 

\subsection{Coupling to an external gauge field}\label{sec:coupling}

Our analysis will focus on the transport properties of the model, after introducing an external electromagnetic field. The coupling of the model with an external vector potential is defined via the Peierls substitution. This means that, both in the Hamiltonian and in the physical observables, any product of fermionic operators $a^{+}_{x,r} a^{-}_{y,r'}$ is replaced by:
\begin{equation}\label{eq:peierls}
a^{+}_{x,r} a^{-}_{y,r'} \quad \longrightarrow\quad  a^{+}_{x,r} a^{-}_{y,r'} e^{i \int_{x\to y} A\cdot d\ell }\;,
\end{equation}
where $A_x\in\mathbb R^3$, with $x\in Q_L:=(\mathbb R/L\mathbb Z)^3$, and 
$\int_{x\to y}A\cdot d\ell$ denotes the line integral:
\begin{equation} 
\int_0^1 A_{x+s(y-x)}\cdot (y-x)\, ds\;.
\end{equation}
We use the following convention for the Fourier transform of $A$: if $p\in \tfrac{2\pi}{L}\mathbb Z^3$, 
$$\hat A_p=\int_{Q_L} A_x e^{ip\cdot x} dx \quad \Longleftrightarrow \quad A_x=\frac1{L^3}\sum_{p\in \frac{2\pi}{L}\mathbb Z^3}\hat A_p e^{-ip\cdot x}.$$
Notice that the many-body interaction is not affected by the presence of the gauge field, due to the fact that the density operator is gauge invariant. 

Let us denote by $\mathcal{H}^{0}_{L}(A)$ the free Hamiltonian ($\lambda=\nu=0$) in the presence of the gauge field. The charge current operator is defined as:
\begin{equation}
\hat J_{j,p}(A) := - L^3\frac{\partial \mathcal{H}^{0}_{L}(A)}{\partial \hat A_{j,p}}\;.
\end{equation}
We also set $\hat J_{j,p} \equiv \hat J_{j,p}(0)$. A straightforward computation gives, letting $\eta_{p}(\delta) := \frac{1- e^{-ip\cdot \delta}}{ip\cdot \delta}=
\int_0^1ds\, e^{-is p\cdot\delta}$, 
\begin{equation}\label{eq:curr}
\hat J_{j,p} = \frac{1}{L^{3}}\sum_{k\in \mathcal{B}_{L}}\hat a^{+}_{k+p} \Big[-i\sum_{\delta\in \Lambda_{L}} e^{-ik\cdot \delta} \delta_{j}  H^{0}(\delta) \eta_{p}(\delta)\Big]\hat a^{-}_{k} 
\equiv \frac{1}{L^{3}}\sum_{k\in \mathcal{B}_{L}} \hat a^{+}_{k+p} \hat {\mathfrak J}_{j}(k,p) \hat a^{-}_{k}, \end{equation}
where $\hat {\mathfrak J}_{j}(k,p)$ should be understood as a $2\times 2$ matrix, and $a^+_k$ (resp. $a^-_k$) as a row (resp. column) vector of components $a^+_{k,1}, a^+_{k,2}$ 
(resp. $a^-_{k,1}, a^-_{k,2}$). Note that, in the thermodynamic limit,  the kernel $\hat {\mathfrak J}_{j}(k,p)$ can be written more explicitly as 
\begin{equation} \hat {\mathfrak J}_{j}(k,p)=\int_0^1ds\,  \partial_{k_j}\hat H^0(k+sp).\end{equation}
In particular, recalling \eqref{eq:linear2gen}, the kernel of the current at $k=p_F^\omega$ and $p=0$ reads: 
\begin{equation} \label{relcurr}\hat {\mathfrak J}_{j}(p_F^\omega,0)=\begin{cases} v_j^0\sigma_j & \text{if $j=1,2$,}\\
\omega v_3^0\sigma_3 & \text{if $j=3$.}\end{cases}\end{equation}
Equivalently, in real-space, 
\begin{equation}\label{eq:real}
J_{j,z} = \frac{1}{L^{3}} \sum_{p\in  \frac{2\pi}{L}\mathbb Z^3} e^{ip\cdot z} \hat J_{j,p}= \sum_{x,y\in \Lambda_{L}} a^{+}_{x} {\mathfrak J}_{j}(y-x, z-x) a^{-}_{y}\;,
\end{equation}
where ${\mathfrak J}_{j}(y, z)=L^{-6}\sum_{k\in \mathcal B_L}\sum_{p\in\frac{2\pi}{L}\mathbb Z^3} \hat{\mathfrak J}_{j}(k,p) e^{ik\cdot y}e^{ip\cdot z}$. In the thermodynamic limit, 
\begin{equation} {\mathfrak J}_{j}(y, z)=-iy_j H^0(y)\int_0^1 ds \, \delta(z-sy),\end{equation}
where $\delta(z-sy)$ is a Dirac delta. 
Note that the current satisfies a lattice continuity equation. In fact:
\begin{equation}\label{eq:contlat}
p\cdot \hat J_{p} =\frac{1}{L^{3}} \sum_{k\in \mathcal{B}_{L}} \hat a^{+}_{k+p} \sum_{\delta \in \Lambda_{L}} e^{-ik\cdot \delta} H^{0}(\delta) (e^{-ip\cdot \delta} - 1) \hat a^{-}_{k}
=\frac{1}{L^{3}} \sum_{k\in \mathcal{B}_{L}} \hat a^{+}_{k+p} (\hat H^{0}(k+p) - \hat H^{0}(k)) \hat a^{-}_{k}\;,
\end{equation}
which can be rewritten as
\begin{equation}\label{eq:cons}
\partial_{x_{0}} \hat \rho_{(x_{0}, p)} = p\cdot \hat J_{(x_0,p)}\;.
\end{equation}
Here $\hat \rho_{(x_0,p)}=e^{x_0\mathcal H_L}\hat \rho_pe^{-x_0\mathcal H_L}$, with $\hat \rho_p=L^{-3}\sum_{k\in\mathcal B_L}\hat a^+_{k+p}\hat a^-_{k}$. Similarly, $\hat J_{(x_0,p)}=e^{x_0\mathcal H_L}\hat J_pe^{-x_0\mathcal H_L}$. 
We collect the density and the components of the lattice current to form a Euclidean $4$-current $(-i\hat \rho_{p}, \hat J_{p})$, whose components are denoted by $\hat J_{\mu,p}$, $\mu \in \{0,1,2,3\}$:
\begin{equation}
\hat J_{\mu, p} = \frac{1}{L^{3}} \sum_{k\in \mathcal{B}_{L}} \hat a^{+}_{k+p} \hat{\mathfrak J}_{\mu}(k,p) \hat a^{-}_{k}\;,
\end{equation} 
where $\hat{\mathfrak J}_{0} = -i\mathds 1_2$ and $\hat{\mathfrak J}_{j}$, $j=1,2,3$, are given by (\ref{eq:curr}). 

\subsection{The lattice chiral current}\label{sec:couplingqq}
We now introduce a lattice current for the quasi-particle flow between the Weyl nodes, playing the role of the `chiral current' for
our lattice model:
\begin{equation}\label{eq:J5def}
\hat J^{5}_{\mu,p} = \frac{Z^{5}_{\mu,\text{bare}}}{L^{3}} \sum_{k\in \mathcal{B}_{L}} \hat a^{+}_{k+p} \hat{\mathfrak J}^{5}_{\mu}(k,p) \hat a^{-}_{k},\end{equation}
where $p\in \mathcal B_L$ and, letting $\sigma_0:=-i\mathds 1_2$, 
\begin{equation}\hat{\mathfrak J}^{5}_{\mu}(k,p) =\begin{cases} \frac{\sin k_{3} + \sin (k_{3} + p_{3})}{2 \sin(p_{F,3}^+)}\sigma_\mu & \text{if $\mu=0,1,2$,}\\
\phantom{ciaciao}\sigma_3 & \text{if $\mu=3$.}\label{eq:J5def.2}\end{cases}
\end{equation}
Note that the kernel $Z_{\mu,\text{bare}}^5\hat{\mathfrak J}_\mu^5(k,p)$ of the chiral current at $k=p_F^\omega$ and $p=0$ reads:
\begin{equation} Z^5_{\mu,\text{bare}}\hat {\mathfrak J}_{\mu}^5(p_F^\omega,0)=\begin{cases} \omega Z^5_{\mu,\text{bare}} \sigma_\mu & \text{if $\mu=0,1,2$,}\\
Z^5_{3,\text{bare}}\sigma_3 & \text{if $\mu=3$.}\end{cases}\label{relcurr.chi}\end{equation}
Comparing with \eqref{relcurr}, we see that 
at the Weyl nodes the different components of the chiral current behave like those of the total current, 
up to an additional `chirality sign' $\omega=\pm$ and different multiplicative factors. This implies that, in the `infrared regime' of $p$ small and $k$ close to the Weyl nodes, 
the `chiral density' $\hat J^{5}_{0,p}$ defined by \eqref{eq:J5def} with $\mu=0$ reduces to 
the difference between the quasi-particle densities around the Weyl nodes, up to a multiplicative pre-factor; similarly, the spatial components of the chiral current 
reduce to the difference between the quasi-particle currents around the Weyl nodes. Making a parallel with QED$_4$, this infrared chiral current coincides 
with the standard QED$_4$ axial current with different velocities. 

The multiplicative normalizations $Z^5_{\mu,\text{bare}}$ are fixed as follows. Let us define the vertex functions 
\begin{eqnarray}
\hat \G_{\mu}(\kk,\pp) &=& \lim_{\beta,L\to\infty}\frac{1}{\beta L^{3}} \langle {\bf T} \hat J_{\mu,\pp}\,; \hat a^{-}_{\kk+\pp} \hat a^{+}_{\kk}\rangle_{\beta,L}\;,\label{Gammanon}\\
\hat \G^{5}_{\mu}(\kk,\pp) &=&\lim_{\beta,L\to\infty} \frac{1}{\beta L^{3}} \langle {\bf T} \hat J^{5}_{\mu,\pp}\,; \hat a^{-}_{\kk+\pp} \hat a^{+}_{\kk} \rangle_{\beta,L},
\label{Gammachir}\end{eqnarray}
which are $2\times 2$ matrices in the color indices of the fermionic operators, for each choice of $\mu\in\{0,1,2,3\}$. Let us also introduce their 
`amputated' versions,
\begin{eqnarray}
\hat \gamma_{\mu}(\kk,\pp) &=&\big[\hat S_2(\kk+\pp)\big]^{-1}\hat \G_{\mu}(\kk,\pp)\big[\hat S_2(\kk)\big]^{-1},\label{amput.1}\\
\hat \gamma^5_{\mu}(\kk,\pp) &=& \big[\hat S_2(\kk+\pp)\big]^{-1}\hat \G^5_{\mu}(\kk,\pp)\big[\hat S_2(\kk)\big]^{-1},\label{amput.2}
\end{eqnarray}
where we recall that $\hat S_2(\kk)$ is the interacting propagator, see \eqref{intpropx} and \eqref{pp_int}. 
Existence of the limits in \eqref{Gammanon}-\eqref{Gammachir}, in the sense of footnote \ref{betaLlim} above, is part of the main results of this paper, stated in the following section. 
We will impose that the amputated vertex functions agree in the infrared limit, up to a sign: that is, if $\pp_F^\omega$ is the $4$-vector $\pp_F^\omega=(0,p_F^\omega)$,
\begin{equation}\label{z6}
\lim_{\substack{\kk\to \pp_{F}^{\omega} \\ \pp\to {\bf 0}\ }} \hat \gamma^{5}_{\mu}(\kk,\pp)\big[\hat \gamma_{\mu}(\kk,\pp)\big]^{-1} = \omega\mathds 1_2,
\end{equation}
where, for technical convenience, we take the limit in such a way that $\kk-\pp_F^\omega,\pp, \kk+\pp-\pp_F^\omega$ are all of the same order as they tend to zero.
The normalization condition \eqref{z6} can be interpreted as the requirement that, at the level of the amputated vertex correlation functions, for momenta close to a Weyl node,
the insertion of the operator $\hat J^{5}_{\mu,\pp}$ is equivalent to the insertion of $\omega \hat J_{\mu,\pp}$, where $\omega$ is the chirality of the Weyl node.

Finally, we denote by $J_{\mu}^{5}(A)$ the chiral current coupled to the external gauge field, defined again via the Peierls substitution
\eqref{eq:peierls}. We have:
\begin{equation}\label{eq:J5mu}
J^{5}_{\mu, z}(A) = Z^{5}_{\mu,\text{bare}}\sum_{x,y\in \Lambda_{L}} a^{+}_{x} {\mathfrak J}^{5}_{\mu}(y-x,z-x)e^{i\int_{x\to y} A \cdot d\ell} a^{-}_{y}\;,
\end{equation}
where $ {\mathfrak J}^{5}_{\mu}(y-x,z-x)$ is the inverse Fourier transform of $\hat {\mathfrak J}^{5}_{\mu}(k,p)$.

\medskip

{\bf Remarks.}

\medskip

\noindent1) The limiting values of $\hat \gamma_{\mu}(\kk,\pp)$
and of $\hat \gamma^5_{\mu}(\kk,\pp)$
 as $\kk\to\pp_F^\omega$ and $\pp\to{\bf 0}$ divided by $Z v_\mu$, with $Z$ the wave function 
renormalization and $v_\mu$ the dressed velocity\footnote{We use the  convention that $v_0=1$.}, see \eqref{pp_int}, have the physical meaning of (dressed) 
charges of the quasi-particles around the Weyl nodes.  
Gauge invariance ensures that the charge of quasi-particles associated with the electromagnetic current
is not renormalized by the interaction, that is, its dressed and bare values coincide, see \eqref{asympt.2_intro} below. In contrast, the charge associated with the chiral 
current is renormalized non-trivially: this should come as no surprise, because the model is not invariant under chiral gauge transformations and, therefore, there is no 
symmetry protecting such a charge from being dressed by the interaction. The choice of $Z_{\mu,\text{bare}}^5$ is used to impose, via \eqref{z6}, the correct physical normalization of the chiral current, 
namely the one guaranteeing that the charges of the quasi-particles around each Weyl node computed either via the electromagnetic or the chiral current are the same.
An analogous phenomenon takes place in QED$_4$, where it is well known that the axial vertex renormalization is different from the vectorial one, see e.g. \cite[eq.(2)]{A1}: the discrepancy between 
the values of the two vertex renormalizations is a manifestation of the chiral anomaly, since a formal use of chiral gauge invariance would naively suggest their identification.

\medskip

\noindent2) The definition of the components of the chiral current is largely arbitrary, as long as their kernels have 
the right asymptotic form at the Weyl nodes, see \eqref{relcurr.chi}, and the right discrete symmetries, discussed in Appendix \ref{appA.1}.
Changing the specific definition of the chiral current would only affect the specific value of the parameters $Z_{\mu,\text{bare}}^5$.
In view of the universality result we prove, this arbitrariness does not affect the quadratic response at dominant order.

\medskip

\noindent3) The introduction of a lattice current for Weyl semimetals, generalizing the proposal
of \cite{2a} (see also \cite{HW98}) to lattice interacting models, is an original contribution of this paper.
Despite the formal similarity between the electromagnetic and the chiral lattice current, it is important to highlight some basic differences.
In the infrared limit ($p$ small and $k$ close to 
the Weyl nodes), they reduce to the electromagnetic and axial currents of QED$_4$, respectively. In QED$_4$, both these currents are conserved, and their
conservation is associated with basic gauge symmetries of the model (total and chiral, respectively). On the contrary, in our lattice realization, only the electromagnetic current is 
associated with a gauge symmetry, from which exact Ward Identities between correlations follow: these imply, in particular, that the dressed charge is not renormalized, see 
\eqref{asympt.2_intro} below, and that the $4$-divergence of the current vanishes even in the presence of an external electromagnetic field, see, e.g., \eqref{eq:WIfin2} below. 
Neither of these properties holds for the lattice chiral current.

\medskip

\subsection{Main result: condensed matter simulation of the chiral anomaly}

We are interested in the response of the expectation of the chiral $4$-current $\hat J^{5}_{\mu,p}$ to an adiabatic external gauge field of the form  $A_{x}(t) = e^{\eta t} A_{x}$, where $\eta > 0$ plays the role of adiabatic parameter. We denote by $\mathcal{H}_{L}(A({t}))$ the Hamiltonian of the interacting system, coupled to the external gauge field via the Peierls substitution (\ref{eq:peierls}). We will consider the time-evolution of the many-body system from $t = -\infty$ to $t = 0$. 

Let $\langle \cdot \rangle_{\beta,L;t} \equiv \tr \cdot \rho(t)$ be the time-dependent state of the system, where $\rho(t)$ is the solution of the von Neumann equation 
$i\partial_{t} \rho(t) = [ \mathcal{H}_{L}(A({t})), \rho(t) ]$, with boundary condition $\rho(-\infty) = \rho_{\beta, L}$. Let $\langle \hat J^{5}_{\mu,p}(A({t})) \rangle_{\beta,L;t}$ 
be the time-dependent average of the chiral lattice current. As discussed in the introduction, we focus our attention on the {\it quadratic} variation of this quantity with respect to
the external field, denoted by  $\langle \hat J^{5}_{\mu,p}(A({t})) \rangle^{(2)}_{\beta,L;t}$:
\begin{equation}\label{eq:J5av}
\langle \hat J^{5}_{\mu,p}(A({t})) \rangle_{\beta,L;t}^{(2)} = \frac{1}{2} \sum_{i,j=1}^3\sum_{p_{1},p_{2}\in \mathcal{B}_{L}} \hat A_{i,p_{1}}(t) \hat A_{j,p_{2}}(t) 
\delta_{p_{1} + p_{2}+p,0} \hat\Pi^{5;\beta,L}_{\mu,i,j}(\pp_{1}, \pp_{2})\;,
\end{equation}
with $\pp_{i} = (\eta, p_{i})$. The quadratic response $\hat\Pi^{5;\beta,L}_{\mu,i,j}(\pp_{1}, \pp_{2})$ is the analogue, in our condensed matter context, 
of the chiral anomaly in QED$_4$, see \eqref{eq:CS} and following discussion for more details. It can be expressed in terms of equilibrium correlation 
functions, via second-order time-dependent perturbation theory; see Appendix \ref{app:2nd}, Eq.(\ref{eq:secondcomm}). A rigorous application of the Wick rotation, proven 
in Appendix \ref{app:2nd}, allows us to express the $\beta,L\to\infty$ limit of this quantity in terms of Euclidean correlation functions. Let $\hat\Pi^{5}_{\mu,i,j}(\pp_1,\pp_2) = 
\lim_{\beta, L\to \infty} \hat\Pi^{5;\beta,L}_{\mu,i,j}(\pp_1,\pp_2)$. 
For $\pp_{1} = (\eta, p_{1})$, $\pp_{2} = (\eta, p_{2})$, $\pp = -\pp_{1} - \pp_{2}$, we have, see Eq.(\ref{eq:G5wick}):
\begin{equation}\label{eq:euclide}
\hat\Pi^{5}_{\mu, i, j}(\pp_{1}, \pp_{2}) = \pmb{\langle} {\bf T}\, \hat J^{5}_{\mu,\pp}\;; \hat J_{i,\pp_{1}}\;; \hat J_{j,\pp_{2}} \pmb{\rangle}_{\infty} + \text{Schwinger terms,}
\end{equation}
where $\pmb{\langle} \cdot \pmb{\rangle}_{\infty} = \lim_{\beta, L\to \infty} (\beta L^{3})^{-1} \langle \cdot \rangle_{\beta, L}$. We refer the reader to 
Eq.(\ref{eq:G5wick}) for the precise form of the Schwinger terms. More generally, we denote by $\hat\Pi_{\mu,\nu,\sigma}^{5}(\pp_{1},\pp_{2})$ the extension of (\ref{eq:euclide}) to space-time current insertions, the labels $\nu, \sigma = 0$ corresponding to insertions of a density operator $\hat J_{0,\pp_{i}}$.

The next theorem gives the explicit expression of the interacting quadratic response of the chiral current, $\hat\Pi_{\mu,\nu,\sigma}^{5}(\pp_{1},\pp_{2})$, 
in the limit of low momenta.

\begin{thm}\label{thm:main} Suppose that the non interacting part of the Hamiltonian satisfies the hypotheses of Section \ref{sec:2.1}, namely: symmetries (i) to (iv)
and Eqs.\eqref{pF+pF-} to \eqref{c0boundsbelow}. 
There exists $\lambda_{0}>0$, independent of the choice of $\zeta$ in $I$, such that, for all $\zeta\in I$, there exist functions $\nu,Z_{\mu,\text{bare}}^5,v_j,{Z}$, 
analytic in $|\lambda|\le \lambda_0$, for which \eqref{pp_int} and \eqref{z6} hold. Moreover, for $\eta>0$, four-vectors $\pp_{1} = (\eta, p_{1})$, $\pp_{2} = (\eta, p_{2})$ such that 
$P:=\max\{|\pp_1|,|\pp_2|\}\le |p_F^+|$, and $\pp = -\pp_{1} - \pp_{2}$:
\begin{equation}\label{eq:main}
\sum_{\mu=0}^3 p_{\mu} \hat\Pi^{5}_{\mu,\nu,\sigma}(\pp_{1}, \pp_{2}) = -\frac{1}{2\pi^{2}}\sum_{\alpha,\beta=0}^3 p_{1,\alpha} p_{2,\beta} \varepsilon_{\alpha\beta\nu\sigma}+ R^{5}_{\nu,\sigma}(\pp_{1}, \pp_{2})\;,
\end{equation}
where $\varepsilon_{\alpha\beta\nu\sigma}$ is the four dimensional Levi-Civita symbol\footnote{\label{LCsymbol}We use the convention that $\varepsilon_{0123}=1$, and $\varepsilon_{\pi_0,\pi_1,\pi_2,\pi_3}=(-1)^{\pi}$ for $\pi=(\pi_0,\pi_1,\pi_2,\pi_2)$ a permutation of $(0,1,2,3)$, with $(-1)^\pi$ the sign of the permutation. 
As usual, $\varepsilon_{\alpha\beta\nu\sigma}=0$ 
whenever the sequence $(\alpha,\beta,\nu,\sigma)$ has at least a repetition.} and
$|R^{5}_{\nu,\sigma}(\pp_{1}, \pp_{2})| \leq C_{\theta,\zeta} P^{2+\theta}$, for any $\theta\in(0,1)$, $\zeta\in I$ and a suitable $C_{\theta,\zeta}>0$. 
\end{thm}
The proof of this theorem, presented in the rest of this paper, provides a constructive algorithm for computing the functions $\nu, Z^5_\mu, v_j$, as well as 
the $\beta,L\to\infty$ limit of the Euclidean correlation functions of the system. The construction and proof of analyticity of the staggered chemical potential $\nu$ 
and of the Fermi velocities $v_j$, $j=1,2,3$, uniformly in $\zeta$, is not new, see \cite{M1,M1bis}. 
An explicit computation at the level of first order 
perturbation theory shows that the interacting Fermi velocities $v_j$, $j=1,2,3$, are non-trivial functions of $\lambda$ and of 
all the other parameters entering the definition of the Hamiltonian, generically in the choice of $H^0$ and $w$. 
In this sense, the Fermi velocities are non-universal quantities: their value depends on the details of the microscopic Hamiltonian. 
The same is true for the longitudinal Kubo conductivity, see \cite{M1}, and for several other physical observables. 

The important new piece of information contained in Theorem \ref{thm:main} is Eq.\eqref{eq:main}, which shows that 
the quadratic response of the chiral current is {\it universal} in the low momentum limit. It plays the same role as the quantized transverse conductivity in 
quantum Hall fluids. 
Remarkably, \eqref{eq:main} is valid for all the choices of the parameter $\zeta$ controlling the distance among the Weyl points; it is valid, in particular, 
arbitrarily close to the critical point $\zeta=\zeta_1$ where the Weyl points merge and the valence and conduction bands touch quadratically, rather than linearly. 
The situation where the Weyl nodes are very close and the 
quasi-particle velocities are very small is very common in the actual experimental realization of Weyl semimetals, see e.g. \cite[Fig.2 and 3]{AMV} and \cite[Fig.1]{DHM13}. In such a situation,
the natural parameter measuring the strength of the interaction is the coupling strength divided by the quasi-particle velocity; surprisingly, even if this parameter becomes huge, 
the interaction still appears not to affect the anomaly coefficient.

Note that the universality of the `chiral anomaly' in \eqref{eq:main} holds, provided that the lattice chiral current verifies the condition \eqref{z6}.
The subtle cancellation underlying our universality result can also be stated in an alternative, yet equivalent, way: let the chiral current be defined as in \eqref{eq:J5def}, but without apriori
fixing $Z^5_{\mu,\text{bare}}$ in any special way; then the divergence of the quadratic response coefficient $\hat\Pi^5_{\mu,\nu,\sigma}$ {\it divided} by the ratio of the axial and vector vertex renormalizations
(i.e., by the left side of \eqref{z6} times the chirality index $\omega$) is universal at leading order.
The fact that the quadratic response of the axial current to the electromagnetic field {\it divided} by the axial renormalization is universal (and not the quadratic response itself, as often 
stated incorrectly) is the very content of the Adler-Bardeen theorem in massless QED$_4$, see the discussion in \cite{A1} after Eqs.(6) and (7). Indeed, a naive perturbative computation of the chiral anomaly in QED$_4$ would apparently give rise to higher order corrections beyond the chiral triangle graph, see \cite{AI}: however, as discussed in \cite{A1}, these corrections are cancelled exactly by the axial vertex renormalization, which is {\it different} from the vectorial one.
%The discussion in \cite{A1,A6} is based on perturbative arguments; a complete, non-perturbative proof for interacting lattice models is provided by our result. We believe that the non-perturbative validity of the Adler-
%Bardeen theorem in the presence of a finite lattice cutoff will prove relevant for the construction of other chiral lattice QFTs.}

Note
also that, of course, the quadratic response of the electromagnetic, rather than axial, current
vanishes, by current conservation.

\medskip

In order to make the connection between the quadratic response $\hat\Pi^5_{\mu,\nu,\sigma}(\pp_1,\pp_2)$ and the transport experiments in Weyl semimetals, 
we choose the vector potential to be of compact support in space, and set $N^{5}_{t} = i\lim_{\beta,L\to\infty}\langle \hat J^{5}_{0,0}(A({t})) \rangle_{\beta,L;t}^{(2)}$.
We now want to compute $\partial_t N^{5}_{t}$, which represents the quasi-particle flow from the first Weyl node to the second, at quadratic order in $A$. 
From the construction of the interacting Gibbs state of the system in the $\beta,L\to\infty$ limit, which Theorem \ref{thm:main} is based on, 
one finds that the time derivative commutes with the $\beta,L\to\infty$ limit. 
Moreover, note that, by differentiating \eqref{eq:J5av} with respect to time, the right side simply gets multiplied by $2\eta$, because its 
time dependence is all in $\hat A_{i,p_{1}}(t)=e^{\eta t}\hat A_{i,p_{1}}(0)$ and $\hat A_{j,p_{2}}(t)=e^{\eta t}\hat A_{j,p_{2}}(0)$. 
Note also that $2\eta$ is minus the zero-th component of $\pp=-(2\eta,0)$ (recall that in our case $p=-p_1-p_2=0$). Therefore, by using 
\eqref{eq:main} in the expression obtained from \eqref{eq:J5av} by differentiating with respect to time and by taking $\beta,L\to\infty$, we find: 
\begin{eqnarray}
\label{eq:CS}
\partial_{t} N^{5}_{t}&=& \frac{1}{4 \pi^{2}} \int dx\, \partial_\alpha A_{i,x}(t)\partial_\beta A_{j,x}(t) \varepsilon_{\alpha\beta i j} + \frak{e}_{A}(t)\label{eq:2.40}\\
&=&  \frac{1}{2 \pi^{2}} \int dx\, E_x(t) \cdot B_x(t) + \frak{e}_{A}(t)\nonumber\end{eqnarray}
where, in the first line, summation over repeated indices is understood ($\alpha,\beta$ run from $0$ to $3$, while $i,j$ from $1$ to $3$), and 
$\partial_0$ denotes the time derivative. The error term $\frak{e}_{A}(t)$, due to the error term $R^5_{\nu,\sigma}$ in (\ref{eq:main}), collects contributions involving a higher number of derivatives on the vector potential. It is subdominant for a vector potential slowly varying in space. 
For the second identity, we used the definitions of electric and magnetic fields: $E_{i,x}=-\partial_t A_{i,x}$ 
and $B_{i,x}=\sum_{j,l=1}^3 \varepsilon_{ijl}\,\partial_j A_{l,x}$. 
Recent experiments \cite{o2,Huang,LiKh,o1,Zhang} in different Weyl semimetals reported 
evidence for an anomalous flow of quasi-particles between the Weyl nodes, which causes a negative, highly anisotropic, magnetoresistance. 
For instance, in \cite{o1}, the resistivity tensor was measured for different values of the angle between $E$ and $B$, and the response proved to be strongest for  
$E$ and $B$ parallel, see e.g. \cite[Fig.\,6A and 6B]{o1}, in agreement with the  
$E \cdot B$ dependence in \eqref{eq:CS}. Comparing measures at different angles is a way to access directly the quadratic response, which is the one related to the anomaly.
We stress that the $E\cdot B$ dependence found in real (interacting) Weyl semimetals is the same predicted originally for non-interacting systems \cite{2a}. This is a first important instance of 
universality, even though a precise, quantitative, experimental verification of the interaction-independence of the chiral anomaly coefficient has still to come.

The dominant term in the right side of \eqref{eq:2.40} is the usual chiral anomaly of QED$_4$, 
with the same universal prefactor, irrespective of the presence of the interaction. The interaction independence of the coefficient is the analogue, in our context, 
of the Adler-Bardeen anomaly non-renormalization theorem \cite{A6}, which our result is a rigorous version of. In our context, the lattice plays the role 
of a fixed ultraviolet regularization of an effective QFT model: chiral and Lorentz symmetry are emerging and approximate but nevertheless
the chiral anomaly satisfies the AB non-renormalization property. We expect that our methods can be used to prove a generalization of the Adler-Bardeen's 
theorem for interacting lattice gauge theories at finite cutoff, such as infrared lattice QED$_4$, see \cite{M1tris,M3}.

\section{Proof of Theorem \ref{thm:main}}\label{sec:3}

\subsection{General strategy and core argument of the proof}\label{sec:3.0}

The proof of Theorem \ref{thm:main} is based on the following strategy: 
\begin{enumerate}[label={\arabic*}]
\item\label{it1} We first compute  the generating function of correlations via RG methods; the output of the RG construction is that the 
correlation functions are expressed in terms of a series expansion in $\lambda$ and in a sequence of effective parameters, which is convergent, provided that $|\lambda|$ is sufficiently small and the
effective parameters are uniformly bounded under the iterations of the RG map. 
\item\label{it2} Next, we show that these effective parameters are uniformly bounded, as desired, provided that the staggered chemical potential $\nu$ is chosen appropriately. 
We also show how to fix the bare parameters $Z_{\mu,\text{bare}}^5$ in such a way that \eqref{z6} is verified. 
\item\label{it3} The RG expansion also allow us to 
identify an explicit dominant contribution to the correlations (for momenta close to the Weyl nodes), which can be written explicitly in terms of the `dressed parameters' of the model.
The subdominant contributions to the correlations admit improved dimensional bounds, which imply better regularity properties in momentum space, as compared to their 
dominant counterparts. 
\item\label{it4} Once that the correlations have been written as a dominant contribution, plus a better-behaved remainder, we are in business for proving \eqref{eq:main}: in fact, 
the left side of \eqref{eq:main} can be written as the contribution from the dominant part, which can be computed explicitly, plus a remainder, whose value at small momenta 
is fixed by Ward Identities.
\end{enumerate}

In the next three subsections, Sections \ref{sec3.1.1} to \ref{sec3.1.3}, containing the {\bf core argument} of the proof, we will describe more technically: (1) the outcome of items \ref{it1} to \ref{it3} for the quadratic response $\hat\Pi^{5}_{\mu,\nu,\sigma}$, 
see in particular \eqref{eq:tritetr_intro} below; (2) the proof of our main result, Eq.\eqref{eq:main}, starting from the representation formula \eqref{eq:tritetr_intro} for $\hat\Pi^{5}_{\mu,\nu,\sigma}$.

The main ingredients involved in these steps are the following: 
\begin{enumerate}[label=(\roman*)]
\item The decomposition of the quadratic response into an explicit, non-differentiable, term, proportional to the usual chiral triangle graph of QED$_4$ with momentum cutoff (the function $I_{\mu,\nu,\sigma}$ 
in \eqref{eq:tritetr_intro} below), plus a correction (the function $\widetilde H^5_{\mu,\nu,\sigma}$ in \eqref{eq:tritetr_intro} below). {\bf Note}: the correction is not subdominant;
%' in the sense that it gives a negligible contribution to the chiral anomaly; 
rather, it gives a contribution to the chiral anomaly comparable with the one from the first term. However, since it comes from the integration of the infrared-irrelevant terms, it is {\bf more 
regular} than the first one: more precisely, it is continuously differentiable in the momenta, and its derivative is H\"older continuous (contrary to the first term, whose derivative is discontinuous at $\pp_1=\pp_2={\bf 0}$); this allows us to expand it in Taylor series up to first order included. Let us stress that the decomposition \eqref{eq:tritetr_intro} is, obviously, non-unique: it is natural in a 
Wilsonian RG computation of the correlation functions, the first term being obtained by neglecting irrelevant terms in the RG sense. Due to the presence of a momentum cutoff, this term breaks gauge invariance, 
as apparent from \eqref{eq:I.2_intro} below. Such a term has the correct tensorial structure (i.e., its divergence with respect to the first index is proportional to $\sum_{\alpha,\beta}p_{1,\alpha}p_{2,\beta}
\varepsilon_{\alpha,\beta,\nu,\sigma}$, see \eqref{eq:main}). However, if we neglected the remainder, it would produce an incorrect prefactor in the right side of \eqref{eq:main} (even in the absence of interactions). 
\item The explicit computation of the first term in the right side of \eqref{eq:tritetr_intro}, i.e., of the chiral triangle graph of QED$_4$ with momentum cutoff, leading to \eqref{eq:I.1_intro}-\eqref{eq:I.2_intro} below. 
\item The use of Ward Identities, guaranteeing two crucial facts: (1) the vectorial vertex renormalization is proportional to the Fermi velocity, see \eqref{asympt.2_intro}; (2)
the first order Taylor coefficients of the correction term to the quadratic response, 
which are well defined thanks to its differentiability, are uniquely fixed by the conservation of the vectorial current, namely by the condition \eqref{eq:G5WI_intro}. 
\item The normalization of the chiral current, i.e., condition \eqref{z6}, which guarantees that the chiral vertex renormalization equals the vectorial one, see \eqref{asympt.1_intro}. 
\end{enumerate}

Let us now discuss the core argument of the proof in more detail.

\subsubsection{The splitting of $\hat\Pi^{5}_{\mu,\nu,\sigma}$ into chiral triangle graph $+$ differentiable correction}\label{sec3.1.1}

As an outcome of the RG analysis mentioned in item \ref{it1}, of the choice of $\nu$ mentioned in
item \ref{it2} and of the splitting of the correlation functions into dominant plus subdominant parts mentioned in item \ref{it3}, we will prove that the quadratic response of the chiral current to the external electromagnetic field can be written as
\begin{equation}\label{eq:tritetr_intro}
\hat\Pi^{5}_{\mu,\nu,\sigma}(\pp_1,\pp_2)= \frac{Z^{5}_{\mu} Z_{\nu} Z_{\sigma}}{Z^3 v_1v_2v_3} I_{\mu,\nu,\sigma}(\lis\pp_1,\lis\pp_2)+\widetilde H^5_{\mu,\nu,\sigma}(\pp_1,\pp_2),
\end{equation}
where: the first term in the right side is defined in terms of the effective parameters $Z_\mu^5, Z_\mu$ (the chiral/non-chiral vertex renormalizations\footnote{These are defined 
in terms of the amputated vertex functions in \eqref{amput.1}-\eqref{amput.2} as follows: $Z_\mu=\lim_{\kk\to \pp_{F}^{\omega}}^{\pp\to {\bf 0}}\hat \gamma_{\mu}(\kk,\pp)$, where the limit is understood in the sense specified after \eqref{z6}; and $Z_\mu^5=\lim_{\kk\to \pp_{F}^{\omega}}^{\pp\to {\bf 0}}\hat \gamma_{\mu}^5(\kk,\pp)$.}), of $Z$ 
(the wave function renormalization, see \eqref{pp_int}), of $v_l$ 
(the dressed velocities, see \eqref{pp_int}), and of the explicit function $I_{\mu,\nu,\sigma}$ (the `chiral triangle graph' of QED$_4$ with momentum cutoff, see \eqref{defImunusigma} below), computed at 
$\lis\pp_1=(p_{1,0}, v_1 p_{1,1}, v_2p_{1,2}, v_3p_{1,3})$, $\lis\pp_2=(p_{2,0}, v_1 p_{2,1}, v_2p_{2,2}, v_3p_{2,3})$; the second term in the right side is a correction term,
which is differentiable in the external momenta in a sufficiently small neighbourhood around the origin, with derivatives that are H\"older continuous of order 
$\theta$, for all $\theta\in(0,1)$. 

\subsubsection{Ward Identities and the chiral triangle graph}\label{sec:WI_intro1}

Before we proceed further, let us specify a few additional properties of the quadratic response $\hat \Pi^5_{\mu,\nu,\sigma}$ and of the first term in the right side of \eqref{eq:tritetr_intro}.

The first key property we wish to emphasize is a remarkable consequence of the conservation \eqref{eq:cons}, namely the Ward Identity 
\begin{equation}\label{eq:G5WI_intro}
\sum_{\nu} p_{1,\nu} \hat\Pi^{5}_{\mu, \nu, \sigma}(\pp_{1}, \pp_{2}) = 
\sum_{\sigma} p_{2,\sigma}\hat\Pi^{5}_{\mu, \nu, \sigma}(\pp_{1}, \pp_{2}) = 0\;.
\end{equation}
We stress that there is no similar identity for the contraction of the $\mu$ index with the external momenta, due to the lack of lattice {\it chiral} gauge invariance:
while the formal infrared limit of both the electromagnetic and chiral currents are conserved, only the first one is associated with a lattice gauge invariance principle.

Of course, gauge invariance implies the validity of Ward Identities for (infinitely many) other correlations. A crucial one is the `vertex Ward Identity' relating the vertex function 
$\hat \Gamma_\mu(\kk,\pp)$ in \eqref{Gammanon} with the 2-point function $\hat S_2(\kk)$ in \eqref{pp_int}:
\begin{equation}\label{eq:vertexWI}
\sum_{\mu=0,1,2,3} p_{\mu} \hat \G_{\mu}(\kk,\pp) = \hat S_{2}(\kk) - \hat S_{2}(\kk+\pp)\;,
\end{equation}
which implies the following relation between the renormalized parameters: 
\begin{equation}
Z_{\mu} = v_{\mu} Z\;,\label{asympt.2_intro}
\end{equation}
with $v_0:=1$. The condition \eqref{asympt.2_intro} can be interpreted by saying that the electric charge transported by the electromagnetic current is not renormalized, 
thanks to lattice Ward Identities. The analogous identity does not hold for the lattice chiral current, because there is no lattice chiral symmetry protecting the `chiral charge' from 
renormalizing. This is the reason why we need to impose the identity of the charges transported by the electromagnetic and chiral currents
via the constants $Z^5_{\mu,\text{bare}}$, as discussed in Section \ref{sec:couplingqq}; in fact, 
by imposing \eqref{z6}, we obtain that, for $\mu=0,1,2,3$, 
\begin{equation}
Z^{5}_{\mu}=Z_{\mu}. \label{asympt.1_intro}
\end{equation}

Finally, an explicit computation of the chiral triangle graph $I_{\mu,\nu,\sigma}(\pp_1,\pp_2)$ shows that 
\begin{eqnarray}\label{eq:I.1_intro}
\sum_{\mu=0}^3(p_{1,\mu} + p_{2,\mu}) I_{\mu,\nu,\sigma}(\pp_{1}, \pp_{2}) &=& \frac{1}{6\pi^{2}} \sum_{\alpha, \beta=0}^3p_{1,\alpha} p_{2,\beta} \varepsilon_{\alpha\beta\nu\sigma} + R_{\nu,\sigma}(\pp_{1},\pp_{2})\\
\sum_{\nu=0}^3p_{1,\nu} I_{\mu,\nu,\sigma}(\pp_{1}, \pp_{2}) &=& \frac{1}{6\pi^{2}} \sum_{\alpha,\beta=0}^3p_{1,\alpha} p_{2,\beta} \varepsilon_{\alpha\beta\mu\sigma} + \widetilde R_{\mu,\sigma}(\pp_{1}, \pp_{2})\;,\label{eq:I.2_intro}
\end{eqnarray}
where $\varepsilon_{\a\b\n\s}$ is the Levi-Civita symbol, see footnote \ref{LCsymbol}, and where 
both $R$ and $\widetilde R$ are smaller than $CP^3/|p_F^+|$ for $P=\max\{|\pp_1|,|\pp_2|\}$ sufficiently small, as compared with $|p_F^+|$. 
Note that the fact that the $4$-divergences in the left side of \eqref{eq:I.2_intro} is non-zero, contrary to \eqref{eq:G5WI_intro}, is ultimately due to the fact that the triangle graph 
$I_{\mu,\nu,\sigma}(\pp_1,\pp_2)$ is computed in the presence of an ultraviolet momentum cutoff, see \eqref{defImunusigma} below, 
which breaks global and chiral gauge symmetries explicitly.

\subsubsection{The core argument: proof of \eqref{eq:main}}\label{sec3.1.3}

In view of \eqref{eq:tritetr_intro}, the left 
side of \eqref{eq:main} can be rewritten as 
\begin{eqnarray}\label{eq:WIfin}
-\sum_{\mu}(p_{1,\mu} + p_{2,\mu}) \hat\Pi^{5}_{\mu,\nu,\sigma}(\pp_{1}, \pp_{2}) =&-&
\sum_{\mu} (p_{1,\mu} + p_{2,\mu})\frac{Z^{5}_{\mu} Z_{\nu} Z_{\sigma}}{Z^3 v_1v_2v_3} I_{\mu,\nu,\sigma}(\lis\pp_1,\lis\pp_2)\\
&-& \sum_{\mu}(p_{1,\mu} + p_{2,\mu}) \widetilde H^{5}_{\mu, \nu,\sigma}(\pp_{1}, \pp_{2}).\nonumber
\end{eqnarray}
Let us focus on the first term in the right side of \eqref{eq:WIfin}. Thanks to \eqref{asympt.1_intro}, \eqref{asympt.2_intro} and \eqref{eq:I.1_intro}, 
recalling that for $a=1,2$ and $\mu=0,1,2,3$ we defined $\lis p_{a,\mu}=v_\mu p_{a,\mu}$, we can rewrite it as
\begin{eqnarray}&& 
-\frac{v_{\nu} v_{\sigma}}{v_{1} v_{2} v_{3}}\big(\sum_{\alpha,\beta}\frac{1}{6\pi^2} v_\alpha{p}_{1,\alpha} v_\beta{p}_{2,\beta} \varepsilon_{\alpha\beta\nu\sigma} + R_{\nu,\sigma}(\pp_{1},\pp_{2})\big)=\\
&& \qquad =-\frac1{6\pi^2} \sum_{\alpha,\beta}p_{1,\alpha} p_{2,\beta} \varepsilon_{\alpha\beta\nu\sigma}-\frac{v_{\nu} v_{\sigma}}{v_{1} v_{2} v_{3}}R_{\nu,\sigma}(\pp_{1},\pp_{2}) \;,\nonumber\end{eqnarray}
where, in passing from the left to the right side, we used the fact that $\varepsilon_{\a\b\n\s}\neq 0$ if and only if $(\a,\b,\n,\s)$ is a permutation of $(0,1,2,3)$ (recall that $v_0=1$). 
Plugging this back into \eqref{eq:WIfin}, and recalling that $R_{\nu,\sigma}(\pp_{1},\pp_{2})=O(P^3)$, with $P=\max\{|\pp_1|,|\pp_2|\}$, see the lines following \eqref{eq:I.2_intro}, we find
\begin{equation}\label{eq:p1p2}
-\sum_{\mu}(p_{1,\mu} + p_{2,\mu}) \hat\Pi^{5}_{\mu,\nu,\sigma}(\pp_{1}, \pp_{2}) = -\frac{1}{6\pi^{2}} \sum_{\alpha,\beta}p_{1,\alpha} p_{2,\beta} \varepsilon_{\alpha,\beta,\nu,\sigma} - \sum_{\mu}(p_{1,\mu} + p_{2,\mu}) \widetilde H^{5}_{\mu,\nu,\sigma}(\pp_{1}, \pp_{2})+O(P^3)\;.
\end{equation}
Similarly, using (\ref{eq:I.2_intro}), we find:
\begin{equation}\label{eq:WIfin2}
\sum_{\nu}p_{1,\nu} \hat\Pi^{5}_{\mu,\nu,\sigma}(\pp_{1}, \pp_{2}) = \frac{1}{6\pi^{2}} \sum_{\alpha,\beta}p_{1,\alpha} p_{2,\beta} \varepsilon_{\alpha\beta\mu\sigma} + \sum_{\nu}p_{1,\nu} \widetilde H^{5}_{\mu,\nu,\sigma}(\pp_{1}, \pp_{2})+O(P^3)\;.
\end{equation}

We now use \eqref{eq:G5WI_intro} and the continuous differentiability of $\widetilde H^5_{\mu,\nu,\sigma}$ to obtain, after having expanded $\widetilde H^5_{\mu,\nu,\sigma}$ at first order in Taylor series around the origin, 
\begin{equation}\label{eq:WI2nd}
0 = \frac{1}{6\pi^{2}} \sum_{\alpha,\beta}p_{1,\alpha} p_{2,\beta} \varepsilon_{\alpha\beta\mu\sigma} + \sum_{\nu} p_{1,\nu} \Big(\widetilde H^{5}_{\mu,\nu,\sigma}({\bf 0}, {\bf 0}) + \sum_{a = 1,2}\, \sum_{\rho} p_{a,\rho} \frac{\partial \widetilde H^{5}_{\mu,\nu,\sigma}}{\partial p_{a,\rho}}({\bf 0}, {\bf 0}) \Big)\;.
\end{equation}
Eq.(\ref{eq:WI2nd}) implies that:
\begin{equation}\label{id.1}
\widetilde H^{5}_{\mu,\nu,\sigma}({\bf 0}, {\bf 0}) = 0\;,\qquad \frac{\partial \widetilde H^{5}_{\mu,\nu,\sigma}}{\partial p_{2,\beta}} ({\bf 0}, {\bf 0}) = - \frac{1}{6\pi^{2}}\varepsilon_{\nu\beta\mu\sigma} \;.
\end{equation}
Similarly, using that $\widetilde H^{5}_{\mu,\nu,\sigma}(\pp_{1},\pp_{2}) = \widetilde H^{5}_{\mu,\sigma,\nu}(\pp_{2},\pp_{1})$,
\begin{equation}\label{id.2}
\frac{\partial \widetilde H^{5}_{\mu,\nu,\sigma}}{\partial p_{1,\beta}}({\bf 0}, {\bf 0}) = -\frac{1}{6\pi^{2}}\varepsilon_{\sigma\beta\mu\nu}\;.
\end{equation}
We now go back to \eqref{eq:WIfin2}, expand $\widetilde H^5_{\mu,\nu,\sigma}$ in Taylor series around the origin up to first order included, and 
use the identities \eqref{id.1}-\eqref{id.2}, thus finding
\begin{eqnarray} &&\label{WIfinfin3}
-\sum_{\mu} (p_{1,\mu} + p_{2,\mu})\hat\Pi^{5}_{\mu,\nu,\sigma}(\pp_{1},\pp_{2}) =-\frac{1}{6\pi^{2}} \sum_{\alpha,\beta}p_{1,\alpha} p_{2,\beta} \varepsilon_{\alpha\beta\nu\sigma}\\
&&\qquad + \frac1{6\pi^2}\sum_{\mu} (p_{1,\mu} + p_{2,\mu}) \sum_{a=1,2}(-1)^a \sum_{\beta} p_{a,\beta}\varepsilon_{\nu\beta\mu\sigma}+ 
\sum_{\mu} (p_{1,\mu} + p_{2,\mu}) \widetilde R^{5}_{\mu,\nu,\sigma}({\bf p}_{1}, {\bf p}_{2})+O(P^3)\;,\nonumber
\end{eqnarray}
where the term $\widetilde R^{5}_{\mu,\nu,\sigma}(\pp_1,\pp_2)$ comes from the Taylor remainder of $\widetilde H^5_{\mu,\nu,\sigma}$.
Thanks to the fact that the derivatives of $\widetilde H^5_{\mu,\nu,\sigma}$ are H\"older continuous of order $\theta$, for any $0<\theta<1$, the remainder 
$\widetilde R^{5}_{\mu,\nu,\sigma}({\bf p}_{1}, {\bf p}_{2})$ is of the order $P^{1+\theta}$ (possibly non-uniformly in $\theta$ and in $|p_F^+|$), 
for $P=\max\{|\pp_1|,|\pp_2|\}$ sufficiently small. 
After combining the first two terms in the right side of \eqref{WIfinfin3}, we get \eqref{eq:main}, with $R^5_{\nu,\sigma}(\pp_1,\pp_2)=O(P^{2+\theta})$.
%Note the crucial role played in this argument by 
%the decomposition of the response function into differentiable plus (explicitly computable) non-differentiable contributions.

\subsubsection{Roadmap}\label{sec:roadmap}

In view of the previous subsection, in order to complete the proof of \eqref{eq:main}, we `just' need to prove the validity of \eqref{eq:tritetr_intro}, with $Z_\mu, Z_\mu^5, Z, v_l$ 
satisfying \eqref{asympt.1_intro}-\eqref{asympt.2_intro}, and with $I_{\mu,\nu,\sigma}$ satisfying \eqref{eq:I.1_intro}-\eqref{eq:I.2_intro}; we also need to establish \eqref{eq:G5WI_intro}. The proof of all these claims will be given below, together with the proof of the other claims in Theorem \ref{thm:main}. We will follow the strategy outlined in 
items \ref{it1} to \ref{it3} at the beginning of this section. 
More in detail, the proof in the next sections is organized as follows: 
\begin{itemize}
\item In Section \ref{sec:3.1}, we represent the generating function of Euclidean correlations in terms of a Grassmann functional integral. In particular, 
in Section \ref{sec:WI} we state the gauge-invariance property of the Grassmann generating function, which implies a hierarchy of Ward Identities for the correlation functions, 
including \eqref{eq:G5WI_intro}.
\item In Section \ref{sec:RG} we describe the iterative RG computation of the Grassmann generating function, whose output is a convergent 
expansion for the generating function, in terms of a sequence of effective coupling constants $Z_{\mu,h}, Z_{\mu,h}^5, Z_h, v_{l,h}$, labelled by the 
step $h$ of the RG iteration (these are nothing but the finite-$h$ analogues of the parameters $Z_\mu,Z_\mu^5, Z, v_l$ in \eqref{eq:tritetr_intro}). 
If these parameters are suitably bounded, uniformly in $h$, and if they admit a limit as the number of RG iterations tends to infinity, the expansion for the 
correlation functions is convergent uniformly in the $\beta,L\to\infty$ limit, for $|\lambda|$ small enough, and the limit of the correlations as $\beta,L\to\infty$ exists. In order 
for the bounds on the radius of convergence to be uniform in $\zeta\in I$, the RG iteration must be defined differently, depending 
on whether the momentum scale under consideration in the RG step is larger or smaller than the separation between the Weyl nodes; the two regimes are described in 
Sections \ref{sec:1stregime} and \ref{sec:2ndregime}, respectively. 
\item In Section \ref{sec:flow} we explain how to fix the bare staggered chemical potential $\nu$ in such a way that the sequence of effective 
coupling constants $Z_{\mu,h}, Z_{\mu,h}^5, Z_h, v_{l,h}$ satisfies the desired bounds. We also show how to fix the bare parameters $Z^5_{\mu,\text{bare}}$ in such a way that 
\eqref{asympt.1_intro} holds. 
\item In Section \ref{sec:sing} we explain how to use the convergent expansion provided by the iterative RG computation of the generating function, in order 
to define a splitting of the main correlation functions of interest into an explicit dominant part plus a remainder, whose Fourier transforms 
admit improved dimensional bounds at low external momenta. In particular, we prove \eqref{eq:tritetr_intro} and \eqref{pp_int}, 
we show that \eqref{asympt.1_intro} implies \eqref{z6} 
and, by using the vertex Ward Identity, we prove \eqref{asympt.2_intro}. 
\item In Section \ref{sec:3.6} we prove \eqref{eq:I.1_intro} and \eqref{eq:I.2_intro}.
\end{itemize}
Further technical details are deferred to the appendices: in Appendix \ref{app:symm} we discuss the symmetries of the Grassmann action and their consequences 
for the effective action obtained at the $h$-th step of the RG iteration; in Appendix \ref{app.dimbounds} we provide further technical details on the `tree expansion' for the 
kernels of the effective action and their dimensional bounds; in Appendix \ref{app:2nd} we prove that the quadratic response coefficients we are interested in can be expressed in terms 
of Euclidean correlation functions. 

\subsection{Grassmann representation}\label{sec:3.1}

In this section we introduce a Grassmann integral formulation of the model, which will be the starting point for our RG analysis. In the following, $C,C', \ldots$, 
denote universal constants (in particular, independent of the distance between the Weyl nodes), whose specific values may change from line to line. 

Let $\g$ be any constant larger than $1$, $\alpha_0$ a positive constant, to be fixed later on, 
and $\chi(s)$ a smooth, even, compactly supported function, such that $\chi(s) = 0$ for 
$|s|>\alpha_0 \g$ and $\chi(s) = 1$ for $|s|<\alpha_0$. For technical convenience, we choose $\chi$ to belong to the Gevrey class $G^s$ with $s=2$, see, e.g., 
\cite[Appendix C]{GMT17}.  
Recall that $\mathcal{M}^{\text{f}}_{\beta} = \frac{2\pi}{\beta}(\mathbb{Z} + \frac{1}{2})$ is the set of fermionic Matsubara frequencies.
Given $N\in \mathbb{N}$, we introduce the $UV$-regularized version of $\mathcal{M}^{\text{f}}_{\beta}$ as:
\begin{equation}
\mathcal{M}^{\text{f}}_{\beta, N} := \Big\{ k_{0} \in \mathcal{M}^{\text{f}}_{\beta} \, \Big| \, \chi(2^{-N} |k_{0}|) > 0 \Big\}\;.
\end{equation}
Let $\mathcal{B}_{L}$ be the finite-volume Brillouin zone, Eq.(\ref{eq:BL}), and let $\mathbb{D}_{\beta, L, N} := \mathcal{M}^{\text{f}}_{\beta, N} \times \mathcal{B}_{L}$. We consider the finite Grassmann algebra generated by the Grassmann variables $\{ \hat \psi^{\pm}_{\kk, r} \}$ with $\kk=(k_0,k) \in \mathbb{D}_{\beta, L, N}$ and $r=1,2$. 

The Grassmann Gaussian integration $\int P_{N}(d\psi) (\cdot)$ is a linear functional acting on the Grassmann algebra as follows. Its action on a given Grassmann monomial $\prod_{j=1}^{n} \hat\psi^{\e_{j}}_{\kk_{j}, q_{j}}$ is zero unless $| \{j: \e_{j} =+ \} | = |\{ j: \e_{j} = - \}|$, in which case:
\be \int P_{N}(d\psi) \hat \psi^-_{\kk,r_1}\hat \psi^+_{\kk'_1,r'_{1}}\cdots
\hat \psi^-_{\kk_{n},r_n}\hat \psi^+_{\kk'_n,r'_{n}}=\det[C_{r_j,r_k'}(\kk_j,\kk'_{k})]_{j,k=1,\ldots,n},
\ee
where $C_{r,r'}(\kk, \kk')= \b L^3 \delta_{\kk,\kk'} \big[\hat g_{\beta, L, N}(\kk)\big]_{r,r'}$ and
\bea\label{eq:prop}
&&\hat g_{\beta,L,N}(\kk) := \frac{\chi_{N}(k_{0})}{-ik_{0} + \hat H^0(k)} \;,\qquad \chi_{N}(k_{0}) := \chi(2^{-N}k_{0})\;.
\eea
We represent the free propagator as:
\begin{equation}
\hat g_{\beta, L, N}(\kk) = \chi_{N}(k_{0}) \begin{pmatrix} -ik_{0} + c(k)+d(k) & a(k) - i b(k) \\ a(k) + i b(k) & -ik_{0} - c(k) + d(k) \end{pmatrix}^{-1}, 
\end{equation}
see \eqref{eq:genrepH0}. Recall that $a(p_{F}^{\omega}) = b(p_{F}^{\omega}) = c(p_{F}^{\omega}) = d(p_F^\omega)=0$ and that $a,b,c,d$ satisfy the 
parity properties spelled out after \eqref{eq:genrepH0}. Moreover, letting $a_R(k)=a(k)-v_1^0k_1$, $b_R(k)=b(k)-v_2^0 k_2$, 
$c_{R,\omega}(k)=c(k)-\omega v_3^0(k_3-p_{F,3}^\omega)-\tfrac{b^0}{2}(k_3-p_{F,3}^\omega)^2$ and $d_R(k)=d(k)$, these functions satisfy the bounds in \eqref{number.1} and \eqref{number.2}. 

The Grassmann Gaussian integration might also be rewritten as:
\begin{equation}\label{freeactioncutoff}
\int P_{N}(d\psi)(\cdot) = \frac{1}{\mathcal{N}_{\beta, L, N}} \int d\psi\, \exp\Big\{ -\frac{1}{\beta L^{3}} \sum_{\kk \in \mathbb{D}_{\beta, L, N}} \sum_{r,r' = 1,2}\hat \psi^{+}_{\kk,r} [\hat g_{\beta, L, N}(\kk)^{-1}]_{r,r'} \hat \psi^{-}_{\kk,r'} \Big\} (\cdot)
\end{equation}
with $\int d\psi(\cdot)$ is the usual Berezin integration over the Grassmann algebra and $\mathcal{N}_{\beta, L, N}$ a normalization factor:
\begin{equation}
\mathcal{N}_{\beta, L, N} = \prod_{\kk\in \mathbb{D}_{\beta, L, N}} (\beta L^{3}) \det \hat g_{\beta, L, N}(\kk)\;.
\end{equation}
Let us introduce the configuration space Grassmann field as follows, for $\xx\in [0, \beta) \times \Lambda_{L}$:
\begin{equation}
\psi^{\pm}_{\xx, r} := \frac{1}{\beta L^{3}} \sum_{\kk\in \mathbb{D}_{\beta, L, N}} e^{\pm i\kk\cdot \xx} \hat \psi^{\pm}_{\kk, r}\;;
\end{equation}
we then have:
\begin{eqnarray}
\int P_{N}(d \psi) \psi^{-}_{\xx, r} \psi^{+}_{\yy, r'} &=& \big[g_{\beta, L, N}(\xx-\yy)\big]_{r,r'}\;,\nonumber\\ 
g_{\beta, L, N}(\xx-\yy) &:=& \frac{1}{\beta L^{3}} \sum_{\kk\in \mathbb{D}_{\beta, L, N}} e^{-i\kk\cdot (\xx - \yy)} \hat g_{\beta, L, N}(\kk)\;.
\end{eqnarray}
Denoting by $g_{\beta, L}(\xx)$ the two-point function of the Fock-space model, it is well-known (and easy to check) that $\lim_{N\to \infty} g_{\beta, L, N}(\xx) = g_{\beta, L}(\xx)$ for $\xx\notin \beta \mathbb{N} \times L\mathbb{N}^{2}$, while for $\xx\in \beta \mathbb{N} \times L\mathbb{N}^{2}$:
\begin{equation}
\lim_{N\to \infty} g_{\beta, L, N}(\xx) = \frac{g_{\beta, L}(x_{0}^{+}, \vec x) + g_{\beta, L}(x_{0}^{-}, \vec x)}{2}\;. 
\end{equation}
Next, we introduce the interaction of the Grassmann field theory as:
\begin{equation}\label{VpsiGrass}
\mathcal{V}(\psi) := -\lambda \int d\xx d\yy \sum_{r,r'} n_{\xx, r} w_{r,r'}(\xx, \yy) n_{\yy, r'} - \nu \int d\xx\, \sum_{r} (-1)^{r} n_{\xx, r} 
\end{equation}
with the notations: $w_{r,r'}(\xx,\yy) := \delta(x_{0} - y_{0}) w_{r,r'}(x, y)$; $\int d\xx(\cdot) := \int_{0}^{\beta} dx_{0} \sum_{x\in \Lambda_{L}}(\cdot )$; and $n_{\xx, r} = 
\psi^{+}_{\xx, r} \psi^{-}_{\xx, r}$. From now on, we assume that $\nu$ is real-analytic in $\lambda$, for $\lambda$ small, and $|\nu|\le C|\lambda|$. A posteriori, we will see that this assumption is compatible with 
the other requirements on $\nu$ we will need. 
More generally, the interaction of the Grassmann field theory in the presence of complex external fields $A_{\mu, \xx}$, $A^{5}_{\nu, \xx}$ (depending now on coordinates $\xx=(x_0,x)$, with $x_0$ an imaginary time in $[0,\beta)$) 
and of a Grassmann external field $\phi^{\pm}_{\xx, r}$ is:
\begin{equation}\label{fullVphi}
\mathcal{V}(A, A^{5}, \phi, \psi) = \mathcal{V}(\psi) + B(A, \psi) + B^{5}(A^{5}, A, \psi) + (\psi^+, \phi^-)+(\phi^+,\psi^-)\;,
\end{equation}
where $(\psi^+,\phi^-):=\sum_{r=1}^2\int d\xx \psi^+_{\xx,r}\psi^-_{\xx,r}$ and 
\begin{eqnarray}\label{eq:BB}
B(A,\psi) &:=& (A_{0}, j_0) + (\psi^{+}, (H^{0} - H^{0}(A))\psi^{-}) \nonumber\\
B^{5}(A^{5}, A, \psi) &:=& \sum_{\mu=0}^3(A^{5}_{\mu}, j^{5}_\mu)\;,
\end{eqnarray}
with the understanding that $(A_0,j_0) = \int d\xx A_{0,\xx} j_{0,\xx}$, with $j_{0,\xx}=-i\sum_{r=1}^2\psi^+_{\xx,r}\psi^-_{\xx,r}$, and similarly for $(A^{5}_{\mu}, j^{5}_\mu)$;
here $j^{5}_{\mu}$ denotes the Grassmann counterpart of the chiral lattice current, recall Eq.(\ref{eq:J5mu}):
\begin{equation}\label{eq:chicurgra}
j^{5}_{\m, \zz}(A) = Z^{5}_{\mu,\text{bare}} \sum_{x,y\in \Lambda_{L}} \sum_{r,r'=1}^2 \psi^{+}_{(z_{0}, x),r}{\mathfrak J}^{5}_{\mu;r,r'}(y-x,z-x)e^{i\int_{x\to y} A_{z_{0}} \cdot d\ell} \psi^{-}_{(z_{0}, y),r'}\;,\nonumber
\end{equation}
where $\int_{x\to y}A_{z_0}\cdot d\ell$ denotes the line integral $\int_0^1 A_{(z_0,x+s(y-x))}\cdot (y-x)\, ds$. 
Finally, we introduce the generating functional of the correlations as:
\begin{equation}\label{eq:W}
\mathcal{W}_{\beta, L}(A, A^{5}, \phi) := \lim_{N\to \infty}\log \int P_{N}(d \psi)e^{\mathcal{V}(A, A^{5}, \phi, \psi)}\;.
\end{equation}
The existence of the ultraviolet limit $N\to \infty$ is essentially model independent, and has already been proven in a number of places, see {\it e.g.} \cite{GM, GMPbis} for the Hubbard model on the honeycomb lattice. It is well-known, see {\it e.g.} Section 5.1 of \cite{GMPhall}, that the Fock-space Euclidean correlation functions can be expressed in terms of the derivatives of the generating functional with respect to the external fields. At non-coinciding points:
\begin{eqnarray}
S^{\beta, L}_{2;r,r'}(\xx, \yy) &=& \frac{\partial^{2} \mathcal{W}_{\beta, L}(A, A^{5}, \phi)}{\partial \phi^{-}_{\yy,r'}\partial \phi^{+}_{\xx,r} }\Big|_{A = A^{5} = \phi = 0},\nonumber\\
\G^{\beta, L}_{\mu;r,r'}(\zz, \xx, \yy) &=& \frac{\partial^{3} \mathcal{W}_{\beta, L}(A, A^{5}, \phi)}{\partial A_{\mu, \zz}  \partial \phi^{-}_{\yy,r'}\partial \phi^{+}_{\xx,r}}\Big|_{A=A^{5} = \phi=0},\label{eq:Gcorr}\\
\G^{5;\beta, L}_{ \mu;r,r'}(\zz,\xx,\yy) &=& \frac{\partial^{3} \mathcal{W}_{\beta, L}(A, A^{5}, \phi)}{\partial A^{5}_{\mu, \zz}\partial \phi^{-}_{\yy,r'} \partial \phi^{+}_{\xx,r} }\Big|_{A=A^{5} = \phi=0},\nonumber\\
\Pi^{5;\beta, L}_{\mu, \nu, \sigma}(\zz, \xx, \yy) &=& \frac{\partial^{3}\mathcal{W}_{\beta, L}(A, A^{5}, \phi)}{\partial A^{5}_{\mu, \zz} \partial A_{\nu, \xx} \partial A_{\sigma, \yy}} \Big|_{A = A^{5} = \phi = 0}\;.\nonumber
\end{eqnarray}
By translation invariance, all the correlation functions in (\ref{eq:Gcorr}) only depend on the relative differences of the configuration-space arguments. 

\medskip

{\bf Remark.} The statement of Theorem \ref{thm:main} involves the Fourier transforms of the correlation functions in \eqref{eq:Gcorr}, 
\begin{eqnarray}\hat S^{\beta, L}_{2;r,r'}(\kk)&=&\int d\yy e^{-i\kk\yy}S^{\beta, L}_{2;r,r'}({\bf 0},\yy),\label{eq:3.15}\\
\hat \G^{\beta, L}_{\mu;r,r'}(\kk,\pp)&=&\int d\yy\,d\zz\, e^{-i\pp\zz-i\kk\yy}\,\Gamma^{\beta,L}_{\mu;r,r'}(\zz,{\bf 0},\yy),\label{eq:3.16}\\
\hat \G^{5;\beta, L}_{\mu;r,r'}(\kk,\pp)&=&\int d\yy\,d\zz\, e^{-i\pp\zz-i\kk\yy}\,\Gamma^{5;\beta,L}_{\mu;r,r'}(\zz,{\bf 0},\yy),\label{eq:3.16^5}\\
\hat\Pi^{5;\beta, L}_{\mu, \nu, \sigma}(\pp_{1}, \pp_{2})&=&\int d\xx\, d\yy e^{-i\pp_1\xx-i\pp_2\yy}\,\Pi^{5;\beta,L}_{\mu,\nu,\sigma}({\bf 0},\xx,\yy),\label{FTPi5}\end{eqnarray}
and their $\beta,L\to\infty$ limits, denoted by $\hat S_{2;r,r'}(\kk)$, $\hat \Gamma_{\mu;r,r'}(\kk,\pp)$, $\hat \Gamma^5_{\mu;r,r'}(\kk,\pp)$, $\hat \Pi^5_{\mu,\nu,\sigma}(\pp_1,\pp_2)$, respectively. 
For the purpose of proving Theorem \ref{thm:main}, we can limit ourselves to computing these functions at sufficiently low momenta (more precisely, 
at $\kk$ sufficiently close to the Weyl nodes and $\pp,\pp_1, \pp_2$ sufficiently close to ${\bf 0}$). In particular, in the proof below, 
we can freely assume that $\hat A_{\mu,\pp}$ and $\hat A_{\mu.\pp}^5$ are supported in a sufficiently small neighbourhood of the origin, and we shall do so in the following. 

\subsubsection{Ward Identities}\label{sec:WI}

Ward identities are nontrivial relations between correlation functions, implied by the conservation of the lattice current (\ref{eq:cons}). The Grassmann integral formulation of the model offers a particularly compact way of representing them. The invariance under $U(1)$-local gauge transformation reads:
\begin{equation}\label{eq:wigen}
\mathcal{W}_{\beta, L}(A + \partial \alpha, A^{5}, \phi e^{i\alpha}) = \mathcal{W}_{\beta, L}(A, A^{5}, \phi)\;,
\end{equation}
for any smooth function $\alpha_{\xx}$ on $\mathbb{R}^{3}$, periodic in $x_{0}$ of period $\beta$ and in $x$ of period $L$ (here $\phi e^{i\alpha}$ 
is a shorthand for $\{e^{+i\alpha_\xx}\phi^+_{\xx,r},e^{-i\alpha_\xx}\phi^-_{\xx,r}\}_{\xx\in [0,\beta)\times \Lambda_L}$). 
Differentiating Eq.(\ref{eq:wigen}) with respect to $\alpha$ and with respect to the external fields, we obtain a hierarchy of Ward identities. As already emphasized in Section \ref{sec:WI_intro1}, two Ward Identities that are of particular importance for us are \eqref{eq:G5WI_intro} and \eqref{eq:vertexWI}: 
the first is obtained by deriving once with respect to $\alpha$, once with respect to $A$, once with respect to $A^5$, then 
setting the external fields $A, A^5, \phi$ to zero, and taking the $\beta,L\to\infty$ limit; the second is obtained by deriving once with respect to $\alpha$,  twice with respect to $\phi$, then setting the external fields $A, A^5, \phi$ to zero, and taking the $\beta,L\to\infty$ limit.

\subsection{Renormalization Group analysis}\label{sec:RG}

In this section we sketch the RG analysis of the model in the presence of external fields $A$ and $A^{5}$. Our analysis follows and extends the 
one in \cite{M1, M1bis}, where RG methods have been used to prove the analyticity of the free energy of a specific lattice model of Weyl semimetals
within the class of models considered in this paper, 
and to compute the corresponding two-point function $\hat S^{\beta, L}_{2}$. In the discussion below, we limit ourselves to describe the general scheme and 
to highlight the main differences compared to \cite{M1, M1bis}, referring to those papers, and in particular to \cite[Sections 2 and 3]{M1bis}, 
for additional technical details. Moreover, 
for simplicity, we focus on generating functional $\mathcal{W}_{\beta, L}(A, A^{5}, 0)$. The generalization to the presence of 
an external fermionic field $\phi$ is straightforward and will not be explicitly worked out, see, e.g., \cite[Section 12]{GeM} or \cite[Section 2.2]{GMPbis} 
for a discussion of the necessary modifications. 

The starting point is a scale decomposition of the free propagator, which separates ultraviolet and infrared degrees of freedom:
\begin{eqnarray}\label{eq:splitg}
&&\qquad\qquad\qquad\qquad\hat g_{\beta, L}(\kk) = \hat g^{(\leq 0)}(\kk) + \hat g^{(>0)}(\kk) \nonumber\\
&&\hat g^{(\leq 0)}(\kk) = \chi_{0}(\kk) \hat g_{\beta, L, N}(\kk)\;,\qquad  g^{(>0)}(\kk) = g_{\beta, L, N}(\kk) - \hat g^{(\leq 0)}(\kk)\;,
\end{eqnarray}
where $\chi_{0}(\kk) := \chi\Big( \sqrt{k_{0}^{2} -\det \hat H^0(k)}\Big)$. We assume that $\alpha_0$ is smaller than $\gamma^{-1}\sqrt{c_2}$, with $c_2$ the constant
in \eqref{number.4}, so that on the support of $\chi_0$ (where, in particular, $0\le -\det \hat H^0(k)\le \alpha_0^2\gamma^2$) the momentum $k$ is closer to the Weyl nodes 
than $c_1$ and $|d(k)|\le\tfrac12 \sqrt{a^2(k)+b^2(k)+c^2(k)}$, see \eqref{number.3}. 
We denote by $g^{(\leq 0)}(\xx)$ and $g^{(>0)}(\xx)$ the inverse Fourier transforms of the momentum-space propagators $\hat g^{(\leq 0)}(\kk)$, $\hat g^{(>0)}(\kk)$. Correspondingly, we use the addition principle of Grassmann variables to rewrite the Grassmann Gaussian field $\hat \psi^{\pm}$ as:
\begin{equation}
\psi^{\pm}_{\xx,r} = \psi^{(\leq 0)\pm}_{\xx,r} + \psi^{(>0)\pm}_{\xx,r}\;,
\end{equation}
where $\psi^{(\leq 0)\pm}$, $\psi^{(>0)\pm}$ are independent Grassmann Gaussian fields, with covariances given by $g^{(\leq 0)}$, $ g^{(>0)}$ respectively. Due to 
the fact that the propagator $\hat g^{(>0)}(\kk)$ is supported away from the Weyl nodes  $\pp_{F}^{\omega} = (0, p_{F}^{\omega})$, the configuration-space 
covariance $g^{(>0)}(\xx)$ decays faster than any power in $\xx$; more precisely, using the assumption that $\chi$ is in the Gevrey class of order $s=2$, 
\begin{equation} \|g^{(>0)}(\xx)\|\le C_0 e^{-\kappa_0\sqrt{|\xx|}},\label{gevreydec}\end{equation}
for some $\kappa_0>0$. This allows to integrate out the ultraviolet degrees of freedom via a simple fermionic cluster expansion, which is largely model independent; see e.g. \cite{GM, GMPbis}. After the integration of $g^{(>0)}$ we get:
\begin{equation}\label{eq:start0}
e^{\mathcal{W}_{\beta, L}(A, A^{5}, 0)} = e^{\mathcal{W}^{(0)}(A, A^{5})}\int P_{\leq 0}(d\psi^{(\leq 0)}) e^{\mathcal{V}^{(0)}(A, A^{5}, \psi^{(\leq 0)})}\;,
\end{equation}
where the various objects appearing in Eq. (\ref{eq:start0}) have the following meaning. The fermionic Gaussian integration $P_{\leq 0}(d\psi^{(\leq 0)})$ has covariance given by $g^{(\leq 0)}(\xx- \yy)$; the effective interaction on scale zero $\mathcal{V}^{(0)}(A, A^{5}, \psi^{(\leq 0)})$ has the form:
\begin{eqnarray}\label{eq:V0}
\mathcal{V}^{(0)}(A, A^{5}, \psi)&=&\sum_{\substack{n, m_{1}, m_{2} \geq 0 \\ n>0}}\ \sum_{\underline{r}, \underline{\mu}, \underline{\nu}}
\int d{\bf X} d{\bf Y} d{\bf Z}\Big[ \prod_{i=1}^{2n} \psi^{+}_{\xx_{2i-1}, r_{2i-1}} \psi^{-}_{\xx_{2i}, r_{2i}} \Big]\cdot\\
&\cdot& \Big[ \prod_{j=1}^{m_{1}} A_{\mu_{j}, \yy_{j}} \Big] \Big[ \prod_{\ell = 1}^{m_{2}} A^{5}_{\nu_{\ell}, \zz_{\ell}} \Big]\, W^{(0)}_{n, m_{1}, m_{2}; \underline{r}, \underline{\mu}, \underline{\nu}}({\bf X}, {\bf Y}, {\bf Z})\;,\nonumber
\end{eqnarray}
with ${\bf X} = \{ \xx_{i} \}_{i=1}^{2n}$, ${\bf Y} = \{ \yy_{j} \}_{j=1}^{m_{1}}$, ${\bf Z} = \{\zz_{\ell}\}_{\ell = 1}^{m_{2}}$; the kernels $W^{(0)}$ are analytic in $\lambda$ 
for $|\lambda|$ small enough, and satisfy the following weighted $L^1$ bound (recall that in our assumptions $\nu=O(\lambda)$):
\begin{equation}
\frac{1}{\beta L^{3}} \int d{\bf Q}\, e^{\frac{\kappa_0}2 \sqrt{\delta({\bf Q})}} |W^{(0)}_{n, m_{1}, m_{2}; \underline{r}, \underline{\mu}, \underline{\nu}}({\bf Q})| \leq C^{n+m_{1}+m_{2}} 
\Big(\prod_{\nu=0}^3 \frac{|Z_{\nu,\text{bare}}^5|}{|p_{F,3}^+|^{1-\delta_{\nu,3}}}\Big)
|\lambda|^{\max\{\delta_{m_1+m_2,0},\, n-1\}}, 
\end{equation}
where ${\bf Q} = ({\bf X}, {\bf Y}, {\bf Z})$, $\delta({\bf Q})$ is the `Steiner diameter' of ${\bf Q}$, i.e., the length of the shortest tree connecting all 
the points in $\bf Q$, and $\kappa_0$ is the same as in \eqref{gevreydec} (the reason for the choice of the strecthed exponential weight 
$e^{\frac{\kappa_0}2 \sqrt{\delta({\bf Q})}}$ is the comparable stretched exponential decay of $g^{(>0)}$).
The generating functional on scale zero $\mathcal{W}^{(0)}(A, A^{5})$ has a similar representation, except that only kernels with $n=0$ (no fermionic external lines) 
contribute.

Next, we perform the integration of the infrared degrees of freedom, associated to the massless field $\psi^{(\leq 0)}$. To this end, we decompose the field 
$\psi^{(\leq 0)}$ in terms of single-scale fields $\psi^{(h)}$, $h\leq 0$, that we integrate iteratively. For $v^0_3$ small, i.e., $\zeta$ close to $\zeta_1$, the value at 
which the Weyl nodes merge, 
we distinguish two scale regimes, depending on whether $c(k'+p_{F}^{\omega}) \simeq \omega v_3^0 k'_{3}$ or $c(k'+p_{F}^{\omega}) \simeq \tfrac{b^0}{2}{k'_{3}}^{2}$, recall \eqref{eq:genrepH0}, \eqref{eq:linear2gen}, \eqref{number.1} and (\ref{number.2}). We define $h_{*}\le 0$ as the crossover scale between the two regimes, at which ${k_3'}^2\sim \gamma^{h_*}\sim v_3^0 \gamma^{h_*/2}$; 
more precisely, we let
\begin{equation}\label{eq:cross}
h_*:=\min\{0,\lfloor 2\log_{\gamma} |v^{0}_{3}|\rfloor\}.
\end{equation}
Let $\varepsilon:= |p_{F}^{+} - p_{F}^{-}|$. Recall that, by assumption, $c_0^{-1} |v^{0}_{3}| \leq \varepsilon \leq c_0|v^{0}_{3}|$, see \eqref{c0boundsbelow}. 
Thus, $\gamma^{h_{*}} \sim \varepsilon^{2}$ for $\varepsilon$ small. In the following, we discuss separately the scales $h>h^{*}$ and 
the scales $h \leq h^{*}$.

\medskip

{\bf Remark.} If $h_*=0$, then the infrared analysis simplifies, in that the first regime, discussed in the next subsection, disappears. 
This is the case when $\zeta$ is far from the value $\zeta_1$ 
at which the Weyl points merge (or, similarly, the case, mentioned after \eqref{pF+pF-}, in which the Weyl nodes are well separated uniformly in $\zeta$). 

\subsubsection{Regime $h > h^{*}$}\label{sec:1stregime}

Assume $h_*<0$ (otherwise the reader can pass directly to next subsection). 
We inductively assume that, for all $h_*\le h\le 0$, the generating function of correlations can be written as:
\begin{equation}
e^{\mathcal{W}_{\beta, L}(A, A^{5}, 0)} = e^{\mathcal{W}^{(h)}(A, A^{5})}\int P_{\leq h}(d\psi^{(\leq h)}) e^{\mathcal{V}^{(h)}(A, A^{5},
\psi^{(\leq h)})}\;.\label{ccoorr}
\end{equation}
The Gaussian Grassmann field $\psi^{(\leq h)}$ has Fourier-space covariance $\hat g^{(\leq h)}(\kk)$ given by:
\begin{equation}\label{eq:gh}
\hat g^{(\leq h)}(\kk) = \frac{\chi_{h}(\kk)}{Z_{h}} A_{h}(\kk)^{-1}
\end{equation}
where:
\begin{equation}\label{eq:Ah}
A_{h}(\kk) = \begin{pmatrix} -ik_{0} + c_h(k)+d_h(k) & a_h(k) - i b_h(k) \\ a_h(k) + ib_h(k) & -ik_{0} - c_h(k)+d_h(k) \end{pmatrix}\;,
\end{equation}
$\chi_{h}(\kk) = \chi\Big(\gamma^{-h} \sqrt{k_{0}^{2} -\det \hat H^0(k)}\Big)$, 
$Z_{h}$ is real-analytic in $\lambda$, such that $|Z_{h} - 1| \leq C|\lambda|$, and, recalling that $p_{F,3}\equiv p_{F,3}^+$, 
\begin{eqnarray}
&& a_h(k)=v_{1,h} k_1+a_R(k)/Z_h, \hskip1.85truecm b_h(k)=v_{2,h} k_2+b_R(k)/Z_h,\\
&& c_h(k)=\big(c(k)+\tfrac12 \zeta_h(k_3^2-p_{F,3}^2)\big)/Z_h, \quad d_h(k)=d(k)/Z_h,\end{eqnarray}
with $v_{l,h}$ real-analytic in $\lambda$, such that $|v_{l,h} - v^{0}_{l}| \leq C|\lambda|$, for $l=1,2$, and 
$|\zeta_{h}| \leq C|\lambda|$.
The effective potential $\mathcal{V}^{(h)}$ has the form:
\begin{eqnarray}\label{eq:Vh}\mathcal{V}^{(h)}(A, A^{5}, \psi) &=& \sum_{\substack{n, m_{1}, m_{2} \geq 0 \\ n>0}}\ \sum_{\underline{q},\underline{r}, \underline{\mu}, \underline{\nu}}\int d{\bf X} d{\bf Y} d{\bf Z}\, \Big[ \prod_{i=1}^{2n}\hat\partial^{q_{2i-1}}\psi^{+}_{\xx_{2i-1}, r_{2i-1}} \hat\partial^{q_{2i}}\psi^{-}_{\xx_{2i}, r_{2i}} \Big]\cdot\\
&\cdot&\Big[ \prod_{j=1}^{m_{1}} A_{\mu_{j}, \yy_{j}} \Big] \Big[ \prod_{\ell = 1}^{m_{2}} A^{5}_{\nu_{\ell}, \zz_{\ell}} \Big]\, W^{(h)}_{n, m_{1}, m_{2}, \underline{q}; \underline{r}, \underline{\mu}, \underline{\nu}}({\bf X}, {\bf Y}, {\bf Z})\;,\nonumber
\end{eqnarray}
which is similar to \eqref{eq:V0}, with the difference that now there are the operators $\hat\partial^{q_i}$, with $i=1,\ldots,2n$, acting on the Grassmann variables; here 
the labels $q_i$ are multi-indices of the form $q_i=(q^0_i,q^1_i,q^2_i,q^3_i)$ with $q^\mu_i\in\{0,1,2,3,4\}$, and $\hat \partial^{q_i}$ is 
a pseudo-differential operator, equal to the identity if $q_i=(0,0,0,0)$, and dimensionally equivalent to the composition of a derivative of order $q_i^0$ in direction $0$, of a derivative of order $q^1_i$ in direction $1$, etc, otherwise\footnote{In Fourier space, the Fourier symbol of $\hat \partial^{q}$ is a function of $k$ behaving like $k_0^{q^0}k_1^{q^1}k_2^{q^2}k_3^{q^3}$ for $k$ small. When writing the analogue of \eqref{eq:Vh} in Fourier space, we can re-absorb the Fourier symbols of the operators $\hat \partial^{q_i}$, with $i=1,\ldots,2n$, into the Fourier symbol of the kernel $W^{(h)}_{n, m_{1}, m_{2}, \underline{q}; \underline{r}, \underline{\mu}, \underline{\nu}}$, and we shall do so; after summation over $\underline{q}$, we will denote the resulting modified Fourier symbol of the kernels by $\hat W^{(h)}_{n, m_{1}, m_{2}; \underline{r}, \underline{\mu}, \underline{\nu}}$.}
(if $h=0$, the only non-vanishing contribution to the right side of \eqref{eq:Vh} is the one with $q_i=(0,0,0,0)$, for all $i=1,\ldots,2n$, 
and \eqref{eq:Vh} reduces to \eqref{eq:V0}). 
The kernels $W^{(h)}_{n, m_{1}, m_{2}, \underline{q}; \underline{r}, \underline{\mu}, \underline{\nu}}$ belong to a suitable weighted $L^1$ space, see \eqref{eq:bdW} below, 
and are analytic in $\lambda$ for $|\lambda|$ small enough, uniformly in $h$ (recall that $\nu$ is assumed to be analytic in $\lambda$, as well, and of order $\lambda$). 
We stress that the representation in \eqref{eq:Vh} is not unique: the claim is that there exists such a representation, with the kernels satisfying natural dimensional estimates, see 
\eqref{eq:bdW} below. The generating functional of correlations on scale $h$, $\mathcal{W}^{(h)}$, admits a similar representation, except that 
only kernels with $n=0$ contribute. These assumptions are true at scale zero, with $v_{l,0} = v_{l}^{0}$, $\zeta_0=0$ and $Z_{0} = 1$.

\medskip

The inductive assumption \eqref{ccoorr} is verified at scale $h = 0$, as an outcome of the integration of the ultraviolet degrees of freedom. 
We now assume that \eqref{ccoorr} is valid at scale $h_*<h\le 0$ and prove it for $h-1$. For this purpose, we intend to integrate out the degrees of freedom on scale $h$, 
after having properly rewritten the effective potential in the right side of \eqref{ccoorr}. As a first step, we split $\mathcal V^{(h)}$ as $\mathcal{V}^{(h)} = \mathcal{L} \mathcal{V}^{(h)} + \mathcal{R} \mathcal{V}^{(h)}$, where $\mathcal{L}$ is the {\it localization operator}, acting on $\mathcal{V}^{(h)}$ as follows\footnote{Whenever it will be convenient, from now on, 
the dependence upon the indices $r_1,r_2,\ldots,$ will be left implicit. $\hat \psi^+_\kk$ (resp. $\hat\psi^-_{\kk}$) will be thought of as a row (resp. columnn) vector with components $\hat\psi^+_{\kk,1},\hat\psi^+_{\kk,2}$ (resp. $\hat\psi^-_{\kk,1},\hat\psi^-_{\kk,2}$). When writing $\hat\psi^+_\kk M(\kk)\hat\psi^-_{\kk}$ with $M(\kk)$ a $2\times2$ matrix, we 
will mean $\sum_{r,r'=1}^2 \hat\psi^+_{\kk,r} M_{r,r'}(\kk)\hat\psi^-_{\kk,r'}$.}:
\begin{eqnarray}\label{eq:ell.0}
&& \mathcal{L} \mathcal V^{(h)}(A,A^5,\psi):= \int \frac{d\kk}{(2\pi)^4} \hat \psi^+_\kk\big(\mathcal L \hat W^{(h)}_{1,0,0}(\kk)\big)\hat \psi^-_\kk\\
&&\qquad +
\int \frac{d\kk}{(2\pi)^4}\int \frac{d\pp}{(2\pi)^4} \Big[\hat A_{\mu,\pp}\hat \psi^+_{\kk+\pp}\big(\mathcal L \hat W^{(h)}_{1,1,0;\mu}(\kk,\pp)\big)\hat \psi^-_\kk+
\hat A_{\mu,\pp}^5\hat \psi^+_{\kk+\pp}\big(\mathcal L \hat W^{(h)}_{1,0,1;\mu}(\kk,\pp)\big)\hat \psi^-_\kk\Big],\nonumber\end{eqnarray}
where $\int \tfrac{d\kk}{(2\pi)^4}$ and $\int\tfrac{d\pp}{(2\pi)^4}$ are shorthands for $(\beta L^3)^{-1}\sum_{\kk\in\mathbb{D}_{\beta, L, N}}$ and $(\beta L^3)^{-1}\sum_{\pp\in \mathbb{D}_{\beta, L}^b}$, respectively (here $\mathbb{D}_{\beta, L}^b=\tfrac{2\pi}\beta\mathbb Z\times \tfrac{2\pi}{L}\mathbb Z^3$), and 
\begin{eqnarray}
\mathcal{L} \hat W^{(h)}_{1,0,0}(\kk) &:=& \hat W^{(h),\infty}_{1,0,0}({\bf 0}) + \sum_{\mu=0,1,2}k_{\mu} \partial_{\mu} \hat{W}^{(h),\infty}_{1,0,0}({\bf 0}) + \frac12{k_3^2}\partial_{3}^{2} \hat W^{(h),\infty}_{1,0,0}({\bf 0})\;,\nonumber\\
\mathcal{L} \hat W^{(h)}_{1,1,0;\mu}(\kk,\pp) &:=& \hat W^{(h),\infty}_{1,1,0;\mu}({\bf 0}, {\bf 0})+(k_3\partial_{k_3}+p_3\partial_{p_3})\hat W^{(h)}_{1,1,0;\mu}({\bf 0},{\bf 0})\;,\label{eq:ell}\\
\mathcal{L} \hat W^{(h)}_{1,0,1;\mu}(\kk, \pp) &:= &\hat W^{(h),\infty}_{1,0,1;\mu}({\bf 0}, {\bf 0})+(k_3\partial_{k_3}+p_3\partial_{p_3})\hat W^{(h),\infty}_{1,0,1;\mu}({\bf 0},{\bf 0})\;,\nonumber
\end{eqnarray}
where $\hat W^{(h),\infty}_{1,0,0}(\kk)$ indicates the $\beta,L\to\infty$ limit of $\hat W^{(h)}_{1,0,0}(\kk)$, and similarly for $\hat W^{(h),\infty}_{1,1,0;\mu}(\kk,\pp)$ and $\hat W^{(h)}_{1,0,1;\mu}(\kk,\pp)$, whose existence follows from the inductive 
construction of the kernels and from the corresponding bounds, uniform in $\beta,L$ described below. 
The {\it renormalization} operator is defined as $\mathcal{R} := 1 - \mathcal{L}$. 
Notice that $\mathcal{R} \mathcal{L} = \mathcal{L} \mathcal{R} = 0$.

\medskip

{\bf Remarks.}

\medskip

\noindent1) There are potentially other terms in the Taylor expansion of $\hat W^{(h),\infty}_{1,0,0}(\kk)$ 
that we could include `for free' in the definition in the first line, namely $k_3 \partial_{3} \hat W^{(h),\infty}_{1,0,0}({\bf 0})$, 
$\sum_{\mu=0,1,2}k_\mu k_3$ $\partial_\mu\partial_{3} \hat W^{(h),\infty}_{1,0,0}({\bf 0})$ and 
$\tfrac16 k_3^3\partial_{3}^3 \hat W^{(h),\infty}_{1,0,0}({\bf 0})$: in fact, as stated in the following lemma,  
these are all zero, by the parity properties of $\hat W^{(h),\infty}_{1,0,0}(\kk)$. Therefore, the reader can think of the right side of the first line in \eqref{eq:ell}
as being the Taylor expansion of $\hat W^{(h),\infty}_{1,0,0}(\kk)$ around $\kk={\bf 0}$, including all terms proportional to $1,\{k_\mu\}_{\mu=0,1,2,3}, 
\{k_\mu k_3\}_{\mu=0,1,2,3}$, and $k_3^3$, 
where some of the terms have not been written explicitly, simply because they are zero. Consequently, 
$\mathcal{R} \hat W^{(h)}_{1,0,0}(\kk)$ is equal to the corresponding 
Taylor remainder of $\hat W^{(h),\infty}_{1,0,0}(\kk)$, which is quadratic in $\{k_\mu\}_{\mu=0,1,2}$ and, at $k_0=k_1=k_2=0$, is {\it quartic} in $k_3$;
plus a finite size correction, proportional to $\hat W^{(h)}_{1,0,0}(\kk)-\hat W^{(h),\infty}_{1,0,0}(\kk)$, which is exponentially small in $\beta,L$ 
as $\beta,L\to\infty$, and can be bounded as discussed in \cite[Appendix B]{GMT20}. We anticipate the fact that the scaling dimension of 
$\hat W^{(h),\infty}_{n,m_1,m_2;\underline \mu,\underline \nu}$ is $\frac72-\frac52n-m_1-m_2$, see \eqref{eq:bdW} (the convention here is that kernels with positive/zero/negative
scaling dimensions correspond to relevant/marginal/irrelevant operators in the RG sense); any additional derivative with respect to $k_\mu$ decreases the 
scaling dimension by $1$, if $\mu=0,1,2$, and by $1/2$, if $\mu=3$. Therefore, from the comments above, it follows that $\mathcal{R} \hat W^{(h)}_{1,0,0}(\kk)$ is irrelevant, with scaling dimension $-1$.

\medskip

\noindent2) 
Similarly, from the definitions in \eqref{eq:ell}, 
it follows that $\mathcal R\hat W^{(h)}_{1,1,0;\mu}(\kk,\pp)$
and $\mathcal R\hat W^{(h)}_{1,0,1;\mu}(\kk,\pp)$ are equal to Taylor remainders that are linear in $k_\mu,p_\mu$ with $\mu=0,1,2$, and quadratic in $k_3,p_3$, 
up to a (subdominant) finite size correction. From the formula of the scaling dimension anticipated in the previous item, it follows that they are irrelevant with scaling dimension $-1$. Note that 
the linear terms $(k_3\partial_{k_3}+p_3\partial_{p_3})\hat W^{(h)}_{1,1,0;\mu}({\bf 0},{\bf 0})$ and 
$(k_3\partial_{k_3}+p_3\partial_{p_3})\hat W^{(h),\infty}_{1,0,1;\mu}({\bf 0},{\bf 0})$ in the second and third lines of \eqref{eq:ell}
are irrelevant with dimension $-1/2$; the reason why we prefer to include them into the local part is to guarantee that the irrelevant part has largest scaling dimension equal to 
$-1$. This guarantees that the improved dimensional bound discussed in the paragraph after \eqref{eq:bdW} has an additional factor $\gamma^{\theta h}$, with $\theta$ any positive constant smaller than $1$ (rather than $1/2$), see also Appendix \ref{app.dimbounds}; 
this fact is useful in the study of the flow of the effective couplings discussed in Section \ref{sec:flow}, see in particular case $\mu=3$ in \eqref{cases1} and case $\mu=0,1,2$ in \eqref{cases2}, where having $\theta>1/2$ has important consequences for the resulting flow of $Z_{\mu,h}$ and $Z_{\mu,h}^5$.

\medskip

The next lemma shows that the lattice symmetries of the model impose non-trivial constraints on the kernels of $\mathcal{L} \mathcal{V}^{(h)}$.
\begin{lemma}\label{lem:sym1} The following identities hold:
\begin{eqnarray}\label{eq:sym1}
&&\hat{W}^{(h),\infty}_{1,0,0}({\bf 0}) =n_{h} \sigma_{3} \;,\quad \partial_{\mu} \hat{W}^{(h),\infty}_{1,0,0}({\bf 0}) = \begin{cases} z_{\mu,h} \sigma_{\mu} & \text{if $\mu=0,1,2$,}\\ 0 & \text{if $\mu=3$}\end{cases}\;, \quad \partial_{3}^{2} \hat{W}^{(h),\infty}_{1,0,0}({\bf 0}) =  z_{3,h} \sigma_{3}\,\nonumber\\
&&
\partial_{0}\partial_{3} \hat{W}^{(h),\infty}_{1,0,0}({\bf 0}) =\partial_{1}\partial_{3} \hat{W}^{(h),\infty}_{1,0,0}({\bf 0}) =\partial_{2}\partial_{3} \hat{W}^{(h),\infty}_{1,0,0}({\bf 0}) =\partial_{3}^3 \hat{W}^{(h),\infty}_{1,0,0}({\bf 0}) = 0\;,\qquad 
\end{eqnarray}
where $n_{h}, z_{i,h}, \tilde z_{3,h} \in \mathbb{R}$ and we denoted $\sigma_0:= -i\mathbbm{1}_{2}$. Moreover, letting $p_{F,3}:=p_{F,3}^+$:
\begin{eqnarray}
&&\hat{W}^{(h),\infty}_{1,1,0;\mu}({\bf 0}, {\bf 0}) =\begin{cases}
Z_{\mu,h} \sigma_{\mu}  & \text{if $\mu=0,1,2$,}\\
0 & \text{if $\mu=3$,}\end{cases}, \qquad \hat{W}^{(h),\infty}_{1,0,1;\mu}({\bf 0}, {\bf 0}) = \begin{cases}
0 &  \text{if $\mu=0,1,2$,}\nonumber\\
Z^{5}_{3, h}\sigma_{3} & \text{if $\mu=3$,}\end{cases}\\
&&\partial_{k_3}\hat{W}^{(h),\infty}_{1,1,0;\mu}({\bf 0}, {\bf 0}) =2\Re\Big(\partial_{p_3}\hat{W}^{(h),\infty}_{1,1,0;\mu}({\bf 0}, {\bf 0})\Big)=\begin{cases} 0 & \text{if $\mu=0,1,2$,}\\
\frac{Z_{3,h}}{p_{F,3}} \sigma_3& \text{if $\mu=3$,}\end{cases}\label{2ndeq:lem3.2}\\
&&\partial_{k_3}\hat{W}^{(h),\infty}_{1,0,1;\mu}({\bf 0}, {\bf 0}) = 2\Re\Big(\partial_{p_3}\hat{W}^{(h),\infty}_{1,0,1;\mu}({\bf 0}, {\bf 0})\Big)= \begin{cases}
\frac{Z_{\mu,h}^5}{p_{F,3}}\sigma_\mu &  \text{if $\mu=0,1,2$,}\\
0 & \text{if $\mu=3$,}\end{cases}\nonumber\end{eqnarray}
where $Z_{\mu, h}, Z_{\mu,h}^5\in \mathbb{R}$, for $\mu=0,1,2,3$, and we denoted $\Re(M)=(M+M^\dagger)/2$ the Hermitian part of a complex matrix $M$. Moreover, 
letting $\Im(M)=-i(M-M^\dagger)/2$ for $M$ a complex matrix, 
\begin{eqnarray} && \Im\Big(\partial_{p_3}\hat{W}^{(h),\infty}_{1,1,0;\mu}({\bf 0}, {\bf 0})\Big)=\begin{cases} 0 & \text{if $\mu=0,1,2$,}\\
\frac{\widetilde Z_{3,h}}{p_{F,3}} \sigma_3& \text{if $\mu=3$,}\end{cases}\label{Zcomplex.1}\\
&& \Im\Big(\partial_{p_3}\hat{W}^{(h),\infty}_{1,0,1;\mu}({\bf 0}, {\bf 0})\Big)= \begin{cases}
\frac{\widetilde Z_{\mu,h}^5}{p_{F,3}}\sigma_\mu &  \text{if $\mu=0,1,2$,}\\
0 & \text{if $\mu=3$,}\end{cases}\label{Zcomplex.2}\end{eqnarray}
with $\widetilde Z_{3, h}, \widetilde Z_{\mu,h}^5\in \mathbb{R}$, for $\mu=0,1,2$.
\end{lemma}
We refer to Appendix \ref{app:conssym} for the proof of this lemma. 
Given the definitions of $\mathcal {L}\mathcal{V}^{(h)}$ and $\mathcal{R} \mathcal{V}^{(h)}$, and the properties of the kernels of 
$\mathcal {L}\mathcal{V}^{(h)}$ spelled out in Lemma \ref{lem:sym1}, as well as the definitions of $n_h, z_{\mu,h}, Z_{\mu,h}, Z^5_{\mu,h}$, 
we now manipulate the right side of \eqref{ccoorr} as follows: 
we rewrite $\mathcal{V}^{(h)} = \mathcal{L} \mathcal{V}^{(h)} + \mathcal{R} \mathcal{V}^{(h)}$, and re-absorb part of 
$\mathcal{L} \mathcal{V}^{(h)}$ in the Grassmann Gaussian integration, thus obtaining: 
\begin{equation}\label{eq:WbetaL}
e^{\mathcal{W}_{\beta, L}(A, A^{5}, 0)} = e^{\mathcal{W}^{(h)}(A, A^{5})}
 \int \widetilde{P}_{\leq h}(d\psi^{(\leq h)}) e^{2^{h} \nu_{h} N_{3}(\psi^{(\leq h)}) + B^{(h)}(A, A^{5}, 
\psi^{(\leq h)})+\mathcal{R} \mathcal{V}^{(h)}(A, A^{5}, \psi^{(\leq h)})}\;,\end{equation}
where, recalling that $p_{F,3}\equiv p_{F,3}^+$, we defined: $2^{h} \nu_{h}:= n_{h}+\frac12 p_{F,3}^2 z_{h,3}$, 
$N_{3}(\psi) := \int \tfrac{d\kk}{(2\pi)^{4}} \hat \psi^{+}_{\kk} \sigma_{3} \hat \psi^{-}_{\kk}$, 
\begin{eqnarray} &&
B^{(h)}(A, A^{5}, \sqrt{Z_{h}}\psi):=\int \frac{d\kk}{(2\pi)^{4}}\int \frac{d\pp}{(2\pi)^{4}} \times\\
&&\qquad\qquad \times \Big[  \sum_{\mu=0}^2 Z_{\mu,h}\hat A_{\mu,\pp} \hat \psi^{+}_{\kk+\pp} \sigma_{\mu} \hat \psi^{-}_{\kk}+
(Z_{3,h}\dfrac{k_3+p_3/2}{p_{F,3}}+i\widetilde Z_{3,h}\dfrac{p_3}{p_{F,3}}) \hat A_{3,\pp}
\hat \psi^{+}_{\kk+\pp}  \sigma_{3} \hat \psi^{-}_{\kk} \nonumber \\
&&\qquad\qquad +  
\sum_{\mu=0}^2 (Z_{\mu,h}^5\dfrac{k_3+p_3/2}{p_{F,3}}+i\widetilde Z^5_{\mu,h}\dfrac{p_3}{p_{F,3}})\hat A_{\mu,\pp}^5 \hat \psi^{+}_{\kk+\pp} \sigma_{\mu} \hat \psi^{-}_{\kk} + Z_{3,h}^5 \hat A^5_{3,\pp}
\hat \psi^{+}_{\kk+\pp} \sigma_{3} \hat \psi^{-}_{\kk}\Big], \nonumber
\end{eqnarray}
and the new Grassmann Gaussian integration $\widetilde{P}_{h}(d\psi^{(\leq h)})$ has covariance given by:
\begin{equation}
\tilde g^{(\leq h)}(\kk) = \frac{\chi_{h}(\kk)}{Z_{h-1}(\kk)} \tilde A_{h-1}(\kk)^{-1}\;,
\end{equation}
where: $Z_{h-1}(\kk) := Z_{h} - \chi_{h}(\kk) z_{0,h}$, 
\begin{equation}
\tilde A_{h-1}(\kk) = \begin{pmatrix} -ik_{0} + \tilde c_{h-1}(\kk) +\tilde d_{h-1}(\kk) & \tilde a_{h-1}(\kk) - i \tilde b_{h-1}(\kk) \\ \tilde a_{h-1}(\kk) + i\tilde b_{h-1}(\kk) & -ik_{0} 
- \tilde c_{h-1}(\kk) +\tilde d_{h-1}(\kk) \end{pmatrix}\;,\label{tildeAh-1}
\end{equation}
and
\begin{eqnarray} 
&& \hskip-1.2truecm\tilde a_{h-1}(\kk)=v_{1,h-1}(\kk) k_1+a_R(k)/Z_{h-1}(\kk), \qquad \tilde b_{h-1}(\kk)=v_{2,h-1}(\kk) k_2+b_R(k)/Z_{h-1}(\kk),\\
&&\hskip-1.2truecm\tilde c_{h-1}(\kk)=\big(c(k)+\tfrac12 \zeta_{h-1}(\kk)(k_3^2-p_{F,3}^2)\big)/Z_{h-1}(\kk), \qquad 
\tilde d_{h-1}(\kk)=d(k)/Z_{h-1}(\kk),\end{eqnarray}
with
\begin{eqnarray}
v_{l,h-1}(\kk)&=&(Z_h v_{l,h}-\chi_h(\kk)z_{l,h})/Z_{h-1}(\kk), \qquad l=1,2,\\
\zeta_{h-1}(\kk)&=&\zeta_h-\chi_h(\kk) z_{3,h}.\end{eqnarray}
Fix $\theta\in(0,1)$; from now on, we denote by $C_\theta$, $C_\theta'$, etc., positive, $\theta$-dependent, constants, possibly divergent as $\theta\to 1^-$. 
We inductively assume that, for all scales $k\geq h$, $\mu=0,1,2,3$, and a positive constant $C_\theta$,  
\begin{equation}\label{eq:indhigh}
|z_{\mu,k}| \leq C_\theta|\lambda| \gamma^{\theta k}.
\end{equation}
We will check the validity of these bounds on scale $k=h-1$. We also assume that, for all $k\ge h$,
\begin{equation} \label{eq:nu}|\nu_{k}| \leq C_\theta|\lambda| \gamma^{\theta k}.\end{equation} 
Eqs.(\ref{eq:indhigh}) imply that
\begin{equation}\label{eq:boundsZvl}
|Z_{h-1}(\kk) - 1|\leq C|\lambda|\;,\qquad |\zeta_{h-1}(\kk)|\leq C|\lambda|, \qquad 
|v_{l,h-1}(\kk) - v^{0}_{l}| \leq C|\lambda|, \ \ l=1,2. 
\end{equation}
We also let \begin{equation} Z_{h-1}:=Z_{h-1}({\bf 0}), \quad \zeta_{h-1}:=\zeta_{h-1}({\bf 0}), \quad \text{and}\quad  v_{l,h-1}:=v_{l,h-1}({\bf 0})\quad \text{for}\quad 
l=1,2,3. \label{Zvl}\end{equation}
Obviously, $Z_{h-1}, \zeta_{h-1}, v_{l,h-1}$ satisfy the same bounds as in \eqref{eq:boundsZvl}. 
In order to prove the inductive assumption, we decompose the Grassmann field as:
\begin{equation}
\psi^{(\leq h)} = \psi^{(\leq h-1)} + \psi^{(h)}\;,
\end{equation}
where the Grassmann field $\hat \psi^{(\leq h-1)}$ has covariance given by $\hat g^{(\leq h-1)}(\kk)$, defined as in (\ref{eq:gh}), (\ref{eq:Ah}), with ${h}$ replaced 
by $h-1$, 
while $\psi^{(h)}$ has covariance given by:
\begin{equation}\label{eq:singlescale0}
\hat g^{(h)}(\kk) := \frac{f_{h}(\kk)}{Z_{h-1}(\kk)} \tilde A_{h-1}(\kk)^{-1}, \end{equation}
where $f_h(\kk)=\chi_h(\kk)-\chi_{h-1}(\kk)$ and we used the fact that, on the support of $\chi_{h-1}$, $Z_{h-1}(\kk)=Z_{h-1}$ and $\tilde A_{h-1}(\kk)=
A_{h-1}(\kk)$. On the support of $f_h$, 
$C^{-1}\gamma^{2h}\le  -\det \tilde A_{h-1}(\kk) \le C \gamma^{2h}$, $|k_{\mu}|\leq C\gamma^{h}$ for $\mu=0,1,2$, and $|k_{3}| \leq C\gamma^{\frac{h}{2}}$. 
Therefore, for these values of $\kk$, using also the definition of $h_*$ and the bounds \eqref{c0boundsbelow}, \eqref{number.1}, \eqref{number.2}, 
\begin{eqnarray}
\tilde a_{h-1}(\kk) &=& v_{1,h-1}(\kk)k_{1}+ O(\gamma^{2h}), \hskip1.8truecm \tilde b_{h-1}(\kk) = v_{2,h-1}(\kk)k_{2} +  O(\gamma^{2h}),\\
\tilde c_{h-1}(\kk)&=&\tfrac12(b^{0}+\zeta_{h-1}(\kk)) k_{3}^{2} + O(\gamma^{h_*}+\gamma^{2h}), \qquad \tilde d_{h-1}(\kk)=O(\gamma^{2h}), 
\end{eqnarray}
which implies $\|\hat g^{(h)}(\kk)\|_{\infty} \leq C  \gamma^{-h}$ and $\|\hat g(\kk) \|_{1} \leq C \gamma^{\frac52 h}$. By a similar discussion, we can also 
dimensionally bound the derivatives of $\tilde a_{h-1},\tilde b_{h-1},\tilde c_{h-1},\tilde d_{h-1}$, thus getting 
$$\| \partial^{\alpha}_{\kk} \hat g^{(h)}(\kk)  \|_{\infty} \leq C_{|\alpha|} \gamma^{-h(1 + \alpha_{0} + \alpha_{1} + \alpha_{2} + \frac{1}{2}\alpha_{3})}\qquad \text{and}\qquad \| \partial_{\kk}^{\alpha} \hat g(\kk) \|_{1} \leq C_{|\alpha|} \gamma^{h(\frac{5}{2} - \alpha_{0} - \alpha_{1} - \alpha_{2} - \frac{1}{2}\alpha_{3})},$$ which, 
in turn, imply the following bound on $g^{(h)}$, the inverse Fourier transform of $\hat g^{(h)}$: 
\begin{equation}\label{eq:singlescale}
\| g^{(h)}(\xx) \|\leq  C_0 \gamma^{\frac{5}{2}h}e^{-\kappa_0 \sqrt{\|\xx\|_h}},
\end{equation}
where 
\begin{equation}\label{normh.reg1}\|\xx\|_h:=\gamma^{h}(|x_{0}| + |x_{1}| + |x_{2}|) + \gamma^{\frac{h}{2}}|x_{3}|.\end{equation}
The constants $C_0,\kappa_0$ can be chosen to be the same as in \eqref{gevreydec}. 

Let us now go back to \eqref{eq:WbetaL}. By using the addition principle for Gaussian Grassmann variables, we can rewrite it as:
\begin{eqnarray}\label{eq:sscale1}
e^{\mathcal{W}_{\beta, L}(A, A^{5}, 0)}& = &e^{\mathcal{W}^{(h)}(A, A^{5})} \int P_{\leq h-1}(d\psi^{(\leq h-1)})\cdot\\
&\cdot& \int  P_{h}(d\psi^{(h)}) e^{\mathcal{R} \mathcal{V}^{(h)}(A, A^{5}, \psi^{(\leq h)}) + 2^{h} \nu_{h} N_{3}(\psi^{(\leq h)}) + B^{(h)}(A, A^{5},\psi^{(\leq h)})}\;,\nonumber
\end{eqnarray}
where $P_{h}(d \psi^{(h)})$ has covariance given by $g^{(h)}$, $P_{\leq h-1}(d \psi^{(\leq h-1)})$ has covariance given by $g^{(\leq h-1)}$, and, 
in the exponent in the second line, $\psi^{(\le h)}=\psi^{(\le h-1)}+\psi^{(h)}$. We now integrate the field $\psi^{(h)}$
and denote the logarithm of the result of the integration in the second line by $\widetilde{\mathcal{W}}^{(h)}(A, A^{5})+ 
\mathcal{V}^{(h-1)}(A, A^{5},\psi^{(\leq h-1)})$, so that, letting $\mathcal{W}^{(h-1)}(A, A^{5}):=\mathcal{W}^{(h)}(A, A^{5})+\widetilde{\mathcal{W}}^{(h)}(A, A^{5})$, 
\begin{equation}\label{eq:sscale2}
e^{\mathcal{W}_{\beta, L}(A, A^{5}, 0)} =  e^{\mathcal{W}^{(h-1)}(A, A^{5})} \int P_{\leq h-1}(d\psi^{(\leq h-1)}) e^{\mathcal{V}^{(h-1)}(A, A^{5},\psi^{(\leq h-1)})}\;,
\end{equation} 
which reproduces the inductive assumption \eqref{ccoorr} at scale $h-1$. 
This iterative integration procedure goes on until we reach scale $h_*$, at which point the procedure is changed, as described in the next subsection. 
As discussed, e.g., in \cite{GeM, GM, GMPbis, M1bis}, the effective potential 
and generating function at scale $h$, obtained via such an iterative procedure, 
can be represented as convergent series over Gallavotti-Nicol\`o (GN) trees, see in particular \cite[Section 2]{M1bis}, where the tree expansion for a model of Weyl semimetal in the 
same regime as the one considered here is discussed in detail. 
The kernels of $\mathcal V^{(h)}$ and $\mathcal W^{(h)}$ are analytic in $\lambda$, for $|\lambda|$ small enough, and  satisfy suitable weighted $L^1$ bounds that, in view of the estimate (\ref{eq:singlescale}) for the single
scale propagator, read as follows: for all $h_*\le h<0$, $n\ge 1$, and $|\lambda|$ small enough, 
if \eqref{eq:indhigh} and \eqref{eq:nu} are valid for all $k>h$, then, see \cite[Lemma 1, Eq.(60)]{M1bis}
\begin{eqnarray}\label{eq:bdW}&&\frac{1}{\beta L^{3}} \int d{\bf Q}\, e^{\frac{\kappa_0}{2}\sqrt{\delta_h({\bf Q})}}\, \big|W^{(h)}_{n, m_1,m_2,\underline{q};\underline{r},\underline{\mu},\underline{\nu}}({\bf Q})\big|\le \\
&&\qquad\qquad  \leq C^{n+m_1+m_2} (\max_{k>h}\mathfrak{Z}_k)^{m_1}(\max_{k>h}\mathfrak{Z}^5_k)^{m_2}\,
|\lambda|^{\max\{\delta_{m_1+m_2,0},\, n-1\}} \gamma^{h ( \frac{7}{2}- \frac{5}{2}n - m_{1} - m_{2} -d(\underline{q}))} \;,\nonumber
\end{eqnarray}
where: $\mathfrak{Z}_k=\max\{\max\limits_{\mu=0,1,2}|Z_{\mu,k}|,\tfrac{|Z_{3,k}|}{|p_{F,3}|}, \tfrac{|\widetilde Z_{3,k}|}{|p_{F,3}|}\}$,  
$\mathfrak{Z}_k^5=\max\{|Z^5_{3,k}|,\max\limits_{\mu=0,1,2}\big\{\tfrac{|Z_{\mu,k}^5|}{|p_{F,3}|}, \tfrac{|\widetilde Z^5_{\mu,k}|}{|p_{F,3}|}\big\}\}$, 
$\delta_h({\bf Q})$ is the Steiner diameter measured by using the norm $\|\xx\|_h$ in \eqref{normh.reg1}, and $d(\underline{q})=\sum_{i=1}^{2n}(q^0_i+q^1_i+q^2_i+\frac12q^3_i$). 
The kernels of the single-scale contribution to the generating function, $\widetilde{\mathcal{W}}^{(h)}(A,A^5)$ satisfy an estimate analogous to \eqref{eq:bdW} with $n=0$. 
The combination 
$\frac{7}{2}- \frac{5}{2}n - m_{1} - m_{2}-d(\underline{q})$ is the {\it scaling dimension} of the kernels with $2n$ Grassmann fields, derivatives of order $\underline{q}$ acting on them,  
and $m_1+m_2$ external fields of type $A$ or $A^5$ in the first regime $h\ge h_*$. 
Note, in particular, that the effective quartic interaction, i.e., the kernel with $n=2$, no derivatives, and $m_1=m_2=0$, is {\it 
irrelevant}, 
with scaling dimension $-3/2$. The irrelevance of the quartic interaction allows us to derived improved bounds on all the contributions to 
the kernels associated with GN trees containing at least one interaction endpoint: this is the same as in models of graphene with short range interactions, 
see \cite{GM} and, in particular, \cite[Theorem 2]{GM}. More precisely, we can split $W^{(h)}_{n, m_1,m_2,\underline{q}} = W^{(h);\text{d}}_{n, m_1,m_2,\underline{q}} + W^{(h);\text{r}}_{n, m_1,m_2,\underline{q}}$, where `d' stands for dominant and `r' for remainder, and 
where the second term collects the contribution of all GN trees with at least one endpoint of type $\nu_k$ or on scale $1$ (in particular, it includes all the contributions from GN trees 
with at least one endpoint associated with a quartic interaction, which is necessarily on scale $1$); the term $W^{(h);\text{r}}_{n, m_1,m_2,\underline{q}}$ admits an 
improved dimensional bound, analogous to \eqref{eq:bdW}, but with an extra factor $\gamma^{\theta h}$ in the right side, for any fixed $\theta$ smaller than $1$
(and the constant $C$ replaced by a $\theta$-dependent constant $C_\theta$). The proof of (\ref{eq:bdW}) and of its 
improved analogue for $W^{(h);\text{r}}_{n, m_1,m_2,\underline{q}}$ are completely analogous to those of Eqs.(60) and (61) in Lemma 1 of \cite{M1bis}, respectively, 
modulo a few minor differences discussed in Appendix \ref{app.dimbounds}. 

By using these bounds, we can also prove the inductive assumptions (\ref{eq:indhigh}). In fact, assume that \eqref{eq:indhigh} and \eqref{eq:nu} are valid for $k\ge h$. 
Then the bound \eqref{eq:bdW} is valid
at scale $h-1$, and similarly for its improved analogue for $W^{(h-1);\text{r}}_{n, m_1,m_2,\underline{q}}$. 
By using the definitions of $z_{\mu,h-1}$ with $\mu=0,1,2,3$, and the bound on $W^{(h-1);\text{r}}_{1, 0,0,\underline{q}}$, we find that 
\eqref{eq:indhigh} is valid with $k=h-1$, as well. 

Concerning the proof of \eqref{eq:nu}, let $\beta^\nu_h:=\nu_{h-1}-2\nu_h$ be the {\it beta function} for $\nu_h$. 
From the bound on $W^{(h-1);\text{int}}_{1, 0,0}$, we find that  
\begin{equation} |\beta^\nu_h|\le C_\theta|\lambda|\gamma^{\theta h},\label{betanubd1st}\end{equation}
for any $\theta\in(0,1)$. 
This does not imply that {\it any} solution of the beta function equation
\begin{equation} \nu_{h-1}=2\nu_h+\beta^\nu_h\label{eq:betanu1st}\end{equation}
with $\nu_{0}=O(\lambda)$ satisfies \eqref{eq:nu}. However, it is enough that we find {\it one} special 
choice of $\nu_0$ (or, equivalently, of the staggered chemical potential $\nu$ in \eqref{eq:H})
for which the solution satisfies such a bound. We temporarily proceeding by assuming the existence of such a good initial datum, 
and we will prove this assumption in Section \ref{sec:flow}. In that section, we will also derive bounds on the effective couplings 
$Z_{\mu,h}, Z^5_{\mu,h}$. 

\subsubsection{Regime $h \le h^{*}$}\label{sec:2ndregime}

In order to integrate the remaining scales, we proceed as follows. We choose $\alpha_0$ in the definition of $\chi$ (see beginning of Section \ref{sec:3.1})
small enough that the support of $\chi_{h_{*}}(\kk)$ consists of two `well-separated' sets\footnote{By `well-separated' we mean that the corridor between these two sets 
centered at $\pp_F^\pm$ has diameter comparable with the diameter of the sets themselves.}, centered at $\pp_{F}^{+}$ and $\pp_F^-$, respectively. 
We decompose the fermionic field in terms of two independent {\it quasi-particle fields}, associated with the two Weyl nodes (as in the previous section, we will often leave the $r$ indices implicit, and think $\psi^+_{\xx,\omega}$ resp. $\psi^-_{\xx,\omega}$ as two-component row resp. column vectors):
\begin{eqnarray}\label{eq:qp}
&&\psi^{(\leq h_{*})\pm}_{\xx} := \sum_{\omega = \pm} e^{\pm i \pp_{F}^{\omega} \xx} \psi^{(\leq h_{*})\pm}_{\xx, \omega}\;,\nonumber\\
&&\psi^{(\leq h_{*})\pm}_{\xx, \omega} := \frac{1}{\beta L^{3}} \sum_{\kk'} e^{\pm i \kk' \xx} \hat \psi^{(\leq h_{*})\pm}_{\kk',\omega}\;,\quad \hat \psi^{(\leq h_{*})\pm}_{\kk', \omega} := \hat \psi^{(\leq h_{*})\pm}_{\pp_{F}^{\omega} + \kk'}\;.
\end{eqnarray}
where in the second line the sum over $\kk'$ runs over the four-dimensional vectors such that $\kk'+\pp_F^\omega$ is in $\mathbb{D}_{\beta, L, N}\cap{\rm{supp}}(\chi_{h_*})$
and $|k'_3|<|p_F^\omega|$. The quasi-particle fields $\hat \psi^{(\leq h_{*})\pm}_{\kk',\omega}$ have covariance $\hat g^{(\leq h_{*})}_{\omega}(\kk') =
\chi_{h_{*},\omega}(\kk')Z_{h_{*}}^{-1} A_{h_{*},\omega}(\kk')^{-1}$, 
with $\chi_{h_{*},\omega}(\kk') := \chi_{h_{*}}(\kk'+\pp_{F}^{\omega})\mathds 1(|k'_3|<|p_F^\omega|)$, 
and $A_{h_{*},\omega}(\kk') = A_{h_{*}}(\kk' + \pp_{F}^{\omega})$. 

As in the previous regime, we integrate the scales $h\le h_{*}$ in a multiscale fashion: at each step, we decompose the quasi-particle Grassmann field $\psi^{(\le h)}_\omega$ 
as the sum of two independent fields, one describing the fluctuations at scale $h$, and the other at smaller scales: $\psi^{(\le h)}_\omega
=\psi^{(\le h-1)}_\omega+\psi^{(h)}_\omega$; we decompose the effective potential as the sum of a localized part plus an `irrelevant' remainder; we combine part 
of the localized part of the effective potential with the Grassmann measure; we integrate out the field at scale $h$; we iterate. 
More precisely, we inductively assume that, for any $h_\beta\le h\le h_*$, with $h_\beta=\lfloor \log_{\gamma}
(\pi/\beta)\rfloor$, the generating functional of the correlations can be written as: 
\begin{equation}\label{ind_2nd}
e^{\mathcal{W}_{\beta, L}(A, A^{5}, 0)} = e^{\mathcal{W}^{(h)}(A, A^{5})}\int P_{\leq h}(d\psi^{(\leq h)}) e^{\mathcal{V}^{(h)}(A, A^{5},\psi^{(\leq h)})}\;.
\end{equation}
Let us explain the meaning of the various objects involved: $P_{\leq h}(d\psi^{(\leq h)})$ denotes a Gaussian Grassmannn integration with covariance 
\begin{equation}
\int P_{\leq h} (d\psi^{(\leq h)}) \hat \psi^{-(\leq h)}_{\omega_{1}, \kk'_{1}, r_{1}} \hat \psi^{+(\leq h)}_{\omega_{2}, \kk'_{2}, r_{2}} = \beta L^{3} \delta_{\omega_{1},\omega_{2}}\delta_{\kk'_{1},\kk'_{2}}\big[\hat g^{(\leq h)}_{\omega_{1}}(\kk'_{1})\big]_{r_1,r_2}\;,
\end{equation}
where 
\begin{equation}\label{gleh2d}
\hat g^{(\leq h)}_{\omega}(\kk') := 
\frac{\chi_{h,\omega}(\kk')}{Z_{h}} A_{h,\omega}(\kk')^{-1}
\end{equation}
and: $\chi_{h,\omega}(\kk') := \chi_{h}(\kk'+\pp_{F}^{\omega})\mathds 1(|k'_3|<|p_F^\omega|)$, $Z_{h}$ is a real-analytic function of $\lambda$, to be inductively defined below, 
\begin{equation}\label{Ah2nd}
A_{h,\omega}(\kk') = \begin{pmatrix} -ik_{0}' + c_{h,\omega}(k')+d_{h,\omega}(k') & a_{h,\omega}(k') - i b_{h,\omega}(k') \\ a_{h,\omega}(k') + ib_{h,\omega}(k') & 
-ik_{0}' - c_{h,\omega}(k')+d_{h,\omega}(k') \end{pmatrix}\;,
\end{equation}
with 
\begin{eqnarray}
&& a_{h,\omega}(k')=v_{1,h} k_1'+a_R(k'+p_F^\omega)/Z_h, \quad b_{h,\omega}(k')=v_{2,h} k_2'+b_R(k'+p_F^\omega)/Z_h,\\
&& c_{h,\omega}(k')=\omega v_{3,h}k_3'+	\tilde c_{R,\omega}(k')/Z_h \hskip1.7truecm d_{h,\omega}(k')=d(k'+p_F^\omega)/Z_h,\end{eqnarray}
and, recalling that $c_{R,\omega}$ was defined by \eqref{eq:linear2gen}, we let $\tilde c_{R,\omega}(k'):=\frac{b^0+\zeta_{h_*}}{2}(k_3')^2+c_{R,\omega}(k'+p_F^\omega)$.
Moreover, the function $\mathcal{V}^{(h)}$ in the right side of \eqref{ind_2nd} is the effective potential, which can be represented in a way analogous to \eqref{eq:Vh}, namely
\begin{eqnarray}
\mathcal{V}^{(h)}(A, A^{5}, \psi) &=& \sum_{\substack{n, m_{1}, m_{2} \geq 0 \\ n>0}} \sum_{\substack{\underline{q},\underline{r}, \underline{\omega}, \\  \underline{\mu}, \underline{\nu}}}\int d{\bf X} d{\bf Y} d{\bf Z}\, \Big[ \prod_{i=1}^{2n} \hat\partial^{q_{2i-1}}\psi^{+}_{\xx_{2i-1}, \omega_{2i-1}, r_{2i-1}} \hat\partial^{q_{2i}}\psi^{-}_{\xx_{2i}, \omega_{2i}, r_{2i}} \Big]\cdot
\nonumber\\
&\cdot&\Big[ \prod_{j=1}^{m_{1}} A_{\mu_{j}, \yy_{j}} \Big] \Big[ \prod_{\ell = 1}^{m_{2}} A^{5}_{\nu_{\ell}, \zz_{\ell}} \Big] W^{(h)}_{n, m_{1}, m_{2},\underline{q};  \underline{\omega}, \underline{r},\underline{\mu}, \underline{\nu}}({\bf X}, {\bf Y}, {\bf Z})\;,\label{eq:Vh2}
\end{eqnarray}
and $\mathcal{W}^{(h)}(A, A^{5})$ is the finite-scale contribution to the generating function, which admits a representation analogous to \eqref{eq:Vh2}, with the difference that only 
terms with $n=0$ contribute to the sums. 

The inductive assumption \eqref{ind_2nd} and following equations is verified at scale $h = h_{*}$, as an outcome of the integration of the first regime, provided we let 
\begin{equation}\label{v3h*}v_{3,h_*}=v_3^0+\zeta_{h_*}p_{F,3}\end{equation}
(note that the representation \eqref{eq:Vh2} at scale $h_*$ follows from \eqref{eq:Vh} at the same scale, in combination with the definition of quasi-particle fields (\ref{eq:qp})).
We now assume that \eqref{ind_2nd} is valid at scale $h$ and prove it for $h-1$. We decompose $\mathcal{V}^{(h)} = \mathcal{L} \mathcal{V}^{(h)} + \mathcal{R} \mathcal{V}^{(h)}$, 
with $\mathcal{L}$ the localization operator, defined as follows:
\begin{eqnarray}\label{eq:ell.2nd}
 \mathcal{L} \mathcal V^{(h)}(A,A^5,\psi)&:=&\sum_{\omega=\pm} \int \frac{d\kk'}{(2\pi)^4} \hat \psi^+_{\kk',\omega}\big(\mathcal L \hat W^{(h)}_{1,0,0;(\omega,\omega)}(\kk')\big)\hat \psi^-_{\kk',\omega}\\
&+&\sum_{\omega=\pm}
\int \frac{d\kk'}{(2\pi)^4}\int \frac{d\pp}{(2\pi)^4} \Big[\hat A_{\mu,\pp}\hat \psi^+_{\kk'+\pp,\omega}\big(\mathcal L \hat W^{(h)}_{1,1,0;(\omega,\omega),\mu}(\kk',\pp)\big)\hat \psi^-_{\kk',\omega}\nonumber\\
&+&
\hat A_{\mu,\pp}^5\hat \psi^+_{\kk'+\pp,\omega}\big(\mathcal L \hat W^{(h)}_{1,0,1;(\omega,\omega),\mu}(\kk',\pp)\big)\hat \psi^-_{\kk',\omega}\Big],\nonumber\end{eqnarray}
where $\int \tfrac{d\kk}{(2\pi)^4}$ and $\int\tfrac{d\pp}{(2\pi)^4}$ are shorthands for $(\beta L^3)^{-1}\sum_{\kk'}$ and $(\beta L^3)^{-1}\sum_{\pp}$,
and 
\begin{eqnarray}\label{eq:ell2nd}
\mathcal{L} \hat W^{(h)}_{1,0,0;(\omega,\omega)}(\kk') &:=& \hat W^{(h),\infty}_{1,0,0;(\omega,\omega)}({\bf 0}) + \sum_{\mu=0}^3 k_{\mu}' \partial_{\mu} \hat{W}^{(h),\infty}_{1,0,0;(\omega,\omega)}({\bf 0})\;,\\
\mathcal{L} \hat W^{(h)}_{1,1,0;(\omega,\omega),\mu}(\kk',\pp) &:=& \hat W^{(h),\infty}_{1,1,0;(\omega,\omega),\mu}({\bf 0}, {\bf 0})\;,\qquad 
\mathcal{L} \hat W^{(h)}_{1,0,1;(\omega,\omega),\mu}(\kk', \pp) := \hat W^{(h),\infty}_{1,0,1;(\omega,\omega),\mu}({\bf 0}, {\bf 0})\;,\nonumber
\end{eqnarray}
where $\hat W^{(h),\infty}_{1,0,0;(\omega,\omega)}(\kk')$ indicates the $\beta,L\to\infty$ limit of $\hat W^{(h)}_{1,0,0;(\omega,\omega)}(\kk')$, and similarly for\\ 
$\hat W^{(h)}_{1,1,0;(\omega,\omega),\mu}$ $(\kk',\pp)$  and  $\hat W^{(h)}_{1,0,1;(\omega,\omega),\mu}(\kk',\pp)$.

\medskip

{\bf Remark.} We anticipate the fact that the scaling dimension of 
$\hat W^{(h),\infty}_{n,m_1,m_2;\underline\omega,\underline \mu,\underline \nu}$ in this second regime is $4-\frac32n-m_1-m_2$, see \eqref{eq:bdW2};
any additional derivative with respect to $k_\mu$ decreases the 
scaling dimension by $1$, so that, from the definitions, it follows that $\mathcal{R} \hat W^{(h)}_{1,0,0}(\kk')$, $\mathcal R\hat W^{(h)}_{1,1,0;(\omega,\omega),\mu}(\kk',\pp)$ 
and $\mathcal R \hat W^{(h)}_{1,0,1;(\omega,\omega),\mu}(\kk',\pp)$ are all irrelevant, with scaling dimension $-1$. The reader may recognize that the scaling dimensions are the 
same as those of QED$_4$: our theory can be seen as a lattice realization of lattice infrared QED$_4$, 
with the important difference that the speed of light is replaced by the anisotropic running velocities $v_{\mu,h}$. Since $v_{3,h}$ (or, equivalently, $v_{3}^0$) vanishes
in the limit as the Weyl nodes merge, 
the dimensional bounds on the effective potentials, see \eqref{eq:bdW2} below, are potentially affected by dangerous $1/v_{3,h}$ factor, which may a priori have an impact 
on the convergence properties of the infrared expansion of the observables of interest. However, by carefully tracking the loss and gain of factors $v_{3,h}$, due to 
the non-uniform dependence of the propagator upon $v_{3,h}$ and to the difference of scaling dimensions between the first and second regimes, one finds that the 
bad and good dependences upon these factors compensates, and lead to the overall factor $|v_3^0|^{n-1+\|\underline q^3\|_1}$ in \eqref{eq:bdW2} below; see \cite[Section 3]{M1bis}
and Appendix \ref{app.dimbounds}.

\medskip

As in the regime $h>h_{*}$, we shall use the localized term $\mathcal{L}\mathcal{V}^{(h)}$ to redefine the Grassmann integration and the coupling of the fermions with the external fields. The next lemma establishes important symmetry properties of the kernels of $\mathcal{L}\mathcal{V}^{(h)}$ (see Appendix \ref{app:conssym} for the proof). 

\begin{lemma}\label{lem:sym2} The following identities hold:
\begin{equation}\label{eq:sym2}
\hat{W}^{(h),\infty}_{1,0,0;(\omega,\omega)}({\bf 0}) = n_{h}\sigma_{3} \;,\qquad \partial_{\mu} \hat{W}^{(h),\infty}_{1,0,0;(\omega,\omega)}({\bf 0}) = z_{\mu,h} \sigma_{\mu,\omega},
\end{equation}
with $n_{h}, z_{\mu,h} \in \mathbb{R}$, and 
\begin{equation} \sigma_{\mu,\omega}:=\begin{cases} \sigma_\mu & \text{if $\mu=0,1,2$,}\\
\omega\sigma_3 & \text{if $\mu=3$.}\end{cases}\label{defsigmamuomega}\end{equation} 
Moreover, 
\begin{equation}
\hat{W}^{(h),\infty}_{1,1,0;(\omega,\omega), \mu}({\bf 0}, {\bf 0}) = Z_{\mu,h} \sigma_{\mu,\omega} ,
\quad \hat{W}^{(h),\infty}_{1,0,1;(\omega,\omega),\mu}({\bf 0}, {\bf 0}) = \omega Z_{\mu,h}^5 \sigma_{\mu,\omega},\label{locApsipsi}
\end{equation}
with $Z_{\mu,h}, Z_{\mu,h}^5\in\mathbb R$.
\end{lemma}

We now: rewrite $\mathcal{V}^{(h)}=\mathcal{L}\mathcal{V}^{(h)}+\mathcal{R}\mathcal{V}^{(h)}$ in the right side of \eqref{ind_2nd}; then write 
$\mathcal{L}\mathcal{V}^{(h)}$ explicitly, in view of Lemma \ref{lem:sym2}, in terms of $n_h$, $z_{\mu,h}$, $Z_{\mu,h}$ and $Z_{\mu,h}^5$; then 
re-absorb the quadratic part of $\mathcal{L}\mathcal{V}^{(h)}$ associated with 
$\sum_{\mu} k_{\mu}' \partial_{\mu} \hat{W}^{(h),\infty}_{1,0,0;\omega}({\bf 0})\equiv \sum_{\mu} k_\mu'\sigma_{\mu,\omega} z_{\mu,h}$ 
in the Gaussian Grassmann integration, thus getting
\begin{equation}\label{eq:WbetaL2}
e^{\mathcal{W}_{\beta, L}(A, A^{5}, 0)} = e^{\mathcal{W}^{(h)}(A, A^{5})}
\int \widetilde{P}_{\leq h}(d\psi^{(\leq h)}) e^{2^{h} \nu_{h} N_{3}(\psi^{(\leq h)}) + B^{(h)}(A, A^{5},\psi^{(\leq h)})+\mathcal{R} \mathcal{V}^{(h)}(A, A^{5}, 
\psi^{(\leq h)})}\;,\end{equation}
where: $2^{h} \nu_{h} \equiv n_{h}$, $N_{3}(\psi) = \sum_{\omega} \int \frac{d\kk'}{(2\pi)^{4}}\, \hat\psi^{+}_{\kk',\omega} \sigma_{3} \hat\psi^{-}_{\kk',\omega}$,
\begin{equation}\label{Bh2nd}
B^{(h)}(A, A^{5}, \psi) =\sum_{\omega,\mu}  \int \frac{d\kk'}{(2\pi)^{4}} \frac{d\pp}{(2\pi)^{4}}\big(Z_{\mu,h}\hat A_{\mu,\pp}+\omega Z_{\mu,h}^5\hat A^5_{\mu,\pp}\big)
\hat \psi^{+}_{\kk'+\pp,\omega} \sigma_{\mu,\omega} \hat \psi^{-}_{\kk',\omega}\;,\end{equation}
and the new Grassmann Gaussian integration $\widetilde P_{\leq h}(d\psi^{(\leq h)})$ has covariance given by:
\begin{equation}
\tilde g^{(\leq h)}_{\omega}(\kk') =  \frac{\chi_{h,\omega}(\kk')}{Z_{h-1,\omega}(\kk')} \tilde A_{h-1,\omega}(\kk')^{-1}\;,
\end{equation}
where: $Z_{h-1,\omega}(\kk')=Z_h-\chi_{h,\omega}(\kk')\, z_{0,h}$,  
\begin{equation}
\tilde A_{h-1,\omega}(\kk') = \begin{pmatrix} -ik_{0}' + \tilde c_{h-1,\omega}(\kk')+\tilde d_{h,\omega}(\kk') & \tilde a_{h-1,\omega}(\kk')-i\tilde b_{h-1,\omega}(\kk') \\ 
\tilde a_{h-1,\omega}(\kk')+i\tilde b_{h-1,\omega}(\kk') &  -ik_{0}' - \tilde c_{h-1,\omega}(\kk')+\tilde d_{h-1,\omega}(\kk') \end{pmatrix}\;,
\end{equation}
and
\begin{eqnarray} 
&& \tilde a_{h-1,\omega}(\kk')=v_{1,h-1,\omega}(\kk') k_1'+\frac{a_R(k'+p_F^\omega)}{Z_{h-1,\omega}(\kk')}, \qquad \tilde b_{h-1,\omega}(\kk')=v_{2,h-1,\omega}(\kk') k_2'+\frac{b_R(k'+p_F^\omega)}{Z_{h-1,\omega}(\kk')},\nonumber\\
&&\tilde c_{h-1,\omega}(\kk')=\omega\, v_{3,h-1,\omega}(\kk') k_3'+\frac{\tilde c_{R,\omega}(k')}{Z_{h-1,\omega}(\kk')},\qquad \tilde d_{h-1,\omega}(\kk')=\frac{d(k'+p_F^\omega)}{Z_{h-1,\omega}(\kk')},\nonumber\end{eqnarray}
with
\begin{equation}
v_{l,h-1,\omega}(\kk')=\frac{Z_h v_{l,h}-\chi_{h,\omega}(\kk')z_{l,h}}{Z_{h-1,\omega}(\kk')}, \qquad l=1,2,3.
\end{equation}
Similarly to the previous regime, we inductively assume that, for all scales $h\le k\le h_*$ and any $\theta\in(0,1)$:
\begin{equation}\label{eq:ind}
 |z_{\mu,k}| \leq C_\theta|\lambda|\, |v_\mu^0| \gamma^{\theta k}, 
\end{equation}
where $v_0^0:=1$. We will check the validity of these bounds on scale $k=h-1$. We also assume that $\nu_k$ satisfies \eqref{eq:nu}, for all $h\le k\le h_*$.
Notice that (\ref{eq:ind}), in combination with the bounds on $Z_h, v_{l,h}$ derived in Section \ref{sec:1stregime}, implies:
\begin{equation}
|Z_{h-1,\omega}(\kk') - 1|\leq C|\lambda|, \qquad |v_{l,h-1,\omega}(\kk') - v_{l}^0| \leq C|\lambda|\,  |v_{l}^0|. \label{eq:Zvlbounds2nd}\end{equation}
We also let 
$$Z_{h-1}:=Z_{h-1,\omega}({\bf 0}) \qquad \text{and}\qquad v_{l,h-1}:=v_{l,h-1,\omega}({\bf 0}),$$
and, of course, these effective parameters satisfy the same bounds as \eqref{eq:Zvlbounds2nd}. 
To check the inductive assumption, we decompose the Grassmann field as:
\begin{equation}
\psi_{\omega}^{(\leq h)} = \psi_{\omega}^{(\leq h-1)} + \psi_{\omega}^{(h)}\;,
\end{equation}
where $\hat \psi^{(\leq h-1)}_\omega$ has covariance given by $\hat g_\omega^{(\leq h-1)}(\kk)$, defined as in (\ref{gleh2d}), (\ref{Ah2nd}), with ${h}$ replaced 
by $h-1$, while $\psi_\omega^{(h)}$ has covariance:
\begin{equation}\label{eq:singlescale2nd}
\hat g^{(h)}_{\omega}(\kk') =  \frac{f_{h,\omega}(\kk')}{Z_{h-1,\omega}(\kk')} \tilde A_{h-1,\omega}(\kk')^{-1}\;,
\end{equation}
with $f_{h,\omega}(\kk') = \chi_{h,\omega}(\kk') - \chi_{h-1,\omega}(\kk')$ and we used the fact that, on the support of $\chi_{h-1,\omega}$, $Z_{h-1,\omega}(\kk')=Z_{h-1}$ and $\tilde A_{h-1}(\kk')=
A_{h-1}(\kk')$. On the support of $f_{h,\omega}$, 
$$C^{-1}\gamma^{2h}\le  -\det \tilde A_{h-1,\omega}(\kk') \le C \gamma^{2h},$$ $|k_{\mu}'|\leq C\gamma^{h}$ for $\mu=0,1,2$, and $|k_{3}'| \leq C\gamma^{h}/|v_3^0|$. 
Therefore, for these values of $\kk'$, recalling the definition of $h_*$, 
\begin{eqnarray}
\tilde a_{h-1}(\kk) &=& v_{1,h-1,\omega}(\kk')k_{1}'+ O(\gamma^{2h}), \hskip1.8truecm \tilde b_{h-1}(\kk) = v_{2,h-1,\omega}(\kk')k_{2}' +  O(\gamma^{2h}),\nonumber\\
\tilde c_{h-1}(\kk)&=&\omega v_{3,h-1,\omega}(\kk')k_{3}' + O(\gamma^{2h-h_*}), \qquad\ \tilde d_{h-1}(\kk)=O(\gamma^{2h}),\nonumber 
\end{eqnarray}
which implies $\|\hat g^{(h)}(\kk)\|_{\infty} \leq C  \gamma^{-h}$ and $\|\hat g(\kk) \|_{1} \leq C \gamma^{3 h}/|v_3^0|$. By a similar discussion, we can also 
dimensionally bound the derivatives of $\tilde a_{h-1},\tilde b_{h-1},\tilde c_{h-1},\tilde d_{h-1}$, thus getting 
\begin{equation}\label{dimgh2nd}
\| \partial^{\alpha}_{\kk'} \hat g^{(h)}_{\omega}(\kk') \|_{\infty} \leq C_{|\alpha|} |v^{0}_{3}|^{|\alpha_{3}|} \gamma^{-h(1 + |\alpha|)}\;,\qquad \| \partial_{\kk'}^{\alpha} \hat g_{\omega}(\kk') \|_{1} \leq C_{|\alpha|}\gamma^{h(3 - |\alpha|)}/|v_3^0|\;,
\end{equation}
where $\alpha=(\alpha_0,\alpha_1,\alpha_2,\alpha_3)$ and $|\alpha|=\sum_\mu|\alpha_\mu|$. These in turn imply the following bound in configuration space:
\begin{equation}\label{eq:ghlow}
\| g^{(h)}_{\omega}(\xx) \| \leq (C_0/|v^{0}_{3}|) \gamma^{3h} e^{-\kappa_0\sqrt{\gamma^h d(\xx)}}
\end{equation}
for suitable $C_0,\kappa_0>0$, where $d(\xx):= |x_{0}| + |x_{1}| + |x_{2}| + |x_{3}|/|v_3^0|$.  

At this point, we integrate the field $\psi^{(h)}$ out, as done in Eqs.(\ref{eq:sscale1}) and (\ref{eq:sscale2}), thus obtaining the analogue of (\ref{eq:sscale2}), where the effective action $\mathcal{V}^{(h-1)}$ can be represented as in \eqref{eq:Vh2}, with $h$ replaced by $h-1$. Also in this case, the kernels of the effective potential can be represented in terms of a convergent GN tree expansion, and are analytic in $\lambda$, for $|\lambda|$ small, {\it uniformly in $\zeta$}, i.e., in the distance between the Weyl nodes. 
They satisfy the following weighted $L^1$ estimates: for all $h_\beta\le h<h_*$, $n\ge 1$, and $|\lambda|$ small enough, uniformly in $\zeta$, if \eqref{eq:ind} and \eqref{eq:nu}
are valid for $h<k\le h_*$, then, see \cite[Lemma 2, Eq.(93)]{M1bis}, 
\begin{eqnarray}&&\frac{1}{\beta L^{3}} \int d{\bf Q}\, e^{\frac{\kappa_0}{2}\sqrt{\gamma^h\delta_*({\bf Q})}}\, \big|W^{(h)}_{n, m_1,m_2,\underline{q};\underline{\omega},\underline{r},\underline{\mu},\underline{\nu}}({\bf Q})\big|\le\label{eq:bdW2} \\
&&\   \leq C^{n+m_1+m_2} |v_3^0|^{n-1+\|\underline q^3\|_1}(\max_{k>h}\mathfrak{Z}_k)^{m_1}(\max_{k>h}\mathfrak{Z}^5_k)^{m_2}\,
|\lambda|^{\max\{\delta_{m_1+m_2,0},\, n-1\}} \gamma^{h (4-3n - m_{1} - m_{2} -\|\underline{q}\|_1)} \;,\nonumber
\end{eqnarray}
where $\delta_*({\bf Q})$ is the Steiner diameter measured by using the norm $d(\xx)$ defined after \eqref{eq:ghlow}, 
$\|\underline q^3\|_1=\sum_{i=1}^{2n}|q^3_i|$, $\|\underline q\|_1=\sum_{i=1}^{2n}\sum_{\mu=0}^3|q^\mu_i|$, and, if $k\le h_*$, 
$\mathfrak{Z}_k=\max_\mu|Z_{\mu,k}|$ and $\mathfrak{Z}_k^5=\max_\mu|Z_{\mu,k}^5|$ (while, if $k>h_*$, $\mathfrak{Z}_k$ and $\mathfrak{Z}^5_k$ were defined after 
\eqref{eq:bdW}). The kernels of the single-scale contribution to the generating function, $\widetilde{\mathcal{W}}^{(h)}(A,A^5)$ satisfy an estimate analogous to \eqref{eq:bdW2} with $n=0$. 
The combination $4-3n - m_{1} - m_{2} -\|\underline{q}\|_1$ is the {\it scaling dimension} of the kernels with $2n$ Grassmann fields, derivatives of order $\underline{q}$ 
acting on them,  
and $m_1+m_2$ external fields of type $A$ or $A^5$ in the second regime $h< h_*$. Also in this case, the quartic interaction ($n=2, m_1=m_2=0, \|\underline{q}\|_1=0$) is irrelevant, 
and its scaling dimension is $-2$. Also in  the second regime, we can obtain improved bounds on all the contributions to 
the kernels associated with GN trees containing at least one interaction endpoint: more precisely,
if we split $W^{(h)}_{n, m_1,m_2,\underline{q}} = W^{(h);\text{d}}_{n, m_1,m_2,\underline{q}} + W^{(h);\text{r}}_{n, m_1,m_2,\underline{q}}$,
where the second term collects the contribution of all GN trees with at least one endpoint of type $\nu_k$ or one endpoint on scale $h_*+1$ (including, in particular, all the contributions from GN trees 
with endpoints associated with quartic interactions), the term $W^{(h);\text{r}}_{n, m_1,m_2,\underline{q}}$ admits an 
improved dimensional bound, analogous to \eqref{eq:bdW}, but with an extra factor in the right side equal to:
$\gamma^{\theta h}$, if $m_1=m_2=0$ or $n\ge 2$; $\gamma^{\theta(h-h_*)}$, if $m_1+m_2\ge 1$ and $n=1$ (again, $\theta$ can be chosen to be any positive exponent smaller than $1$, at the cost of replacing the constant $C$ in the right side of \eqref{eq:bdW2} by $C_\theta$, which possibly diverges as $\theta\to 1^-$). 
The proof is completely analogous to that of Eq.(94) of Lemma 2 in \cite{M1bis}, modulo a few minor differences discussed in Appendix \ref{app.dimbounds}.

The definition of $z_{\mu,h-1}$ and the bound on $W^{(h-1);\text{r}}_{n, m_1,m_2,\underline{q};\underline{\omega},\underline{r},\underline{\mu},\underline{\nu}}$, 
readily imply the validity of (\ref{eq:ind}) at scale $k=h-1$, as desired. Concerning \eqref{eq:nu}, also in this second regime we find that $\nu_h$ satisfies the beta function equation
\eqref{eq:betanu1st}, with $\beta^\nu_h$ satisfying \eqref{betanubd1st} even for $h<h_*$. The choice of the initial datum $\nu_{0}$ (or, equivalently, of the staggered chemical potential $\nu$) 
for which $\nu_k$ satisfies \eqref{eq:nu} for all $h\le 0$ will be discussed in the next section, where we will also derive bounds on the flow of the effective couplings 
$Z_{\mu,h}, Z^5_{\mu,h}$. 

The iterative RG integration of the second regime goes on until we reach scale $h_\beta$. At that point, we integrate `in one shot' all the remaining Grassmann degrees of freedom,
thus obtaining the desired generating function of correlations, in the form of a sum of single-scale contributions from $h=h_\beta$ to $h=0$ 
(the covariance of $\psi^{(\le h_\beta)}_\omega$ admits the same dimensional bounds as the one of 
$\psi^{(h_\beta+1)}_\omega$; therefore, the result of the integration of $\psi^{(\le h_\beta)}_\omega$ admits the same qualitative bounds as the one at scale $h_\beta+1$, 
so that the last iteration step does not give any additional difficulty). 

The uniformity in $\beta,L$ and $\zeta$ of the convergence of the GN tree expansion for the single-scale contributions to the generating function, as well as the
dimensional bounds \eqref{eq:bdW} and \eqref{eq:bdW2}, imply that the correlation functions at fixed space-time
positions are analytic functions of $\lambda$ and $\{\nu_h\}_{h\le 0}$, 
in a sufficiently small neighbourhood of the origin, uniformly in $\beta,L$ and $\zeta$. Elaborating on this and on the fact that 
all the Taylor coefficients of such convergent expansions admit a limit as $\beta,L\to\infty$ implies the existence of the $\beta,L\to\infty$ limit of the Euclidean correlations, as
well as the analyticity of the limits, stated in Theorem \ref{thm:main}, provided that the sequence $\{\nu_h\}_{h\le 0}$ satisfies the promised bounds, see \eqref{eq:nu}, and is itself 
analytic in $\lambda$. This will be proved in the next section, together with the bounds on the flow of the effective vertex functions $Z_h, Z_h^{5}$. 

\subsection{The flow of effective couplings}\label{sec:flow}

In this section we discuss the RG flow of the effective parameters $\nu_h$, $Z_{\mu,h}$, $Z^5_{\mu,h}$ (as well as the one of $\tilde Z_{3,h}$ and $\tilde Z^5_{\mu,h}$ with 
$\mu=0,1,2$, which appear in the first regime). In particular, we explain how to fix the bare staggered chemical potential $\nu$ in \eqref{eq:H} in such a way that the running staggered 
chemical potential $\nu_h$ satisfies \eqref{eq:nu} at all scales $h\le 0$. We also explain how to fix $Z^5_{\mu,\text{bare}}$, see \eqref{eq:J5def}, in such a way that their  
dressed counterpart, $Z^5_{\mu,h}$, flow to any prescribed $4$-tuple of values as $h\to-\infty$. The bounds discussed in the following use a few properties of the 
GN trees contributing to the effective potential, discussed in \cite[Sections 2 and 3]{M1bis}, see also Appendix \ref{app.dimbounds}. 

\subsubsection{Flow of $\nu$.} Starting from the beta function equation for $\nu_h$, \eqref{eq:betanu1st}, we find 
\begin{equation} \nu_0 = 2^k \nu_k-\sum_{k<h\le 0}2^{h-1}\beta^\nu_h,\end{equation}
for any $k<0$. Here $\nu_0$ is an analytic function of $\lambda$ and $\nu$ that, assuming $\nu$ to be of order $\lambda$, is of the form $\nu_0=\nu+O(\lambda^2)$. 
If we now require that $\nu_k$ tends to zero as $k\to-\infty$, we get  
\begin{equation} \nu_0 = -\sum_{h\le 0}2^{h-1}\beta^\nu_h,\end{equation}
and, more in general, 
\begin{equation} \nu_k = -\sum_{h\le k}2^{h-k-1}\beta^\nu_h.\label{eq:betanu_reloaded}\end{equation}
Recall that $\beta^\nu_h$ is a function of $\lambda$ and of the effective parameters $\nu_{h'}$ on scales $h'\ge h$. Therefore, 
we can regard the right side of \eqref{eq:betanu_reloaded} as a function of the whole sequence $\underline\nu:=\{\nu_k\}_{k\le 0}$, which we denote by 
$T_k(\underline\nu,\lambda)$, so that \eqref{eq:betanu_reloaded} can be read as a fixed point equation $\nu_k=T_k(\underline\nu,\lambda)$ for the sequence $\underline\nu$. 
By proceeding as done in many other papers before, see e.g. \cite[Section 6.4.2]{GMT20}, we look for a fixed point in the Banach space of sequences $\underline\nu$ such that 
$\|\underline\nu\|_\theta:=\sup_{h\le 0}|\nu_h|\gamma^{-\theta h}\le K|\lambda|$, for $\theta=3/4$ (say) and $K$ a suitable (sufficiently large) constant. Following the same strategy as 
\cite[Section 6.4.2]{GMT20} (in a much simpler setting), the reader can check that 
the map $T:\underline\nu\to \{T_k(\underline\nu,\lambda)\}_{k\le 0}$ is a contraction on such a Banach space, which implies the existence of a unique fixed point in that space. 
The value of $\nu_0$ of such a fixed point sequence corresponds to the `right' initial datum to assign in order for \eqref{eq:nu} to be satisfied at all scales. Finally, note that 
fixing $\nu_0$ is equivalent to fixing $\nu$ (recall that $\nu_0=\nu_0(\nu,\lambda)$ is analytic in $\nu,\lambda$ and that $\nu_0(\nu,\lambda)=\nu_0+O(\lambda^2)$; appealing to the 
 implicit function theorem, we can analytically invert $\nu_0$ with respect to $\nu$). The value of the bare staggered chemical potential $\nu$ fixed via this strategy is 
 the one stated in Theorem \ref{thm:main}. 

\subsubsection{Flow of $Z_{\mu,h}$.} In the {\it first regime}, recalling that $Z_{\mu,h}$ is defined via \eqref{2ndeq:lem3.2}, 
we write 
\begin{equation} 
Z_{\mu,h-1}=Z_{\mu,h}+\beta_{\mu,h},\label{flowZ}
\end{equation}
where $\beta_{\mu,h}$ includes the contributions from GN trees that have at least one endpoint of type $\nu_h$ or one endpoint on scale $1$ of order $\lambda$. 
Therefore, $\beta_{\mu,h}$ is bounded in the same way as $W^{(h);\text{r}}_{1, 1,0,\underline{q}}$, with $\|\underline q\|_1=\|\underline q^3\|_1$ equal to $0$ or $1$, 
depending on whether $\mu=0,1,2$ or $\mu=3$, respectively; see the paragraph after \eqref{eq:bdW}. We thus get, for any $0<\theta<1$,
\begin{equation}
|\beta_{\mu,h}|\le C_\theta |\lambda|\gamma^{\theta h}\begin{cases}\sup_{h\le k\le 0}|Z_{\mu,k}| & \text{if $\mu=0,1,2$}\\ 
\gamma^{-h/2}\sup_{h\le k\le 0}(|Z_{3,k}|+|\tilde Z_{3,k}|) & \text{if $\mu=3$}\end{cases}\label{cases1}
\end{equation}
where $Z_{\mu,0}$ for $\mu=0,1,2,3$ and $\tilde Z_{3,0}$ are analytic functions of $\lambda$, bounded as $|Z_{\mu,0}-v_\mu^0|\le C|\l|\,|v_\mu^0|$ and $|\tilde Z_{3,0}|\le C|\lambda|\, |v_3^0|$. 

\medskip

{\bf Remark.}  The importance of having a gain factor $\gamma^{\theta h}$ with $\theta$ any positive constant smaller than $1$ (rather than $1/2$, as other simpler choices of the
localization operator would have implied), is apparent from \eqref{cases1}. In fact, choosing $\theta$ larger than $1/2$ makes all the components of the beta function summable in $h$, 
uniformly in $v_3^0$. A posteriori, this motivates the definitions \eqref{eq:ell}, see also the remarks following it. Similar considerations are valid for the beta function for $Z^5_{\mu,h}$, 
see in particular \eqref{cases2} below.

\medskip

 Similarly, the flow equation for $\tilde Z_{3,h}$ is  
\begin{equation} 
\tilde Z_{3,h-1}=\tilde Z_{3,h}+\tilde\beta_{3,h},
\end{equation}
with 
\begin{equation}
|\tilde \beta_{3,h}|\le C_\theta|\lambda|\gamma^{\theta h}\gamma^{-h/2}\sup_{h\le k\le 0}(|Z_{3,k}|+|\tilde Z_{3,k}|). 
\end{equation}
From these bounds on the beta function, we readily find 
\begin{equation}|Z_{\mu,h}-Z_{\mu,0}|\le C|\lambda|\, |Z_{\mu,0}|, \qquad |\tilde Z_{3,h}|\le C|\lambda|\, |Z_{3,0}|,\label{coronavI}\end{equation}
for any $h\ge h_*$, uniformly in $h$. 

\medskip

In the {\it second regime}, the flow of $Z_{\mu,h}$ is controlled by a flow equation of the same form as \eqref{flowZ}, where 
$\beta_{\mu,h}$  is bounded in the same way as $W^{(h);\text{r}}_{1, 1,0,\underline{0}}$, see the paragraph after \eqref{eq:bdW2}: 
\begin{equation}
|\beta_{\mu,h}|\le C|\lambda|\gamma^{\theta (h-h_*)}\sup_{h\le k\le h_*} |Z_{\mu,k}|.
\end{equation}
Using this bound on the beta function and \eqref{coronavI}, we readily find that 
\begin{equation}|Z_{\mu,h}-Z_{\mu,0}|\le C|\lambda|\, |Z_{\mu,0}|,\end{equation}
for any $h\le 0$, that $Z_\mu:=\lim_{h\to-\infty}Z_{\mu,h}$ exists and is analytically close to $Z_{\mu,0}$, and that the limit is reached at an exponential rate: 
$|Z_{\mu,h}-Z_\mu|\le C|\l|\, |Z_{\mu}|\gamma^{\theta (h-h_*)}$. 

\subsubsection{Flow of $Z_{\mu,h}^5$.}\label{sec:3.4.3} The discussion is very similar to that of the flow of $Z_{\mu,h}$. In the {\it first regime}, recalling that $Z^5_{\mu,h}$ is defined via \eqref{2ndeq:lem3.2}, 
we write 
\begin{equation} 
Z^5_{\mu,h-1}=Z^5_{\mu,h}+\beta^5_{\mu,h},\label{flowZ5}
\end{equation}
where $\beta^5_{\mu,h}$ includes the contributions from GN trees that have at least one endpoint of type $\nu_h$ or one endpoint on scale $1$ of order $\lambda$. 
Therefore, $\beta^5_{\mu,h}$ is bounded in the same way as $W^{(h);\text{r}}_{1, 0,1,\underline{q}}$, with $\|\underline q\|_1=\|\underline q^3\|_1$ equal to $0$ or $1$, 
depending on whether $\mu=3$, or $\mu=0,1,2$, respectively; see the paragraph after \eqref{eq:bdW}. We thus get, for any $0<\theta<1$,
\begin{equation}
|\beta^5_{\mu,h}|\le C|\lambda|\gamma^{\theta h}\begin{cases}
\gamma^{-h/2}\sup_{h\le k\le 0}(|Z^5_{\mu,k}|+|\tilde Z^5_{\mu,k}|) & \text{if $\mu=0,1,2$}\\
\sup_{h\le k\le 0} |Z^5_{3,k}| & \text{if $\mu=3$}\\ 
\end{cases}\label{cases2}
\end{equation}
where $Z^5_{\mu,0}$ for $\mu=0,1,2,3$ and $\tilde Z^5_{\mu,0}$ for $\mu=0,1,2$ are analytic functions of $\lambda$, bounded as $|Z^5_{\mu,0}-Z^5_{\mu,\text{bare}}|\le C|\l|\,
|Z^5_{\mu,\text{bare}}|$ and $|\tilde Z^5_{\mu,0}|\le C|\lambda|\, |Z^5_{\mu,\text{bare}}|$. Similarly, the flow equation for $\tilde Z^5_{\mu,h}$ for $\mu=0,1,2$ is  
\begin{equation} 
\tilde Z^5_{\mu,h-1}=\tilde Z^5_{\mu,h}+\tilde\beta^5_{\mu,h},
\end{equation}
with 
\begin{equation}
|\tilde \beta^5_{\mu,h}|\le C|\lambda|\gamma^{\theta h}\gamma^{-h/2}\sup_{h\le k\le 0}(|Z^5_{\mu,k}|+|\tilde Z^5_{\mu,k}|). 
\end{equation}
From these bounds on the beta function, we readily find 
\begin{equation}\label{coronavIR}|Z^5_{\mu,h}-Z^5_{\mu,\text{bare}}|\le C|\lambda|\, |Z^5_{\mu,\text{bare}}|, \qquad |\tilde Z^5_{\mu,h}|\le C|\lambda|\, |Z^5_{\mu,\text{bare}}|,
\end{equation}
for any $h\ge h_*$, uniformly in $h$. 

\medskip

In the {\it second regime}, the flow of $Z^5_{\mu,h}$ is controlled by a flow equation of the same form as \eqref{flowZ5}, where 
$\beta^5_{\mu,h}$  is bounded in the same way as $W^{(h);\text{r}}_{1, 0,1,\underline{0}}$, see the paragraph after \eqref{eq:bdW2}: 
\begin{equation}
|\beta^5_{\mu,h}|\le C|\lambda|\gamma^{\theta (h-h_*)}\sup_{h\le k\le h_*} |Z^5_{\mu,k}|.
\end{equation}
Using this bound on the beta function and \eqref{coronavIR}, we readily find that 
\begin{equation}|Z^5_{\mu,h}-Z^5_{\mu,\text{bare}}|\le C|\lambda|\, |Z_{\mu,\text{bare}}|,\end{equation}
for any $h\le 0$, that $Z^5_\mu:=\lim_{h\to-\infty}Z^5_{\mu,h}$ exists and is analytically close to $Z^5_{\mu,\text{bare}}$, and that the limit is reached at an exponential rate: 
$|Z^5_{\mu,h}-Z^5_\mu|\le C|\l|\, |Z^5_{\mu}|\gamma^{\theta (h-h_*)}$. Note that these facts imply that the relation between $Z_\mu^5$ and $Z^5_{\mu,\text{bare}}$ can be 
inverted into $Z^5_{\mu,\text{bare}}=(1+O(\lambda))Z^5_{\mu}$, 
thus allowing us to fix the dressed chiral renormalizations as desired. In particular, in the following we will need $Z^5_\mu\equiv Z_\mu$; therefore, we will fix the bare chiral parameters
in such a way that this condition is satisfied. 

\subsection{Asymptotic infrared behaviour of the correlation functions}\label{sec:sing}
In this section we use the RG construction described in the previous sections to study the $\beta,L\to\infty$ limit of the 
correlation functions in Equations \eqref{eq:3.15} to \eqref{FTPi5}. 
Our goal is to isolate the dominant part, at large distances and/or momenta close to the Weyl nodes, which has an explicit structure in terms of the dressed 
parameters 
\begin{equation}
\label{dressedRCC}
Z:=\lim_{h\to-\infty}Z_h, \qquad v_{l}:=\lim_{h\to-\infty} v_{l,h}, \qquad Z_{\mu}:=\lim_{h\to-\infty} Z_{\mu,h}, \qquad Z_{\mu}^5:=\lim_{h\to-\infty} Z_{\mu,h}^5,
\end{equation}
plus a subdominant part, which has better regularity properties in momentum space. 

\medskip

{\it The quadratic response coefficient to the chiral current.} We start by analyzing $\hat\Pi^{5}_{\mu,\nu,\sigma}(\pp_{1}, \pp_{2})$, which we rewrite as
\begin{equation}\label{eq:GammaR}
\hat\Pi^{5}_{\mu, \nu, \sigma}(\pp_{1}, \pp_{2}) = \hat\Pi^{5;\text{d}}_{\mu, \nu, \sigma}(\pp_{1}, \pp_{2}) + 
 \hat\Pi^{5;\text{r}}_{\mu, \nu, \sigma}(\pp_{1}, \pp_{2}),
\end{equation}
where the first term in the right side is, by definition, the sum of the dominant parts of the single-scale contributions to the generating function $\widetilde{\mathcal W}^{(h)}(A,A_5)$
of order $m_1=2$ in $A$ and $m_2=1$ in $A^5$, with $h<h_*$, in the $\beta,L\to\infty$ limit. More precisely, using 
the same convention of Fourier transform as in \eqref{FTPi5}, 
\begin{equation}\label{eq:JJJ}
 \hat\Pi^{5;\text{d}}_{\mu, \nu, \sigma}(\pp_{1}, \pp_{2})
:= \sum_{h<h_*} 
\hat W^{(h),\infty;\text{d}}_{0, 2, 1;\mu, \nu, \sigma}(\pp_{1}, \pp_{2})\;,\end{equation}
where $\hat W^{(h),\infty;\text{d}}_{0, 2, 1;\mu, \nu, \sigma}(\pp_{1}, \pp_{2})=\lim_{\beta,L\to\infty}\hat W^{(h);\text{d}}_{0, 2, 1;\mu, \nu, \sigma}(\pp_{1}, \pp_{2})$, and 
$\hat W^{(h),\infty;\text{d}}_{0, 2, 1;\mu, \nu, \sigma}(\pp_{1}, \pp_{2})$ collects the contributions from 
GN trees with one endpoint of type $A^5\psi^+\psi^-$ and two endpoints of type $A\psi^+\psi^-$, all on scales smaller than $h_*$. In terms of the decomposition 
\eqref{eq:euclide}, the Schwinger terms are all included in the remainder term $\hat\Pi^{5;\text{r}}_{\mu, \nu, \sigma}(\pp_{1}, \pp_{2})$, while 
$\hat\Pi^{5;\text{d}}_{\mu, \nu, \sigma}(\pp_{1}, \pp_{2})$
corresponds to the `infrared contribution' (i.e., the dominant one from the scales smaller than $h_*$) to $\pmb{\langle} {\bf T}\, \hat J^{5}_{\mu,\pp}\,; \hat J_{\nu,\pp_{1}}\,; \hat J_{\sigma,\pp_{2}} \pmb{\rangle}_\infty$. 

Thanks to the bound (\ref{eq:bdW2}) (or, better to its analogue with $n=0$) 
and to the boundedness of $\mathfrak{Z}:=\sup_{k\le 0}\mathfrak{Z}_k$ and of $\mathfrak{Z}^5:=\sup_{k\le 0}\mathfrak{Z}_k^5$, 
proved in the previous section, we see that the $h$-th term in the sum in the right side of \eqref{eq:JJJ} is bounded by $C\mathfrak{Z}^5(\mathfrak{Z})^2\gamma^h/|v_3^0|$. 
Therefore, the sum in the right  side of \eqref{eq:JJJ} is absolutely convergent. 
Moreover, each term in the sum is continuous with respect to $\pp_{1}$, $\pp_{2}$, uniformly in $h$. Hence, \eqref{eq:JJJ} is continuous as a function of the external momenta.

Note, however, that the bound (\ref{eq:bdW2}) does not imply {\it differentiability} of \eqref{eq:JJJ} in the external momenta. In fact, it implies that 
the derivatives in $\pp_1$ or $\pp_2$ of $\hat W^{(h),\infty;\text{d}}_{0, 2, 1;\mu, \nu, \sigma}(\pp_{1}, \pp_{2})$ are bounded by 
$C\mathfrak{Z}^5(\mathfrak{Z})^2|v_3^0|^{-1}$, which is clearly non summable over $h$. 

\medskip

Let us now consider the remainder term, 
\begin{equation}\label{eq:JJJr}
 \hat\Pi^{5;\text{r}}_{\mu, \nu, \sigma}(\pp_{1}, \pp_{2})
:= \sum_{h<h_*} 
\hat W^{(h),\infty;\text{r}}_{0, 2, 1;\mu, \nu, \sigma}(\pp_{1}, \pp_{2})+\sum_{h\ge h_*}\hat W^{(h),\infty}_{0, 2, 1;\mu, \nu, \sigma}(\pp_{1}, \pp_{2}) \;.\end{equation}
The summands and their derivatives are bounded via \eqref{eq:bdW}, \eqref{eq:bdW2}, and its improved version for the remainder term, discussed 
after \eqref{eq:bdW2}. These bounds imply that \eqref{eq:JJJr} is not only continuous in the external momenta, but also differentiable: in fact, they imply that 
the derivatives in $\pp_1$ or $\pp_2$ of $\hat W^{(h),\infty;\text{r}}_{0, 2, 1;\mu, \nu, \sigma}(\pp_{1}, \pp_{2})$ with $h<h_*$ are bounded by 
$C_\theta\mathfrak{Z}^5(\mathfrak{Z})^2|v_3^0|^{-1}\gamma^{\theta (h-h_*)}$, and that those of 
$\hat W^{(h),\infty}_{0, 2, 1;\mu, \nu, \sigma}(\pp_{1}, \pp_{2})$ with $h\ge h_*$ are bounded by 
$C\mathfrak{Z}^5(\mathfrak{Z})^2\gamma^{-h/2}$. Therefore, the derivatives in $\pp_1$ or $\pp_2$ of \eqref{eq:JJJr} are finite, and bounded by
$C\mathfrak{Z}^5(\mathfrak{Z})^2 |v_3^0|^{-1}$. Moreover, they are continuous in a sufficiently small neighbourhood of the origin; more precisely, the bound proportional 
to $\gamma^{\theta(h-h_*)}$ on the derivatives of $\hat W^{(h),\infty;\text{r}}_{0, 2, 1;\mu, \nu, \sigma}(\pp_{1}, \pp_{2})$ with $h<h_*$ implies that 
the derivatives in $\pp_1,\pp_2$ of \eqref{eq:JJJr} are H\"older continuous of exponent $\theta>0$, for any $\theta<1$. 

\medskip

Let us now go back to $\hat\Pi^{5;\text{d}}_{\mu, \nu, \sigma}(\pp_{1}, \pp_{2})$, which we want to further decompose into 
an explicitly computable contribution plus an additional differentiable remainder. We write: 
\begin{equation}\label{eq:JJJrelnonrel}
\hat\Pi^{5;\text{d}}_{\mu, \nu, \sigma}(\pp_{1}, \pp_{2}) = \hat\Pi^{5;\text{rel}}_{\mu, \nu, \sigma}(\pp_{1}, \pp_{2}) + 
 \hat\Pi^{5;\text{nr}}_{\mu, \nu, \sigma}(\pp_{1}, \pp_{2}),\end{equation}
where the label `rel' stands for relativistic and `nr' for non-relativistic. The first term in the right side is defined via the same 
GN tree expansion as the one for $\hat\Pi^{5;\text{d}}_{\mu, \nu, \sigma}(\pp_{1}, \pp_{2})$, with the difference that the values of the GN trees are computed by replacing: 
\begin{enumerate}
\item the 
single scale propagators $\hat g^{(h)}_\omega(\kk')$ in \eqref{eq:singlescale2nd} by its asymptotic, relativistic, counterpart, namely by 
\begin{equation}\label{uff.juv}
\hat g^{(h); \text{rel}}_{\omega}(\kk') = \frac{f_{h,\omega}(\kk')}{Z}  \begin{pmatrix} -ik_{0}' + v_{3} \omega k'_{3} & v_{1} k'_{1} - i v_{2} k'_{2} \\ v_{1} k'_{1} + iv_{2} k'_{2}  & -ik_{0}' - v_{3}\omega k'_{3} \end{pmatrix}^{-1}\;,
\end{equation}
with $Z, v_{l}$ as in \eqref{dressedRCC};
\item the parameters $Z_{\mu,h}$ and $Z^5_{\mu,h}$ associated with the endpoints of type $A_\mu\psi^+\psi^-$ and $A^5_\mu\psi^+\psi^-$, respectively, by their $h\to-\infty$ limits 
$Z_{\mu}$ and $Z_{\mu}^5$, respectively, see \eqref{dressedRCC}. 
\end{enumerate}
From the bounds on $Z_h, v_{l,h}, Z_{\mu,h}, Z_{\mu,h}^5$ derived in Sections \ref{sec:2ndregime} and \ref{sec:flow}, we know that 
\begin{eqnarray}\label{eq:zz} &&
|Z_{h} - Z| \leq C|\lambda| \gamma^{\theta h}\;,\hskip2.65truecm |v_{l,h} - v_{l}| \leq C|\lambda| |v_{l}| \gamma^{\theta h}\;,\\
&& |Z_{\mu,h} - Z_{\mu}| \leq C|\lambda|\, |Z_{\mu}|\gamma^{\theta (h-h_*)}, \qquad  |Z_{\mu,h}^5 - Z^5_{\mu}| \leq C|\lambda|\,|Z_{\mu}^5| \gamma^{\theta (h-h_*)},\label{eq:3.127ek}
\end{eqnarray} 
from which it follows that the difference between $\hat g^{(h)}_\omega(\kk')$ and $\hat g^{(h);\text{rel}}_\omega(\kk')$ admits an improved dimensional bound 
with an extra factor $\gamma^{\theta h}$, as compared with the bounds \eqref{dimgh2nd}. This, combined with the bounds in \eqref{eq:3.127ek}, 
implies that the non-relativistic term $\hat\Pi^{5;\text{nr}}_{\mu, \nu, \sigma}(\pp_{1}, \pp_{2})$ in the right side of \eqref{eq:JJJrelnonrel} is a sum of single-scale contributions, each of which is bounded by 
$C_\theta\mathfrak{Z}^5(\mathfrak{Z})^2|v_3^0|^{-1}\gamma^{h}\gamma^{\theta(h-h_*)}$, and whose derivatives with respect to $\pp_1$ or $\pp_2$ are bounded by 
$C_\theta\mathfrak{Z}^5(\mathfrak{Z})^2|v_3^0|^{-1}\gamma^{\theta (h-h_*)}$. Hence, by summing over $h<h_*$, we find that the non-relativistic term in the right 
side of \eqref{eq:JJJrelnonrel} is differentiable in $\pp_1,\pp_2$, with $\theta$-H\"older continuous derivatives, in a sufficiently small neighbourhood of $\pp_1=\pp_2={\bf 0}$. 

In conclusion, we can rewrite the response function $\hat\Pi^{5}_{\mu,\nu,\sigma}(\pp_{1},\pp_{2})$ as:
\begin{equation}\label{eq:split}
\hat\Pi^{5}_{\mu,\nu,\sigma}(\pp_{1},\pp_{2}) = \hat\Pi^{5;\text{rel}}_{\mu,\nu,\sigma}(\pp_{1},\pp_{2})+ \hat H^{5}_{\mu,\nu,\sigma}(\pp_{1},\pp_{2})\;,
\end{equation}
where $\hat H^{5}_{\mu,\nu,\sigma}(\pp_{1},\pp_{2})=\hat\Pi^{5;\text{r}}_{\mu,\nu,\sigma}(\pp_{1},\pp_{2})+\hat\Pi^{5;\text{nr}}_{\mu,\nu,\sigma}(\pp_{1},\pp_{2})$, which is
differentiable with respect to the external momenta, with derivatives that are H\"older continuous of exponent $\theta$, for any $0<\theta<1$, 
in a sufficiently small neighbourhood of the origin. 
Instead, the first term coincides with the contribution due to a {\it non-interacting} relativistic model, with 
dressed parameters (equal to the limiting parameters $Z_{\mu}, v_{l}, Z_{\mu}, Z_{\mu}^5$) and an ultraviolet cutoff on scale $h_*$. 
Eq.\eqref{eq:split} is almost the desired representation for the quadratic response, see \eqref{eq:tritetr_intro}. 
However, in order to prove \eqref{eq:tritetr_intro} starting from \eqref{eq:split} we need 
to compute the relativistic term $ \hat\Pi^{5;\text{rel}}_{\mu,\nu,\sigma}(\pp_{1},\pp_{2})$ more explicitly, which will be done in the next section. 

\medskip

{\it The dressed propagator and the vertex functions.} We now consider the dressed propagator $\hat S_2(\kk)$ and the vertex functions $\hat\Gamma_{\mu}(\kk,\pp)$ and 
$\hat\Gamma_{\mu}^5(\kk,\pp)$, 
which are obtained by an RG construction analogous to the one described in Section \ref{sec:RG}, in the presence of the Grassmann source field $\phi$, see 
\eqref{fullVphi} and \eqref{eq:W}. We decompose $\hat S_2(\kk)$ (resp. $\hat\Gamma_{\mu}(\kk,\pp)$, resp.
$\hat\Gamma_{\mu}^5(\kk,\pp)$) in a way analogous to \eqref{eq:GammaR}: that is, we distinguish a dominant part, analogous to 
\eqref{eq:JJJ}, which consists of the sum of the single-scale contributions from the GN trees with one endpoint of type 
$\phi^+\psi^-$ and one of type $\psi^+\phi^-$ (resp. one endpoint of type $A\psi^+\psi^-$, one of type $\phi^+\psi^-$, one of type $\psi^+\phi^-$, resp.
one endpoint of type $A^5\psi^+\psi^-$, one of type $\phi^+\psi^-$, one of type $\psi^+\phi^-$), all on scales smaller than $h_*$, 
from a remainder term, analogous to \eqref{eq:JJJr}, which includes all the other contributions (those from GN trees with root on scale smaller than $h_*$ and at least one endpoint on scale $h_*+1$, and those from GN with root on scale $h_*$ or larger). The dominant term is further decomposed in analogy with \eqref{eq:JJJrelnonrel} into a `relativistic' contribution, obtained via the substitutions spelled in items (i) and (ii) after \eqref{eq:JJJrelnonrel}, plus an additional remainder. The relativistic contribution is explicit: for 
the dressed propagator with $\kk$ close to $\pp_F^\omega$, letting $\kk'=\kk-\pp_F^\omega$, it is equal to 
\begin{equation}\label{S2rel}\hat S_2^{\text{rel}}(\kk)=\sum_{h<h_*}\hat g^{(h);\text{rel}}_\omega(\kk')\equiv \chi_{h_*-1,\omega}(\kk')\hat g_{\omega}^{\text{rel}}(\kk'),\end{equation}
where $\chi_{h,\omega}=\sum_{k\le h}f_{h,\omega}$, and $\hat g_{\omega}^{\text{rel}}(\kk)$ is defined in the same way as \eqref{uff.juv}, with $f_{h,\omega}(\kk)$ replaced by $1$, 
that is, 
\begin{equation}\label{gomegarel}\hat g_{\omega}^{\text{rel}}(\kk')=\frac1{Z}\Big(\sum_{\mu=0}^3v_{\mu} k_\mu' \sigma_{\mu,\omega} \Big)^{-1}, \end{equation}
where $v_{0}=1$ and 
$\sigma_{\mu,\omega}$ was defined in \eqref{defsigmamuomega} (note, a posteriori, that the cutoff function $\chi_{h_*-1,\omega}$ can be dropped from the right side of 
\eqref{S2rel} for $\kk'$ small enough, because $\chi_{h_*-1,\omega}\equiv 1$ in a sufficiently small neighbourhood of the origin). Similarly, 
for the vertex functions with $\kk$ close to $\pp_F^\omega$, letting again $\kk'=\kk-\pp_F$, the relativistic contribution reads
\begin{eqnarray} \hat\Gamma^{\text{rel}}_\mu(\kk+\pp)=\hat g^{\text{rel}}_\omega(\kk'+\pp)Z_{\mu}\sigma_{\mu,\omega} \hat g^{\text{rel}}_\omega(\kk'),\\
\hat\Gamma^{5;\text{rel}}_\mu(\kk+\pp)=\hat g^{\text{rel}}_\omega(\kk'+\pp)Z_{\mu}^5\omega\sigma_{\mu,\omega} \hat g^{\text{rel}}_\omega(\kk').\end{eqnarray}
On the other hand, the remainder terms (the analogue of \eqref{eq:JJJr} and the analogue of the non-relativistic term in \eqref{eq:JJJrelnonrel}) admit an improved dimensional bound: 
if $\kk-\pp_F^\omega,\pp$ and $\kk+\pp-\pp_F^\omega$ are all small and of the order $\gamma^h$, the remainder terms have an additional factor $\gamma^{\theta h}$,
as compared to the corresponding dominant, relativistic, term. 

Summarizing, letting $\kk'=\kk-\pp_F^\omega$ and assuming that $|\kk'|, |\kk' + \pp|, |\pp|$ are all in $[c\gamma^h,C\gamma^h]$, for some $h<h_*$ and $c,C>0$, 
we find that $\hat S_2(\kk'+\pp_F^\omega)$ satisfies \eqref{pp_int}, with 
$\|R_\omega(\kk')\|\le C_{\theta,h_*}
\gamma^{\theta h}$, while
\begin{eqnarray}\label{eq:S21}
\hat\Gamma_{\mu}(\kk,\pp)&=& Z_{\mu} \hat S_{2}(\kk+\pp) \sigma_{\mu,\omega} \hat S_{2}(\kk) (1 + \hat R_{\mu}(\kk,\pp))\;,\\
\hat\Gamma_{\mu}^5(\kk,\pp)&=& Z_{\mu}^5 \hat S_{2}(\kk+\pp) \omega\sigma_{\mu,\omega} \hat S_{2}(\kk) (1 + \hat R_{\mu}^5(\kk,\pp))\;,\label{yaeq}
\end{eqnarray}
with $\|\hat R_{\mu}(\kk,\pp)\|, \|R^5_\mu(\kk,\pp)\| \leq C_{\theta,h_*}\gamma^{\theta h}$. 

By plugging \eqref{eq:S21} and \eqref{pp_int} into the vertex Ward Identity \eqref{eq:vertexWI}, we obtain Eq.\eqref{asympt.2_intro}, that is,
$Z_{\mu} = v_{\mu} Z$. As already mentioned, this identity says that the charge carried by the electromagnetic 
current is not renormalized by the interaction.

There is no analogue of the vertex Ward Identity for the chiral vertex $\hat\Gamma_{\mu}^5(\kk,\pp)$. Therefore, the condition 
\eqref{z6} must be enforced by properly fixing $Z_{\mu,\text{bare}}^5$. By using \eqref{eq:S21} and \eqref{yaeq}, we find that 
the left side of \eqref{z6}
equals $(Z^5_{\mu}/Z_{\mu})\omega\mathds 1_2$, so that \eqref{z6} reduces to \eqref{asympt.1_intro}, $Z^5_\mu=Z_\mu$. 
This condition is enforced by fixing the bare parameters $Z_{\mu,\text{bare}}^5$ in the way discussed at the end of Section \ref{sec:3.4.3}. 

\subsection{The relativistic contribution to the quadratic response $\hat\Pi^5_{\mu,\nu,\sigma}$}\label{sec:3.6}

In this section we first prove \eqref{eq:tritetr_intro}, starting from \eqref{eq:split}. Next, we prove \eqref{eq:I.1_intro}-\eqref{eq:I.2_intro}. 
For this purpose, we need to compute the relativistic 
contribution to the quadratic response $\hat\Pi^5_{\mu,\nu,\sigma}$ more explicitly.
The definitions imply that, for $\pp_1,\pp_2$ sufficiently close to ${\bf 0}$, recalling \eqref{S2rel} and \eqref{gomegarel}, 
\begin{eqnarray}\label{eq:tribi}
&&\hskip-.4truecm\hat\Pi^{5;\text{rel}}_{\mu,\nu,\sigma}(\pp_1,\pp_2)= -Z^{5}_{\mu} Z_{\nu} Z_{\sigma} \sum_{\o = \pm} 
\int \frac{d\kk}{(2\pi)^{4}}\,\chi_{h_*-1,\omega}(\kk)\, \chi_{h_*-1,\omega}(\kk+\pp_1)\chi_{h_*-1,\omega}(\kk+\pp_1+\pp_2)\cdot\nonumber\\
&&\hskip-.2truecm\cdot  \Tr \{ \hat g_{\omega}^{\text{rel}}(\kk) \omega\sigma_{\mu,\omega} \hat g^{\text{rel}}_{\omega}(\kk + \pp_{1} + \pp_{2}) \sigma_{\sigma,\omega} \hat g^{\text{rel}}_{\omega}(\kk+\pp_{1}) \sigma_{\nu,\omega}  \} + \Big[(\pp_{1}, \nu)\leftrightarrow (\pp_{2}, \sigma)\Big].
\end{eqnarray}
For the purpose of computing the dominant contribution to 
$\hat\Pi^{5;\text{rel}}_{\mu,\nu,\sigma}(\pp_1,\pp_2)$, we can freely replace $\chi_{h_*-1,\omega}(\kk)$ by $\chi(\gamma^{-h_*}\|\kk\|_v)$, where
$\|\kk\|_v^2=\sum_\mu v_\mu^2k_\mu^2$. 
In fact, if we denote by $\hat\Pi^{5;*}_{\mu,\nu,\sigma}(\pp_1,\pp_2)$ the 
analogue of the right side of \eqref{eq:tribi} with $\chi_{h_*-1,\omega}(\kk)$ replaced by $\chi(\gamma^{-h_*}\|\kk\|_v)$, it is easy to check that the difference 
$\hat\Pi^{5;\text{rel}}_{\mu,\nu,\sigma}(\pp_1,\pp_2)-\hat\Pi^{5;*}_{\mu,\nu,\sigma}(\pp_1,\pp_2)$ is continuously differentiable in $\pp_1,\pp_2$ in a small neighbourhood of the origin. After rescaling $k_\mu\to k_\mu/v_\mu$, we find, letting $\chi_*(\kk):=\chi(\gamma^{-h_*}|\kk|)$,
$\lis\pp_1=(p_{1,0}, v_1p_{1,1}, v_2p_{1,2}, v_3 p_{1,3})$, and similarly for $\lis\pp_2$, 
\begin{eqnarray}\label{eq:tritri}
\hat\Pi^{5;*}_{\mu,\nu,\sigma}(\pp_1,\pp_2)&=& -\frac{Z^{5}_{\mu} Z_{\nu} Z_{\sigma}}{Z^3 v_1v_2v_3} 
\int \frac{d\kk}{(2\pi)^{4}}\frac{\chi_{*}(\kk)\, \chi_{*}(\kk+\lis\pp_1)\chi_{*}(\kk+\lis\pp_1+\lis\pp_2)}{|\kk|^2\,|\kk+\lis\pp_1|^2\,|\kk+\lis\pp_1+\lis\pp_2|^2}\cdot\\
&&\hskip-3.truecm \cdot\sum_{\mu_1,\mu_2,\mu_3=0}^3 k_{\mu_1}(k_{\mu_2}+\lis p_{1,\mu_2}+\lis p_{2,\mu_2})(k_{\mu_3}+\lis p_{1,\mu_3})\sum_{\o = \pm}  \Tr \{ \sigma_{\mu_1,\omega}^\dagger (\omega\sigma_{\mu,\omega})\sigma_{\mu_2,\omega}^\dagger  \sigma_{\sigma,\omega} \sigma_{\mu_3,\omega}^\dagger \sigma_{\nu,\omega}  \}\nonumber \\
&+& \Big[(\pp_{1}, \nu)\leftrightarrow (\pp_{2}, \sigma)\Big],\nonumber
\end{eqnarray}
The sum over $\omega$ of the trace of $\{ \sigma_{\mu_1,\omega}^\dagger (\omega\sigma_{\mu,\omega})\sigma_{\mu_2,\omega}^\dagger  \sigma_{\sigma,\omega} \sigma_{\mu_3,\omega}^\dagger \sigma_{\nu,\omega}  \}$ can be conveniently written in terms of four-dimensional Euclidean gamma matrices. In fact, after performing the 
trace-preserving transformations $$\sigma_{\mu,+}\to i\sigma_{\mu,+}=\begin{cases} \mathds 1_2 &\text{if $\mu=0$}\\ i\sigma_\mu & \text{if $\mu>0$}\end{cases}\quad\text{and}\quad \sigma_{\mu,-}\to i\sigma_3\sigma_{\mu,-}\sigma_3=\begin{cases} \mathds 1_2 &\text{if $\mu=0$}\\ -i\sigma_\mu & \text{if $\mu>0$}\end{cases},$$ we recognize that 
\begin{equation}\sum_{\o = \pm}  \Tr \{ \sigma_{\mu_1,\omega}^\dagger (\omega\sigma_{\mu,\omega})\sigma_{\mu_2,\omega}^\dagger  \sigma_{\sigma,\omega} \sigma_{\mu_3,\omega}^\dagger \sigma_{\nu,\omega}  \}=- \Tr \{ \gamma_{\mu_1}\gamma_{\mu}\gamma_5\gamma_{\mu_2} \gamma_{\sigma} \gamma_{\mu_3} \gamma_{\nu}  \},
\end{equation}
where 
\begin{equation}\label{eq:gammamat}
\gamma_{0} = \begin{pmatrix} 0 & \mathbbm{1} \\ \mathbbm{1} & 0 \end{pmatrix}\;,\qquad \gamma_{j} = \begin{pmatrix} 0 & i\sigma_{j} \\  -i\sigma_{j} & 0 \end{pmatrix}\quad \text{for}\quad j = 1,2,3,
\end{equation}
and 
\begin{equation}\label{eq:gamma5}
\gamma_{5} = \gamma_{0} \gamma_{1} \gamma_{2} \gamma_{3} = \begin{pmatrix} \mathbbm{1} & 0 \\ 0 & -\mathbbm{1} \end{pmatrix}\;.
\end{equation}
Note that the gamma matrices are Hermitian and satisfy the anticommutation rules $\{\gamma_\mu,\gamma_\nu\}=2\delta_{\mu,\nu}$, for all $\mu,\nu\in\{0,1,2,3,5\}$. 
In conclusion, 
\begin{equation}\label{eq:tritetr}
\hat\Pi^{5;\text{rel}}_{\mu,\nu,\sigma}(\pp_1,\pp_2)= \frac{Z^{5}_{\mu} Z_{\nu} Z_{\sigma}}{Z^3 v_1v_2v_3} I_{\mu,\nu,\sigma}(\lis\pp_1,\lis\pp_2)+R^5_{\mu,\nu,\sigma}(\pp_1,\pp_2),
\end{equation}
where $R^5_{\mu,\nu,\sigma}(\pp_1,\pp_2)$ is smooth in $\pp_1,\pp_2$ in a sufficiently small neighbourhood of the origin, and, letting $\slashed{\kk}:=\sum_{\mu=0}^3\gamma_\mu k_\mu$, 
\begin{equation}\label{defImunusigma}
I_{\mu,\nu,\sigma}(\pp_1,\pp_2)= \int \frac{d\kk}{(2\pi)^{4}}\, \Tr\Big\{ \frac{\chi_{*}(\kk)}{\slashed{\kk}} \gamma_{\mu}\gamma_{5} \frac{\chi_{*}(\kk + \pp_{1}+\pp_2)}{\slashed{\kk} + \slashed{\pp}_{1} + \slashed{\pp}_{2}} \gamma_{\sigma} \frac{\chi_{*}(\kk + \pp_{1})}{\slashed{\kk} + \slashed{\pp}_{1}} \gamma_{\nu} \Big\} + \Big[(\pp_{1}, \nu)\leftrightarrow (\pp_{2}, \sigma)\Big],
\end{equation}
which is nothing but the chiral triangle graph of QED$_4$ in the presence of an ultraviolet cutoff $\chi_*$. By plugging \eqref{eq:tritetr} into \eqref{eq:split}, we get \eqref{eq:tritetr_intro}, 
with $\widetilde H^5_{\mu,\nu,\sigma}=\hat H^5_{\mu,\nu,\sigma}+R^5_{\mu,\nu,\sigma}$. 

\subsubsection{The chiral triangle graph with momentum cutoff}\label{app:tri}
Let us now prove (\ref{eq:I.1_intro})-\eqref{eq:I.2_intro}. For ease of notation, in the present subsection we drop the label $*$ from $\chi_*$, and denote it simply by $\chi$.
We start from \eqref{eq:I.1_intro}, whose left side equals, letting $\qq:=\pp_1+\pp_2$, 
\bea   \sum_{\mu=0}^3 q_{\mu} I_{\m,\n,\s}(\pp_1,\pp_2)&=& \int\frac{d\kk}{(2\pi)^4}\Tr\Big\{\frac{\c(\kk)}{\slashed{\kk}} \slashed{\qq} \g_5\frac{\c(\kk+\qq)}{\slashed{\kk}+\slashed{\qq}}\g_\n\frac{\c(\kk+\pp_2)}{\slashed{\kk}+\slashed{\pp}_{2}}\g_\s\Big\}+[(\n,\pp_1)\otto(\s,\pp_2)]
\nonumber\\
&\equiv& T_{\n\s}(\pp_1,\pp_2)+T_{\s\n}(\pp_2,\pp_1)\;.\label{eq:1.1}\eea
Using the anti-commutativity of the fifth gamma matrix,
\be T_{\n\s}(\pp_1,\pp_2)=\int\frac{d\kk}{(2\p)^4}\Tr\Big\{\frac{\c(\kk)}{\slashed{\kk}}\slashed{\qq}\frac{\c(\kk+\qq)}{\slashed{\kk}+\slashed{\qq}}\g_\n\g_5\frac{\c(\kk+\pp_2)}{\slashed{\kk}+\slashed{\pp}_{2}}\g_\s\Big\}\;.\ee
Now we use the following rewriting:
\be \frac{\c(\kk)}{\ks}\qs\frac{\c(\kk+\qq)}{\ks+\qs}=\Big[\frac{\c(\kk)}{\slashed{\kk}}-\frac{\c(\kk+\qq)}{\slashed{\kk}+\slashed{\qq}}\Big]+\frac{\c(\kk)}{\slashed{\kk}}C(\kk,\qq)\frac{\c(\kk+\qq)}{\slashed{\kk}+\slashed{\qq}}\;,\label{eq:WI}\ee
where, for $\kk$ and $\kk+\qq$ in the support of $\chi(\kk) \chi(\kk+\qq)$: 
\be C(\kk,\qq)=\ks(\c^{-1}(\kk)-1)-(\ks+\qs)(\c^{-1}(\kk+\qq)-1)\;.\label{eq:C}\ee
The contribution to $T_{\m\n}(\pp_1,\pp_2)$ due to the difference in square brackets in \eqref{eq:WI} is:
\be \int\frac{d\kk}{(2\p)^4}\Tr\Big\{\frac{\c(\kk)}{\ks}\g_\n\g_5\frac{\c(\kk+\pp_2)}{\ks+\ps_2}\g_\s\Big\}-
\int\frac{d\kk}{(2\p)^4}\Tr\Big\{\frac{\c(\kk+\qq)}{\ks+\qs}\g_\n\g_5\frac{\c(\kk+\pp_2)}{\ks+\ps_2}\g_\s\Big\}\;.\label{eq:1.6}\ee
Changing integration variable $\kk\to \kk-\pp_2$ in the second integral, and using the cyclicity of the trace we get: 
\be \eqref{eq:1.6}=\int\frac{d\kk}{(2\p)^4}\Tr\Big\{\frac{\c(\kk)}{\ks}\g_\n\g_5\frac{\c(\kk+\pp_2)}{\ks+\ps_2}\g_\s\Big\}-
\int\frac{d\kk}{(2\p)^4}\Tr\Big\{\frac{\c(\kk)}{\ks}\g_\s\frac{\c(\kk+\pp_1)}{\ks+\ps_1}\g_\n\g_5\Big\},
\label{eq:1.7}\ee
which gives zero contribution to \eqref{eq:1.1}. Therefore, 
\be \sum_{\mu=0}^3 q_\m I_{\m,\n, \sigma}(\pp_1,\pp_2)= T^C_{\n\s}(\pp_1,\pp_2)+ T^C_{\s\n}(\pp_2,\pp_1)\;, \label{eq:1.8}\ee
with 
$$T^C_{\n\s}(\pp_1,\pp_2):=\int\frac{d\kk}{(2\p)^4}\Tr\Big\{\frac{\c(\kk)}{\ks}C(\kk,\qq)\frac{\c(\kk+\qq)}{\ks+\qs}\g_\n\g_5\frac{\c(\kk+\pp_2)}{\ks+\ps_2}\g_\s\Big\}\;.$$
Using \eqref{eq:C} we find
\bea T^C_{\n\s}(\pp_1,\pp_2)&=&\int\frac{d\kk}{(2\p)^4}\Big[(1-\c(\kk))\c(\kk+\qq)\c(\kk+\pp_2)\Tr\Big\{\frac{1}{\ks+\qs}\g_\n\g_5\frac{1}{\ks+\ps_2}\g_\s\Big\}\nonumber\\
&-&(1-\c(\kk+\qq))\c(\kk)\c(\kk+\pp_2)\Tr\Big\{\frac{1}{\ks}\g_\n\g_5\frac{1}{\ks+\ps_2}\g_\s\Big\}\Big]
\;.\nonumber\eea
We now shift the integration variable $\kk\to \kk-\pp_2$ in the term in the first line and use the cyclicity of the trace, thus finding
\bea T^C_{\n\s}(\pp_1,\pp_2)&=&\int\frac{d\kk}{(2\p)^4}\Big[(1-\c(\kk-\pp_2))\c(\kk+\pp_1)\c(\kk)\Tr\Big\{\frac{1}{\ks}\g_\s\g_5\frac{1}{\ks+\ps_1}\g_\n\Big\}\nonumber\\
&-&(1-\c(\kk+\qq))\c(\kk)\c(\kk+\pp_2)\Tr\Big\{\frac{1}{\ks}\g_\n\g_5\frac{1}{\ks+\ps_2}\g_\s\Big\}\Big]\;,\label{eq:1.10}\eea
so that, after plugging back this expression into \eqref{eq:1.8} and exchanging names $(\pp_1,\n)\otto(\pp_2,\s)$ in one of the 
terms contributing to $T^C_{\n\s}(\pp_1,\pp_2)$, we get
\be \sum_{\mu=0}^3q_\m I_{\m,\nu,\sigma}(\pp_1,\pp_2)= \tilde T^C_{\n\s}(\pp_1,\pp_2)+ \tilde T^C_{\s\n}(\pp_2,\pp_1)\;,\label{eq:1.8bis}\ee
with 
\be\label{tildeTC} \tilde T^C_{\n\s}(\pp_1,\pp_2):= 
\int\frac{d\kk}{(2\p)^4}\c(\kk)\c(\kk+\pp_2)(\c(\kk+\qq)-\c(\kk-\pp_1))\Tr\Big\{\frac{1}{\ks}\g_\n\g_5\frac{1}{\ks+\ps_2}\g_\s\Big\}\;.\ee
Note that $\tilde T^C_{\n\s}({\bf 0},{\bf 0})=0$. 
Let us now expand $\tilde T^C_{\n\s}(\pp_1,\pp_2)$ in Taylor series around $(\pp_1,\pp_2)=({\bf 0}, {\bf 0})$ and let us focus on the terms of order $1$ and $2$ in the momenta, to be denoted by $[\tilde T^C_{\n\s}(\pp_1,\pp_2)]^{(1)}$ and $[\tilde T^C_{\n\s}(\pp_1,\pp_2)]^{(2)}$, respectively. It is easy to check that, for $P=\max\{|\pp_1|,|\pp_2|\}$ sufficiently small, 
as compared to $\gamma^{h_*}$, the Taylor remainder of order $3$ is bounded by\footnote{In order to prove this, we bound the Taylor remainder of order 3 by $P^3 D_3$, with $D_3$ an upper bound on the third derivative with respect to $\pp_1,\pp_2$ of the right side of \eqref{Iform}. Next, we 
note that, due to the structure of the right side of \eqref{tildeTC}, at least one of such derivatives acts on one of the cutoff 
functions $\chi$; therefore, for $|\pp_1|,|\pp_2|$ small enough (as compared with the support of $\chi$), $D_3$ can be dimensionally bounded by (const.)$\int_{S_*}d\kk\, |\kk|^{-5}$, 
where $S_*=\{\kk: c\gamma^{h_*}\le |\kk|\le C\gamma^{-h_*}\}$.\label{explanTaylor}} $C\gamma^{-h_*}P^3$. Moreover, it is straightforward to check that $[\tilde T^C_{\n\s}(\pp_1,\pp_2)]^{(1)}=0$, by parity.
Let us now consider $[\tilde T^C_{\n\s}(\pp_1,\pp_2)]^{(2)}$, which consists of several terms:
\be [\tilde T^C_{\n\s}(\pp_1,\pp_2)]^{(2)}=A_{\n\s}(\pp_1,\pp_2)+B_{\n\s}(\pp_1,\pp_2)+C_{\n\s}(\pp_1,\pp_2)+D_{\n\s}(\pp_1,\pp_2)\;,\label{eq:blah}\ee
where, using the convention that repeated indices are summed from $0$ to $3$, 
\bea && A_{\n\s}(\pp_1,\pp_2)=\frac12 (q_\m q_\l-p_{1,\m}p_{1,\l})\int\frac{d\kk}{(2\p)^4}\c^2(\kk)\partial_\m\partial_\l\c(\kk)\Tr\Big\{\frac{1}{\ks}\g_\n\g_5\frac{1}{\ks}\g_\s\Big\}\nonumber\\
 && B_{\n\s}(\pp_1,\pp_2)=p_{2,\m}(q_\l+p_{1,\l}) \int\frac{d\kk}{(2\p)^4}\c(\kk)\partial_\m\c(\kk)\partial_\l\c(\kk)\Tr\Big\{\frac{1}{\ks}\g_\n\g_5\frac{1}{\ks}\g_\s\Big\}\nonumber\\
 && C_{\n\s}(\pp_1,\pp_2)=(q_\mu+p_{1,\mu})\int\frac{d\kk}{(2\p)^4}\frac{\c^2(\kk)\partial_\m\c(\kk)}{|\kk|^2}\Tr\Big\{\frac{1}{\ks}\g_\n\g_5 \ps_2\g_\s\Big\}\nonumber\\
 && D_{\n\s}(\pp_1,\pp_2)=-2(q_\mu+p_{1,\mu})p_{2,\l}\int\frac{d\kk}{(2\p)^4}\frac{\c^2(\kk)\partial_\m\c(\kk) k_\l}{|\kk|^2}\Tr\Big\{\frac{1}{\ks}\g_\n\g_5 \frac1{\ks}\g_\s\Big\}\;.\nonumber
 \eea
Now, $A_{\n\s}(\pp_1,\pp_2)=B_{\n\s}(\pp_1,\pp_2)=D_{\n\s}(\pp_1,\pp_2)=0$ by simple parity reasons: in fact, after the computation of the trace, letting $\chi'$ be the derivative 
of $\chi$ with respect to $|\kk|$, $\hat k_\mu=k_\mu/|\kk|$, and $\epsilon_{\a\n\b\s}$ the Levi-Civita symbol (see footnote \ref{LCsymbol}), 
\bea && A_{\n\s}(\pp_1,\pp_2)=2 (q_\m q_\l-p_{1,\m}p_{1,\l})\e_{\a\n\b\s}\int\frac{d\kk}{(2\p)^4}\c^2(\kk)\big[\c''(\kk)\hat k_\m\hat k_\l+\frac{\c'(\kk)}{|\kk|}(\d_{\m\l}-\hat k_\m\hat k_\l)\big]\frac{k_\a k_\b}{|\kk|^4}\;,\nonumber\\
&& B_{\n\s}(\pp_1,\pp_2)=4p_{2,\m}(q_\l+p_{1,\l}) \e_{\a\n\b\s} \int\frac{d\kk}{(2\p)^4}\c(\kk)(\c'(\kk))^2\hat k_\m\hat k_\l\frac{k_\a k_\b}{|\kk|^4}\;,\nonumber\\
 && D_{\n\s}(\pp_1,\pp_2)=-8(q_\mu+p_{1,\mu})p_{2,\l}\e_{\a\n\b\s}\int\frac{d\kk}{(2\p)^4}\c^2(\kk)\c'(\kk) \frac{\hat k_\m k_\l k_\a k_\b}{|\kk|^6}\;,\nonumber\eea
which are all zero by the anti-symmetry in $\a\otto\b$. 
The only non-trivial term we are left with is 
\be C_{\n\s}(\pp_1,\pp_2)=4(q_\mu+p_{1,\mu})p_{2,\b}\e_{\a\n\b\s}\int\frac{d\kk}{(2\p)^4}\frac{\c^2(\kk)\c'(\kk)\hat k_\m k_\a}{|\kk|^4}\;,\label{eq:1.15}\ee
where, using that the angular integration in the 4D integral over $\kk$ gives $2\p^2$, we can rewrite
\be \int\frac{d\kk}{(2\p)^4}\frac{\c^2(\kk)\c'(\kk)\hat k_\m k_\a}{|\kk|^4}=\frac{\d_{\m\a}}{4}\int\frac{d\kk}{(2\p)^4}\frac{\c^2(\kk)\c'(\kk)}{|\kk|^3}=\frac{\d_{\m\a}}{4}\frac{1}{(2\p)^4}2\p^2 \Big(-\frac13\Big)=-\frac{\d_{\m\a}}{96\p^2}\nonumber\ee
where in the second equality we used that $\int_0^{\infty} \c^2(\rho)\c'(\r)d\r=-1/3$, where, with some abuse of notation, we denoted $\chi(|\kk|)\equiv \chi(\kk)$. 
Plugging this back into \eqref{eq:1.15} and using again the anti-symmetry in $\a\otto\b$, we find (recalling that $\qq=\pp_1+\pp_2$)
\be C_{\n\s}(\pp_1,\pp_2)=-\frac{1}{12\p^2}p_{1,\a}p_{2,\b}\e_{\a\n\b\s}\;,\ee
so that, putting things together,
\be q_\m I_{\m,\n,\s}(\pp_1,\pp_2)=-\frac{1}{6\p^2}p_{1,\a}p_{2,\b}\e_{\a\n\b\s}+O(\gamma^{-h_*}P^3)\;,\label{aff}\ee
which proves \eqref{eq:I.1_intro} (note that the order of the indices in $\varepsilon_{\a\b\n\s}$ in the right side of \eqref{eq:I.1_intro} is different from the one in the right side of 
\eqref{aff}, which explains the different sign). 

\medskip

Let us now prove (\ref{eq:I.2_intro}). We compute:
\bea   p_{1,\n} I_{\m,\n,\s}(\pp_1,\pp_2)&=&\int\frac{d\kk}{(2\p)^4}\Tr\Big\{\frac{\c(\kk)}{\ks}\g_\m\g_5\frac{\c(\kk+\qq)}{\ks+\qs}\ps_1\frac{\c(\kk+\pp_2)}{\ks+\ps_2}\g_\s\Big\}\nonumber\\
&+&\int\frac{d\kk}{(2\p)^4}\Tr\Big\{\frac{\c(\kk)}{\ks}\g_\m\g_5\frac{\c(\kk+\qq)}{\ks+\qs}\g_\s\frac{\c(\kk+\pp_1)}{\ks+\ps_1}\ps_1\Big\}\;.\eea
In the first term we rename $\kk\to-\kk-\qq$ and use the cyclicity of the trace; in the second we rename $\kk\to-\kk-\pp_1$, use the cyclicity of the trace and the anti-commutation properties of $\g_5$; we get:
\bea  -p_{1,\n} I_{\m,\n,\s}(\pp_1,\pp_2)&=&\int\frac{d\kk}{(2\p)^4}\Tr\Big\{\frac{\c(\kk+\qq)}{\ks+\qs}\g_\m\g_5\frac{\c(\kk)}{\ks}\ps_1\frac{\c(\kk+\pp_1)}{\ks+\ps_1}\g_\s\Big\}\nonumber\\
&+&\int\frac{d\kk}{(2\p)^4}\Tr\Big\{\frac{\c(\kk-\pp_2)}{\ks-\ps_2}\g_\s\g_5\frac{\c(\kk)}{\ks}\ps_1\frac{\c(\kk+\pp_1)}{\ks+\ps_1}\g_\m\Big\}\;.\eea
Now, in both integrals we rewrite $\frac{\c(\kk)}{\ks}\ps_1\frac{\c(\kk+\pp_1)}{\ks+\ps_1}$ by using \eqref{eq:WI}:
\be \frac{\c(\kk)}{\ks}\ps_1\frac{\c(\kk+\pp_1)}{\ks+\ps_1}=\Big[\frac{\c(\kk)}{\ks}-\frac{\c(\kk+\pp_1)}{\ks+\ps_1}\Big]+\frac{\c(\kk)}{\ks}C(\kk,\pp_1)\frac{\c(\kk+\pp_1)}{\ks+\ps_1}\;.\label{eq:WI2}\ee
It is easy to see that the contribution to $p_{1,\nu}I_{\m,\n,\s}(\pp_1,\pp_2)$ coming from the term in square brackets in \eqref{eq:WI2} vanishes. Therefore, we are left with 
\bea   -p_{1,\n}I_{\m,\n,\s}(\pp_1,\pp_2)&=&\int\frac{d\kk}{(2\p)^4}\Tr\Big\{\frac{\c(\kk+\qq)}{\ks+\qs}\g_\m\g_5\frac{\c(\kk)}{\ks}C(\kk,\pp_1)\frac{\c(\kk+\pp_1)}{\ks+\ps_1}\g_\s\Big\}\nonumber\\
&+&\int\frac{d\kk}{(2\p)^4}\Tr\Big\{\frac{\c(\kk-\pp_2)}{\ks-\ps_2}\g_\s\g_5\frac{\c(k)}{\ks}C(\kk,\pp_1)\frac{\c(\kk+\pp_1)}{\ks+\ps_1}\g_\m\Big\}\eea
that, using the explicit form of $C(\kk,\pp_1)$ becomes: 
\bea   -p_{1,\n} I_{\m,\n,\s}(\pp_1,\pp_2)&=&\int\frac{d\kk}{(2\p)^4}\c(\kk+\qq)(1-\c(\kk))\c(\kk+\pp_1)\Tr\Big\{\frac{1}{\ks+\qs}\g_\m\g_5\frac{1}{\ks+\ps_1}\g_\s\Big\}\nonumber\\
&-&\int\frac{d\kk}{(2\p)^4}\c(\kk+\qq)\c(\kk)(1-\c(\kk+\pp_1))\Tr\Big\{\frac{1}{\ks+\qs}\g_\m\g_5\frac{1}{\ks}\g_\s\Big\}\nonumber\\
&+&\int\frac{d\kk}{(2\p)^4} \c(\kk-\pp_2)(1-\c(\kk))\c(\kk+\pp_1)\Tr\Big\{\frac{1}{\ks-\ps_2}\g_\s\g_5\frac{1}{\ks+\ps_1}\g_\m\Big\}\nonumber\\ 
&-&\int\frac{d\kk}{(2\p)^4}\c(\kk-\pp_2)\c(\kk)(1-\c(\kk+\pp_1))\Tr\Big\{\frac{1}{\ks-\ps_2}\g_\s\g_5\frac{1}{\ks}\g_\m\Big\}\;.\nonumber\eea
If we now rename $\kk\to \kk-\pp_1$ in the first line, and $\kk\to \kk+\pp_2$ in the third and fourth lines, we can rewrite this as
\bea  && -p_{1,\n}I_{\m,\n,\s}(\pp_1,\pp_2)=\int\frac{d\kk}{(2\p)^4}\c(\kk)\c(\kk+\pp_2)(\c(\kk+\qq)-\c(\kk-\pp_1))\Tr\Big\{\frac{1}{\ks}\g_\s\g_5\frac{1}{\ks+\ps_2}\g_\m\Big\}\nonumber\\
&&\qquad +\int\frac{d\kk}{(2\p)^4} \c(\kk)\c(\kk+\qq)(\c(\kk+\pp_1)-\c(\kk+\pp_2))\Tr\Big\{\frac{1}{\ks}\g_\s\g_5\frac{1}{\ks+\qs}\g_\m\Big\}.\label{Iform}\eea
The expression in the right side vanishes at $(\pp_1,\pp_2)=({\bf 0}, {\bf 0})$. Also in this case, we expand it in Taylor series around the origin and focus 
on the terms of order $1$ and $2$ in the momenta, the Taylor remainder of order $3$ being smaller than $C\gamma^{-h_*}P^3$, for $P=\max\{|\pp_1|,|\pp_2|\}$ small enough, see footnote \ref{explanTaylor}. 
Also in this case, it is straightforward to check that the term of order $1$ vanishes, by parity. 
After having computed the trace, we find that the term of order 2 can be rewritten as
\be  p_{1,\n} [I_{\m,\n,\s}(\pp_1,\pp_2)]^{(2)}=-4\big[(q_\nu+p_{1,\nu}) p_{2,\beta} +(p_{1,\nu}-p_{2,\nu})q_\b\big]\e_{\a\b\m\s}\int\frac{d\kk}{(2\p)^4}\c^2(\kk)\c'(\kk) \frac{\hat k_\n k_\a}{|\kk|^4}.\ee
Recalling that $\qq=\pp_1+\pp_2$ and computing the integral over $\kk$ we finally get: 
\be  p_{1,\n} I_{\m,\n,\s}(\pp_1,\pp_2)=\frac{1}{6\pi^2}p_{1,\a}p_{2,\b} \e_{\a\b\m\s}+O(\gamma^{-h_*}P^3),\label{uff}\ee
which proves \eqref{eq:I.2_intro}. As anticipated in \eqref{sec:roadmap}, this concludes the proof of Theorem \ref{thm:main}. 
\qed

\appendix

\section{Symmetries}\label{app:symm}

In this appendix we first reformulate the symmetries (ii) to (iv) of Section \ref{sec:2.1}, as well as the Hermitian conjugation symmetry, in terms of the Grassmann variables used in 
the RG construction of the generating functional of correlations. Next, we discuss the implications of the symmetries, including the proof of Lemmas \ref{lem:sym1} and \ref{lem:sym2}. 

\subsection{Symmetries of the Grassmann action}\label{appA.1}

Consider the Grassmann action $S_0(\psi,A)+\mathcal V(\psi)+\sum_\mu(A^5_\mu, j^5_\mu)$, where $\mathcal V(\psi)$ was defined in \eqref{VpsiGrass}, 
$(A^5_\mu, j^5_\mu)$ in \eqref{eq:BB}, and, letting $\int \frac{d\kk}{(2\pi)^4}$ is a shorthand for $\frac1{\beta L^3}\sum_{\kk\in \mathbb{D}_{\beta, L, N}}$, 
\begin{equation}S_0(\psi,{\bf A}):=-\int \frac{d\kk}{(2\pi)^4}\hat \psi^{+}_{\kk} [\hat g_{\beta, L, N}(\kk)]^{-1} \hat \psi^{-}_{\kk}+(A_0,j_0)+(\psi^+,(H_0-H_0(A))\psi^-),\end{equation}
with $\hat g_{\beta, L, N}(\kk)$ as in 
\eqref{freeactioncutoff}, and $(A_0,j_0)+(\psi^+,H_0-H_0(A)\psi^-)$ as in \eqref{eq:BB}. 
The reader can easily check that all the terms in Grassmann action are separately invariant under the following symmetries. 
\begin{enumerate}
\item {\it Hermitian conjugation.} 
\begin{equation}\psi^\pm_{\xx}\to \pm\psi^{\mp,T}_{(-x_0,x)}, \qquad A_\mu(\xx)\to (-1)^{\d_{\m,0}}A_\mu(-x_0,x),
\qquad A_\mu^5(\xx)\to (-1)^{\d_{\m,0}}A_\mu^5(-x_0,x),\nonumber\end{equation}
and $c\to c^*$, where $c$ is a generic numerical constant appearing in the action. In Fourier space, these transformations read: $\hat \psi^\pm_{\kk}\to \pm\hat\psi^{\mp,T}_{(-k_0,k)}$ (or, in the quasi-particle representation, 
$\hat \psi^\pm_{\o,\kk}\to \pm\hat\psi^{\mp,T}_{\o,(-k_0,k)}$), $\hat A_{\mu,\pp}\to (-1)^{\d_{\m,0}} \hat A_{\mu,(p_0,-p)}$, $\hat A^5_{\mu,\pp}\to (-1)^{\d_{\m,0}} \hat A^5_{\mu,(p_0,-p)}$, 
and $c\to c^*$. 
\item {\it Inversion symmetry.} 
\begin{eqnarray} && \psi^-_{\xx}\to \sigma_3\psi^-_{(x_0,-x)},\hskip3.2truecm \psi^+_\xx\to \psi^+_{(x_0,-x)}\sigma_3,\nonumber\\
&& A_\mu(\xx)\to (-1)^{1-\d_{\m,0}}A_\mu(x_0,-x), \qquad A_\mu^5(\xx)\to (-1)^{\d_{\m,0}}A_\mu^5(x_0,-x).\nonumber
\end{eqnarray}
In Fourier space, these transformations read: 
$\hat\psi^-_{\kk}\to \sigma_3\hat\psi^-_{(k_0,-k)}$, $\hat\psi^+_\kk\to \hat \psi^+_{(k_0,-k)}\sigma_3$
(or, in the quasi-particle representation, $\hat\psi^-_{\o,\kk}\to \sigma_3\hat\psi^-_{-\o,(k_0,-k)}$, $\hat\psi^+_{\o,\kk}\to \hat \psi^+_{-\o,(k_0,-k)}\sigma_3$), 
$\hat A_{\mu,\pp}\to (-1)^{1-\d_{\m,0}}\hat A_{\mu,(p_0,-p)}$, and $\hat A^5_{\mu,\pp}\to (-1)^{\d_{\m,0}}\hat A^5_{\mu,(p_0,-p)}$
\item {\it Reflection about a horizontal plane.} 
\begin{eqnarray} &&\psi^\pm_{\xx}\to \psi^\pm_{(x_0,x_1,x_2,-x_3)},\qquad  A_\mu(\xx)\to (-1)^{\d_{\m,3}}A_\mu(x_0,x_1,x_2,-x_3), \nonumber\\
&& \hskip2.truecm A_\mu^5(\xx)\to (-1)^{1-\d_{\m,3}}A_\mu^5(x_0,x_1,x_2,-x_3).\nonumber\end{eqnarray} 
In Fourier space, these transformations read: 
$\hat\psi^\pm_{\kk}\to \hat\psi^\pm_{(k_0,k_1,k_2,-k_3)}$
(or, in the quasi-particle representation, $\hat\psi^\pm_{\o,\kk}\to \hat\psi^\pm_{-\o,(k_0,k_1,k_2,-k_3)}$), 
$\hat A_{\mu,\pp}\to (-1)^{\d_{\m,3}}\hat A_{\mu,(p_0,p_1,p_2,-p_3)}$, and $\hat A^5_{\mu,\pp}\to (-1)^{1-\d_{\m,3}}\hat A^5_{\mu,(p_0,p_1,p_2,-p_3)}$.
\item {\it Reflection about a vertical plane + color exchange.} 
\begin{eqnarray} && \psi^-_{\xx}\to -\sigma_1\psi^-_{\bar \xx},\hskip3.22truecm \psi^+_\xx\to \psi^+_{\bar\xx}\sigma_1,\nonumber\\
&& A_\mu(\xx)\to (-1)^{\d_{\m,0}+\d_{\m,1}}A_\mu(\bar\xx),\qquad A_\mu^5(\xx)\to (-1)^{\d_{\m,0}+\d_{\m,1}}A^5_\mu(\bar\xx),\nonumber\end{eqnarray}
where $\bar\xx=(-x_0,-x_1,x_2,x_3)$. In Fourier space, these transformations read: 
$\hat\psi^-_{\kk}\to -\sigma_1\hat\psi^-_{\bar\kk}$, $\hat\psi^+_\kk\to \hat \psi^+_{\bar\kk}\sigma_1$
(or, in the quasi-particle representation, $\hat\psi^-_{\o,\kk}\to -\sigma_1\hat\psi^-_{\o,\bar\kk}$, $\hat\psi^+_{\o,\kk}\to \hat \psi^+_{\o,\bar\kk}\sigma_1$), 
$\hat A_{\mu,\pp}\to (-1)^{\d_{\m,0} + \d_{\m,1}}\hat A_{\mu,\bar\pp}$, and $\hat A^5_{\mu,\pp}\to (-1)^{\d_{\m,0} + \d_{\m,1}}\hat A^5_{\mu,\bar\pp}$.
\end{enumerate}

\subsection{Consequences of the symmetries}\label{app:conssym}

The symmetries listed above are preserved by the multiscale RG construction described in Section \ref{sec:RG}. This implies, in particular, that the kernels of the effective potential
on scale $h$ are invariant under the symmetries, and their local parts as well. Let us then discuss the implications of the symmetries on the structure of the local parts of the effective action, separately for the two scale regimes $h\leq h_{*}$ and $h > h_{*}$. We will thus prove the symmetry properties listed in Lemmas \ref{lem:sym1} and \ref{lem:sym2}.  

\subsubsection{The local $\psi^+\psi^-$ term.} {\it Regime $h \geq h_{*}$.} Let us consider a quadratic term, symmetric under the symmetries (i), (ii), (iii), (iv) of Section \ref{appA.1}, of the form 
\begin{equation}
\int \frac{d\kk}{(2\pi)^{4}}\, \hat \psi^{+}_{\kk} M \hat \psi^{-}_{\kk}\;,
\end{equation}
where $M$ is a complex $2\times 2$ matrix. The symmetries (i), (ii), (iii), (iv) imply that:
\begin{equation}\label{s10}
M = M^{\dagger} = \sigma_{3} M \sigma_{3} = -\sigma_{1} M \sigma_{1}\;.
\end{equation}
If we now expand $M$ in the `Pauli basis', $M=a_0\mathds 1+a_1\s_1+a_2\s_2+a_3\s_3$, we see that \eqref{s10} implies that $M=a_3\s_3$, with $a_3\in \mathbb R$. This proves 
the first identity in the first line of \eqref{eq:sym1}.

\medskip

\noindent{\it Regime $h < h_{*}$}. Next, let us consider a quadratic term, symmetric under the symmetries (i), (ii), (iii), (iv) of Section \ref{appA.1}, of the form 
\begin{equation}
\sum_{\omega = \pm}\int \frac{d\kk}{(2\pi)^4}\hat \psi^+_{\o,\kk} M_\o \hat\psi^-_{\o,\kk}\;,
\end{equation}
with $M_\o$ a complex $2\times 2$ matrix. Imposing symmetry (iii) we find that $M_\o=M_{-\o}\equiv M$. The symmetries (i), (ii), (iii), (iv) imply that:
\be M=M^\dagger=\s_3 M\s_3=-\s_1 M\s_1.\label{s1}\ee
Writing $M=a_0\mathds 1+a_1\s_1+a_2\s_2+a_3\s_3$, we see that \eqref{s1} implies that $M=a_3\s_3$, with $a_3\in \mathbb R$. This proves the first identity in \eqref{eq:sym2}. 

\subsubsection{The local $\psi^+\partial_\mu\psi^-$ terms.} 

\noindent{\it Regime $h \geq h_{*}$.} Consider a quadratic term, invariant under the symmetries (i), (ii), (iii), (iv) of Section \ref{appA.1}, of the form 
\begin{equation}
\int \frac{d\kk}{(2\pi)^4}\hat \psi^+_{\kk} k_\mu M_{\mu} \hat\psi^-_{\kk}\;,
\end{equation}
where $M_{\mu}$ are complex $2\times 2$ matrices, with $\mu=0,1,2,3$. Symmetry (iii) implies that $M_3=-M_3=0$, while it does not have any implications for $M_\mu$ with $\mu=0,1,2$. 
Imposing the validity of symmetries (i), (ii), (iv), we find: 
\bea 
&& M_0=-M_0^\dagger=\s_3 M_0\s_3=\s_1 M_0 \s_1,\label{s20}\\
&& M_1=M_1^\dagger=-\s_3 M_1\s_3=\s_1 M_1 \s_1,\\
&& M_2=M_2^\dagger=-\s_3 M_2\s_3=-\s_1 M_2 \s_1,\label{s50}\eea
If we now expand $M_\mu$ with $\mu=0,1,2$ in the `Pauli basis', $M_\mu=a_0^\mu\mathds 1+a_1^\mu\s_1+a_2^\mu\s_2+a_3^\mu\s_3$, we see that \eqref{s20}--\eqref{s50} imply that $M_0=a_0^0\mathds 1$, $M_1=a_1^1\s_1$, $M_2=a_2^2\s_2$, 
with $a_0^0\in i\mathbb R$ and $a_1^1,a_2^2\in\mathbb R$. 
This proves the second identity in the first line of \eqref{eq:sym1}.

\medskip

\noindent{{\it Regime $h<h_{*}$} Let us consider a quadratic term, symmetric under the symmetries (i), (ii), (iii), (iv) of Section \ref{appA.1},  of the form 
\begin{equation}
\sum_{\omega = \pm}\int \frac{d\kk}{(2\pi)^4}\hat \psi^+_{\o,\kk} k_\mu M_{\mu,\o} \hat\psi^-_{\o,\kk}\;,
\end{equation}
where  
$M_{\mu,\o}$ are complex $2\times 2$ matrices, with $\mu=0,1,2,3$.
Imposing symmetry (iii) we find that $M_{\mu,\o}=(-1)^{\d_{\m,3}}M_{\mu,-\o}$, so we let $M_{\mu,\o}\equiv M_\mu$, for $\mu=0,1,2$, and $M_{3,\o}\equiv \o M_3$. Imposing the validity of symmetries (i), (ii), (iv), we find: 
\bea 
&& M_0=-M_0^\dagger=\s_3 M_0\s_3=\s_1 M_0 \s_1,\label{s2}\\
&& M_1=M_1^\dagger=-\s_3 M_1\s_3=\s_1 M_1 \s_1,\\
&& M_2=M_2^\dagger=-\s_3 M_2\s_3=-\s_1 M_2 \s_1,\\
&& M_3=M_3^\dagger=\s_3 M_3\s_3=-\s_1 M_3 \s_1.\label{s5}\eea
If we now expand $M_\mu$ in the `Pauli basis', $M_\mu=a_0^\mu\mathds 1+a_1^\mu\s_1+a_2^\mu\s_2+a_3^\mu\s_3$, we see that \eqref{s2}--\eqref{s5} imply that $M_0=a_0^0\mathds 1$, $M_1=a_1^1\s_1$, $M_2=a_2^2\s_2$, $M_3=a_3^3\s_3$, with $a_0^0\in i\mathbb R$ and $a_1^1,a_2^2,a_3^3\in\mathbb R$. This proves 
the second identity in \eqref{eq:sym2}. 

\subsubsection{The local $\psi^{+} \partial_\mu\partial_{3} \psi^{-}$ and $\psi^{+} \partial_{3}^3 \psi^{-}$ terms.}

{\it Regime $h \geq h_{*}$.} Let us consider the following quadratic terms, symmetric under the symmetries (i), (ii), (iii), (iv) of Section \ref{appA.1}:
\begin{equation}
\int \frac{d\kk}{(2\pi)^{4}}\, \hat \psi^{+}_{\kk} k_\mu k_{3} Q_{\mu} \psi^{-}_{\kk}\qquad \text{and}\qquad \int \frac{d\kk}{(2\pi)^{4}}\, \hat \psi^{+}_{\kk} k_{3}^3 \widetilde Q_{3} \psi^{-}_{\kk}\;,
\end{equation}
where $Q_{\mu}$ with $\mu=0,1,2,3$ and $\widetilde Q_3$ are complex $2\times 2$ matrices. Symmetry (iii) implies that $Q_0=Q_1=Q_2=\widetilde Q_3=0$, which proves the second line of \eqref{eq:sym1}. On the other hand, symmetries (i), (ii), (iv) imply that 
\be 
Q_{3} = Q_{3}^{\dagger} = \sigma_{3} Q_{3} \sigma_{3} = -\sigma_{1} Q_{3} \sigma_{1}\;,
\end{equation}
from which we find $Q_{3} = b \sigma_{3}$ with $b\in \mathbb{R}$. This proves the third identity in the first line of \eqref{eq:sym1}.

\subsubsection{The local $A\psi^+\psi^-$ and $A\psi^+\partial_3\psi^-$ terms.} 

\noindent{{\it Regime $h\geq h_{*}$}.} Let us consider the following terms, quadratic in $\psi$ and linear in $A$, symmetric under the symmetries (i), (ii), (iii), (iv) of Section \ref{appA.1}:
$$\int \frac{d\pp}{(2\pi)^4}\int \frac{d\kk}{(2\pi)^4}\, \hat A_{\mu,\pp}\hat \psi^+_{\kk+\pp}\G_{\mu} \hat\psi^-_{\kk} \quad \text{and}\quad \int \frac{d\pp}{(2\pi)^4}\int \frac{d\kk}{(2\pi)^4}\, \hat A_{\mu,\pp}\hat \psi^+_{\kk+\pp}(k_3\Gamma^3_\mu+p_3\widetilde\Gamma^3_\mu) \hat\psi^-_{\kk},$$ 
where $\G_{\mu}, \G_\mu^3, \widetilde\Gamma^3_\mu$ are complex $2\times 2$ matrices. Symmetry (iii) implies that $\Gamma_3=0$ and $\Gamma^3_\mu=\widetilde\Gamma^3_\mu=0$ for $\mu=0,1,2$. Symmetries (i), (ii), (iv) imply that $\Gamma_\mu$ satisfy the same as \eqref{s20}--\eqref{s50}, with $M_\mu$ replaced by $\G_\mu$. Therefore, $\G_0=ic_0\mathds 1$, $\G_1=c_1\s_1$, $\G_2=c_2\s_2$, with 
$c_1,c_1,c_2\in\mathbb R$, which proves the first identity in the first line of \eqref{2ndeq:lem3.2}. Symmetry (i) implies that 
$\Gamma^3_3=(\Gamma^3_3)^\dagger$, while $\widetilde\Gamma^3_3+(\widetilde\Gamma^3_3)^\dagger=\Gamma^3_3$. Moreover, symmetries (ii) and (iv) imply that 
$\Gamma^3_3=\sigma_3\Gamma^3_3\sigma_3=-\sigma_1\Gamma^3_3\sigma_1$ and $\widetilde\Gamma^3_3=\sigma_3\widetilde\Gamma^3_3\sigma_3=-\sigma_1\widetilde\Gamma^3_3\sigma_1$. In conclusion, $\Gamma^3_3=a\sigma_3$ and $\widetilde\Gamma^3_3=b\sigma_3$, with $a\in\mathbb R$, $b\in\mathbb C$ and $\Re(b)=a/2$. This proves the second line of \eqref{2ndeq:lem3.2} and \eqref{Zcomplex.1}. 

\medskip

\noindent{{\it Regime $h< h_{*}$}.} Let us now consider a term quadratic in $\psi$ and linear in $A$, symmetric under  the symmetries (i), (ii), (iii), (iv) of Section \ref{appA.1}, of the form 
$$ \sum_{\omega = \pm}\int \frac{d\pp}{(2\pi)^4}\int \frac{d\kk}{(2\pi)^4}\hat A_{\mu,\pp}\hat \psi^+_{\o,\kk+\pp} \G_{\mu,\o} \hat\psi^-_{\o,\kk},$$ 
where $\G_{\mu,\o}$ are complex $2\times 2$ matrices. 
Imposing symmetry (iii) we find that $\G_{\mu,\o}=(-1)^{\d_{\m,3}}\G_{\mu,-\o}$, so we let $\G_{\mu,\o}\equiv \G_\mu$, for $\mu=0,1,2$, and $\G_{3,\o}\equiv \o \G_3$. 
Imposing the validity of symmetries  (i), (ii), (iv), we find the same as \eqref{s2}--\eqref{s5}, with $M_\mu$ replaced by $\G_\mu$. Therefore, $\G_0=ic_0\mathds 1$, $\G_1=c_1\s_1$, $\G_2=c_2\s_2$, $\G_3=c_3\s_3$, with 
$c_1,c_1,c_2,c_3\in\mathbb R$. This proves the first identity in \eqref{locApsipsi}. 

\subsubsection{The local $A_5\psi^+\psi^-$ and $A_5\psi^+\partial_3\psi^-$ terms.} 
\noindent{{\it Regime $h\geq h_{*}$.}} Let us consider the following terms, quadratic in $\psi$ and linear in $A^5$, symmetric under the symmetries (i), (ii), (iii), (iv) of Section \ref{appA.1}:
$$\int \frac{d\pp}{(2\pi)^4}\int \frac{d\kk}{(2\pi)^4}\, \hat A^5_{\mu,\pp}\hat \psi^+_{\kk+\pp}\G^5_{\mu} \hat\psi^-_{\kk} \quad \text{and}\quad \int \frac{d\pp}{(2\pi)^4}\int \frac{d\kk}{(2\pi)^4}\, \hat A_{\mu,\pp}\hat \psi^+_{\kk+\pp}(k_3\Gamma^{5;3}_\mu+p_3\widetilde\Gamma^{5;3}_\mu) \hat\psi^-_{\kk},$$ 
$\G_{\mu}^5, \G^{5;3}_\mu, \widetilde{\G}^{5;3}_\mu$ are complex $2\times 2$ matrices. Imposing the symmetry (iii), we get $\Gamma^{5}_{\mu} = 0$ for $\mu \neq 3$, and
$\G^{5;3}_3=\widetilde\G^{5;3}_3=0$. Symmetries (i), (ii), (iv) imply that $\G^5_3=(\G^5_3)^\dagger=\s_3 \G^5_3\s_3=-\s_1 \G^5_3 \s_1$, 
which gives $\G_3^5=a\s_3$, with $a\in\mathbb R$. This proves the second identity in the first line of \eqref{2ndeq:lem3.2}.
Symmetry (i) implies that 
$\Gamma^{5;3}_0=-(\Gamma^{5;3}_0)^\dagger$ and $\widetilde\Gamma^{5;3}_0-(\widetilde\Gamma^{5;3}_0)^\dagger=\Gamma^{5;3}_0$, while
$\Gamma^{5;3}_\mu=(\Gamma^{5;3}_\mu)^\dagger$ and $\widetilde\Gamma^{5;3}_\mu+(\widetilde\Gamma^{5;3}_\mu)^\dagger=\Gamma^{5;3}_\mu$ for $\mu=1,2$. 
Moreover, symmetries (ii) and (iv) imply that:
$\Gamma^{5;3}_0=\sigma_3\Gamma^{5;3}_0\sigma_3=\sigma_1\Gamma^{5;3}_0\sigma_1$ and the same $\widetilde{\Gamma}^{5;3}_0$; 
$\Gamma^{5;3}_1=-\sigma_3\Gamma^{5;3}_1\sigma_3=\sigma_1\Gamma^{5;3}_1\sigma_1$ and the same $\widetilde{\Gamma}^{5;3}_1$; 
$\Gamma^{5;3}_2=-\sigma_3\Gamma^{5;3}_2\sigma_3=-\sigma_1\Gamma^{5;3}_2\sigma_1$ and the same $\widetilde{\Gamma}^{5;3}_2$. 
In conclusion:
$\Gamma^{5;3}_0=ia_0\mathds 1_2$ and $\widetilde\Gamma^{5;3}_0=ib_0\mathds 1_2$;
$\Gamma^{5;3}_1=a_1\sigma_1$ and $\widetilde\Gamma^{5;3}_1=b_1\sigma_1$;
$\Gamma^{5;3}_2=a_2\sigma_2$ and $\widetilde\Gamma^{5;3}_2=b_2\sigma_2$;
where $a_\mu\in\mathbb R$, $b_\mu\in\mathbb C$, and $\Re(b_\mu)=a_\mu/2$, for all $\mu=0,1,2$. 
This proves the third line of \eqref{2ndeq:lem3.2} and \eqref{Zcomplex.2}. 

\medskip

\noindent{\it{Regime $h < h_{*}$.}} Let us consider a term quadratic in $\psi$ and linear in $A^5$, symmetric under the symmetries (i), (ii), (iii), (iv) of Section \ref{appA.1}, of the form 
$$\int \frac{d\pp}{(2\pi)^4}\int \frac{d\kk}{(2\pi)^4}\hat A_{\mu,\pp}^5\hat \psi^+_{\o,\kk+\pp} \G_{\mu,\o}^5 \hat\psi^-_{\o,\kk},$$ 
where $\G_{\mu,\o}^5$ are complex $2\times 2$ matrices. 
Imposing symmetry (iii) we find that $\G_{\mu,\o}^5=(-1)^{1-\d_{\m,3}}$ $\G_{\mu,-\o}^5$, so we let $\G_{\mu,\o}^5\equiv \o\G_\mu^5$, for $\mu=0,1,2$, and $\G_{3,\o}^5\equiv \G_3^5$. 
Imposing the validity of symmetries (i), (ii), (iv), we find the same as \eqref{s2}--\eqref{s5}, with $M_\mu$ replaced by $\G_\mu^5$. Therefore, $\G_0^5=ic_0\mathds 1$, $\G_1^5=c_1\s_1$, $\G_2^5=c_2\s_2$, $\G_3^5=c_3\s_3$, with 
$c_1,c_1,c_2,c_3\in\mathbb R$. This proves the second identity in \eqref{locApsipsi}. 

\section{Dimensional bounds on the kernels of the effective potential}\label{app.dimbounds}

In this appendix we provide some additional details on the proofs of \eqref{eq:bdW}, \eqref{eq:bdW2} and of their improved analogues, discussed after 
\eqref{eq:bdW} and \eqref{eq:bdW2}, respectively. As anticipated above, the proofs of these bounds are completely analogous to those of Lemma 1 and Lemma 2 in
\cite{M1bis}, where a specific model of Weyl semimetal within the class of models considered in this paper was analyzed. Since there a few minor differences in the bounds \eqref{eq:bdW}, \eqref{eq:bdW2} and in their improved analogues, as compared to 
those stated in \cite[Lemma 1 and 2]{M1bis}, here we highlight where these differences rely and explain their origin. 

A first macroscopic difference is that \cite{M1bis} explicitly treats only the free energy (i.e., $A=A^5=0$), which obviously has an impact on the definition of the iterative step. 
In addition to this, there are other more technical differences, which we discuss separately for the first and second regimes. 

\medskip

{\it First regime.} The RG construction in the first regime is virtually the same as the one described in \cite[Section 2]{M1bis}, modulo a slightly different definition of localization
(compare \cite[Eq.(43)]{M1bis} with the definition in \eqref{eq:ell.0}-\eqref{eq:ell}) that, in particular, takes into account the presence of the external fields $A$ and $A^5$. 
At each scale $h_*\le h\le 0$, the effective potential can be expressed in terms of a GN tree expansion virtually equivalent to the one described in \cite[Section 2.1]{M1bis}.
As already mentioned, the proof of \eqref{eq:bdW} and of its improved analogue (see discussion after \eqref{eq:bdW}) goes along the same lines as the proof of 
Eqs.(60) and (61) of \cite{M1bis}, respectively. However, there are a few technical differences that deserve comments: 
\begin{itemize}
\item The $L^1$ norm in the left side of \eqref{eq:bdW} has a stretched exponential weight, contrary to the one used in \cite[Eq.(60)]{M1bis}. However, the inclusion 
of the stretched exponential weight involving the tree distance can be accomodate without extra difficulties, thanks to the stretched exponential decay of the propagator 
\eqref{eq:ghlow}. See \cite{AGG,GMR} for two recent works on fermionic RG where similar exponentially weighted $L^1$ norms are studied via the same kind of methods of this 
paper. 
\item Eq.\eqref{eq:bdW} takes into account the possibile presence of the derivative labels $\underline q$, which was neglected for simplicity in \cite[Eq.(60)]{M1bis}
(the proof of the bound in the presence of such indices remains unaltered). 
\item Most importantly, the improved dimensional bound on $W^{(h);\text{r}}_{n,m_1,m_2,\underline q}$ has an additional factor $\gamma^{\theta h}$, as compared with 
\eqref{eq:bdW}, where $\theta$ is {\it any} positive constant smaller than $1$; see discussion in the paragraph after \eqref{eq:bdW}. This should be compared with the dimensional 
gain $\gamma^{h/8}$ stated in \cite[Eq.(61)]{M1bis}. Actually, in \cite[Eq.(61)]{M1bis}, the factor $\gamma^{h/8}$ can be replaced `for free' by $\gamma^{\theta h}$ with $\theta$ a positive constant strictly smaller than $1/2$, simply because in that case the largest scaling dimension of an irrelevant operator is $-1/2$ (see the comment ``{\it Note that, if $v$ is not an endpoint, $5\frac{|P_v|}{4}-\frac72+z(P_v)\ge \frac12$ by the definition of $\mathcal R$''} after \cite[Eq.(70)]{M1bis}, and note that, thanks to this fact, the 
factors $\gamma^{-(h_v-h_{v'})/4}$ in \cite[Eqs.(71), (73), (74)]{M1bis} can be replaced by $\gamma^{-\theta(h_v-h_{v'})}$
with $0<\theta<1/2$, thus leading to the claimed improvement). In our case, thanks to the definition of $\mathcal L$, see \eqref{eq:ell.0}-\eqref{eq:ell}, and to the cancellation 
properties stated in the Remark after \eqref{eq:ell} and in Lemma \ref{lem:sym1}, the largest scaling dimension of an irrelevant operator is $-1$, rather than $-1/2$; therefore, 
by the same line of reasoning, the GN trees with at least one endpoint on scale $1$ admit a bound with an additional `short memory factor' $\gamma^{\theta h}$ with 
$\theta$ a positive constant strictly smaller than $1$ (rather than `just' $1/2$). 
\end{itemize}

{\it Second regime.} The RG construction in the first regime is virtually the same as the one described in \cite[Section 3]{M1bis}, modulo a slightly different definition of localization
(compare \cite[Eq.(88)]{M1bis} with the definition in \eqref{eq:ell.2nd}-\eqref{eq:ell2nd}) that, in particular, takes into account the presence of the external fields $A$ and $A^5$. 
At each scale $h<h_*$, the effective potential can be expressed in terms of a GN tree expansion virtually equivalent to the one described in \cite[Section 3.1]{M1bis}.
As already mentioned, the proof of \eqref{eq:bdW2} and of its improved analogue (see discussion in the paragraph after \eqref{eq:bdW2}) goes along the same lines as the proof of 
Eqs.(93) and (94) of \cite{M1bis}, respectively. Also in this case, there are a few technical differences that deserve comments: 
\begin{itemize}
\item As in the first regime, in \eqref{eq:bdW2} we use a weighted $L^1$ norm, rather than the standard one, and we take into account the possibile presence of the derivative labels $\underline q$; as commented above, the proof of the bound remains essentially unaltered by these two modifications.
\item Concerning \eqref{eq:bdW2} and its improved version, an important difference with respect to \cite[Eqs.(93)-(94)]{M1bis} is the presence of the 
pre-factor $|v_3^0|^{n-1+\|\underline q^3\|_1}$. Even though such factor does not appear in \cite[Eqs.(93)-(94)]{M1bis}, 
its presence is actually proved in \cite{M1bis} (in the case $\|\underline q^3\|_1=0$, but the proof in the general case is essentially unchanged), 
see l.1 after \cite[Eq.(107)]{M1bis}.
\item Concerning the improved dimensional bound on $W^{(h);\text{r}}_{n,m_1,m_2,\underline q}$, in the paragraph after \eqref{eq:bdW2} we claim that it has 
an additional factor $\gamma^{\theta h}$ if $m_1=m_2=0$ or $n\ge 2$, or $\gamma^{\theta(h-h_*)}$ if $m_1+m_2\ge 1$ and $n=1$, with $\theta$ any positive constant strictly smaller than $1$; this has to be compared with the factor $\gamma^{\frac18(h-h_*)}$ stated in \cite[Eq.(94)]{M1bis}. First of all, for the same reasons explained above for the 
first regime (see in particular the third item of the dotted list), the factor $\gamma^{\frac18(h-h_*)}$ stated in \cite[Eq.(94)]{M1bis} can be straightforwardly improved to 
$\gamma^{\theta(h-h_*)}$, for any $0<\theta<1$, because the largest scaling dimension of an irrelevant operator is $-1$. We still need to discuss the origin of an additional 
gain factor $\gamma^{\theta h_*}$ for the terms with $m_1=m_2=0$ or $n\ge 2$. The point is that these terms are defined in terms of GN trees with at least one endpoint on scale 
$h_*+1$ with $m_1=m_2=0$ or $n\ge 2$; in turn, each such endpoint has a kernel generated by the tree expansion of the first regime, and comes from GN trees with at least one endpoint on scale $1$: therefore, it is bounded via the improved dimensional bound discussed in the paragraph after \eqref{eq:bdW}, which is characterized by
an additional factor $\gamma^{\theta h_*}$ as compared to the `basic' bound \eqref{eq:bdW}, as desired. 
\end{itemize}

\section{The quadratic response}\label{app:2nd}
 In this appendix we compute the quadratic response coefficient in the right side of \eqref{eq:J5av} and prove that its $\beta,L\to\infty$ limit can be expressed in terms 
 of Euclidean correlation functions. 
  
\subsection{Second-order time-dependent perturbation theory}
Consider a time-dependent vector potential $A_{x}(t) = e^{\eta t} A_x$ with $\eta > 0$. The evolution of the state for $t\le 0$ is determined by the von Neumann equation,
\begin{equation}
i\partial_{t} \rho(t) = [ \mathcal{H}_L(A({t})), \rho(t) ]\;,\qquad \rho(-\infty) = \rho_{\beta, L}, 
\end{equation}
with $\rho_{\beta, L}$ as in \eqref{defGibbs}.
Given an observable $\mathcal{O}$, we denote by $\mathcal{O}(A({t}))$ its coupling to the external gauge field, via the Peierls substitution.
We are interested in computing $\tr\, \mathcal{O}(A({t})) \rho(t)\equiv \langle \mathcal O(A(t))\rangle_{\beta,L;t}$ at second order in the external field, in the case that $\mathcal O=\hat J^5_{\mu,p}$. We set:
\begin{eqnarray}
\langle \mathcal{O}(A({t})) \rangle_{\beta,L;t} - \langle \mathcal{O} \rangle_{\beta,L} &=& (\langle \mathcal{O}(A({t})) \rangle_{\beta,L;t} - \langle \mathcal{O}(A({t})) \rangle_{\beta,L}) + (\langle \mathcal{O}(A({t})) \rangle_{\beta,L} - \langle \mathcal{O} \rangle_{\beta, L}) \nonumber\\
&\equiv& \text{I} + \text{II}\;.
\end{eqnarray}
We compute $\text{I}$ and $\text{II}$ separately, at second order in the external field. To this end, it is convenient to study the evolution of the state in the interaction picture. Let $\rho^{\text{int}}(t) := e^{i\mathcal{H}_L t} \rho(t) e^{-i\mathcal{H}_Lt}$. Then:
\begin{equation}\label{eq:vN}
i\partial_{t} \rho^{\text{int}}_{t} = [ \mathcal{H}_{L,t}(A({t})) - \mathcal{H}_L, \rho^{\text{int}}(t) ]\;.
\end{equation}
where $\mathcal{H}_{L,t}(A({t})) = e^{i\mathcal{H}_L t} \mathcal{H}_L(A({t})) e^{-i\mathcal{H}_Lt}$. In general, in this appendix, the subscript $t$ stands for the evolution at time $t$ in the interaction picture: for any observable $O$, we let $O_t=e^{i\mathcal H_L t}Oe^{-i\mathcal H_Lt}$ (we also let $\mathcal O_t(A(t))\equiv [\mathcal O(A(t))]_t=e^{i\mathcal H_L t}\mathcal O(A(t))e^{-i\mathcal H_Lt}$, etc.)

\medskip

\noindent{\underline{\it Term $\text{I}$}.} A simple computation gives:
\begin{eqnarray} \text{I}=
\langle \mathcal{O}(A({t})) \rangle_{\beta,L;t} - \langle \mathcal{O}(A({t})) \rangle_{\beta, L}&=&-i\int_{-\infty}^{t} ds\, \tr\, \mathcal{O}_{t}(A({t})) [ \mathcal{P}_{s}(A({s})), \rho^{\text{int}}(s) ]\\
&=&-i\int_{-\infty}^{t} ds\, \tr [\mathcal{O}_{t}(A({t})), \mathcal{P}_{s}(A({s}))]\rho^{\text{int}}(s),\nonumber
\end{eqnarray}
where $\mathcal{P}(A(t)) := \mathcal{H}_{L}(A({t})) - \mathcal{H}_L$. and in the second identity we used the cyclicity of the trace. A one-step iteration of this formula gives: 
\begin{eqnarray} \text{I}
 &=& -i\int_{-\infty}^{t} ds\, \langle \mathcal{O}_{t}(A({t})), \mathcal{P}_{s}(A({s}))] \rangle_{\beta, L} \\
&-& \int_{-\infty}^{t} ds \int_{-\infty}^{s} ds_{1} \tr [ [\mathcal{O}_{t}(A({t})), \mathcal{P}_{s}(A({s}))], \mathcal{P}_{s_{1}}(A({s_{1}}))] \rho^{\text{int}}(s_{1})\;.\nonumber
\end{eqnarray}
We denote by $\text{I}^{(2)}$ the second order contribution in $A$. If we write $\mathcal{O}_t^{(k)}(A)$ for the $k$-th order of $\mathcal O_t(A)$ in $A$, and similarly for $\mathcal P_t^{(k)}(A)$, we get: 
\begin{eqnarray} && \text{I}^{(2)} =-i\int_{-\infty}^{t} ds\, \langle [\mathcal{O}_{t}, \mathcal{P}^{(2)}_{s}(A({s}))] \rangle_{\beta, L} -i\int_{-\infty}^{t} ds\, \langle [\mathcal{O}^{(1)}_{t}(A({t})), \mathcal{P}^{(1)}_{s}(A({s}))] \rangle_{\beta, L}\label{appC.I2}\\
&&\hskip.65truecm - \int_{-\infty}^{t} ds \int_{-\infty}^{s} ds_{1} \langle  [ [\mathcal{O}_{t}, \mathcal{P}^{(1)}_{s}(A({s}))], \mathcal{P}^{(1)}_{s_{1}}(A({s_{1}}))] \rangle_{\beta, L}\
\equiv\ \text{I}^{(2)}_1+\text{I}^{(2)}_2+\text{I}^{(2)}_3.\nonumber
\end{eqnarray}
Letting $\int \frac{dp}{(2\pi)^{3}}$ be a shorthand for $L^{-3} \sum_{p \in\frac{2\pi}L\mathbb Z^3}$, we have
\begin{eqnarray}
\mathcal{P}^{(1)}(A) &=& -\sum_{k=1,2,3} \int \frac{dp}{(2\pi)^{3}}\, \hat A_{k,p} \hat J_{k,p}, \nonumber\\ \mathcal{P}^{(2)}(A) &=& -\frac12\sum_{k,k' = 1,2,3} \int \frac{dp_{1}}{(2\pi)^{3}} 
\int\frac{dp_{2}}{(2\pi)^{3}}\,  \hat A_{k,p_{1}} \hat A_{k',p_{2}} \hat \Delta_{k, k'}(p_{1},p_{2})\;.\nonumber
\end{eqnarray}
where $\hat J_{k,p}$ is defined as in \eqref{eq:curr} and the second line should be understood as the definition of $\hat \Delta_{k, k'}(p_{1},p_{2})$. 
Let us now set $\mathcal{O} = \hat J^{5}_{\mu,p}$, in which case we write: 
\begin{equation}\label{eq:OJ5}
\mathcal{O}^{(1)}(A) = \sum_{k=1,2,3} \int \frac{dp_1}{(2\pi)^{3}}\, \hat A_{k,p_1} \hat \Delta^{5}_{\mu,k}(p,p_1), 
\end{equation}
which should be understood as the definition of $\hat\Delta^{5}_{\mu,k}(p,q)$. With these definitions, the three terms in the right side of \eqref{appC.I2} take the following explicit form: 
\begin{eqnarray}
\text{I}^{(2)}_{1} &=& \frac{i}{2} \sum_{k,k' = 1,2,3} \int \frac{dp_{1}}{(2\pi)^{3}}\int \frac{dp_{2}}{(2\pi)^{3}}\,  \hat A_{k,p_{1}} \hat A_{k',p_{2}} \int_{-\infty}^{t} ds\, e^{2\eta s} \langle  [ \hat J^{5}_{\mu,p,t}, \hat \Delta_{k,k',s}(p_{1}, p_{2})  ] \rangle_{\beta, L}\nonumber\\
&\equiv& \frac{i}{2} \sum_{k,k'} \int \frac{dp_{1}}{(2\pi)^{3}} \, e^{2\eta t} \hat A_{k,p_{1}} \hat A_{k',-p-p_{1}} \int_{-\infty}^{0} ds\, e^{2\eta s} \langle  [ \hat J^{5}_{\mu,p}, \hat \Delta_{k,k',s}(p_{1}, -p - p_{1})  ] \rangle_{\beta, L}\;,\label{I^2_1}
\end{eqnarray}
where in the last step we used space and time translation invariance of the Gibbs state,
\begin{equation}
\text{I}^{(2)}_{2} 
=i \sum_{k,k' } \int \frac{dp_{1}}{(2\pi)^{3}} \, e^{2\eta t} \hat A_{k,p_1} \hat A_{k',-p-p_1} \int_{-\infty}^{0} ds\, e^{\eta s} \langle  [ \hat \Delta^{5}_{\mu,k}(p,p_1), \hat J_{k',-p-p_{1},s} ] \rangle_{\beta, L},\label{I^2_2}
\end{equation}
and
\begin{equation}
\text{I}^{(2)}_{3}=- \sum_{k,k'} \int \frac{dp_{1}}{(2\pi)^{3}} \,  e^{2\eta t} \hat A_{k,p_{1}} \hat A_{k',-p-p_{1}}\int_{-\infty}^{0} ds \int_{-\infty}^{s} ds_{1}\, e^{\eta s + \eta s_{1}} \langle  [ [ \hat J^{5}_{\mu,p}, \hat J_{k,p_{1},s} ], \hat J_{k',-p-p_{1},s_{1}} ] \rangle_{\beta, L}\;.\label{I^2_3}
\end{equation}
\noindent{\underline{\it Term $\text{II}$}.} For $\mathcal O=\hat J^5_{\mu,p}$ we write:
$$\mathcal O^{(2)}(A)=\frac12\sum_{k,k'=1,2,3} \int \frac{dp_1}{(2\pi)^{3}}\int\frac{dp_2}{(2\pi)^3} \hat A_{k,p_1} \hat A_{k',p_2}\hat \Delta^{5}_{\mu,k,k'}(p,p_1,p_2), $$
which should be understood as the definition of $\hat\Delta^{5}_{\mu,k,k'}(p,p_1,p_2)$, from which we get: 
\begin{equation}\label{II^2}
\text{II}^{(2)}= \frac12\sum_{k,k'} \int \frac{dp_1}{(2\pi)^{3}}\, e^{2\eta t} \hat A_{k,p_1} \hat A_{k',-p-p_1}\langle\hat \Delta^{5}_{\mu,k,k'}(p,p_1,-p-p_1)\rangle_{\beta,L} \;.  
\end{equation}
In conclusion, $\langle\mathcal O(A(t))\rangle^{(2)}_{\beta,L;t}=\eqref{I^2_1}+\eqref{I^2_2}+\eqref{I^2_3}+\eqref{II^2}$, from which we find that 
the quadratic response coefficient in the right side of \eqref{eq:J5av}, recalling that $\pp_1=(\eta,p_1)$ and $\pp_2=(\eta,p_2)$, is given explicitly by 
\begin{eqnarray}\label{eq:secondcomm}&&
\hat \Pi^{5;\beta,L}_{\mu,k,k'}(\pp_1,\pp_2)=\\
&&\quad =-\frac1{L^3}\int_{-\infty}^{0} ds_1 \int_{-\infty}^{s_1} ds_{2}\, e^{\eta s_1 + \eta s_{2}}\Big\{
\langle  [ [ \hat J^{5}_{\mu,p}, \hat J_{k,p_{1},s_1} ], \hat J_{k',p_{2},s_{2}} ] \rangle_{\beta, L}+\big[(k,p_1)\otto(k',p_2)\big]\Big\}\nonumber\\
&&\hskip.38truecm +\ \frac{i}{L^3}\int_{-\infty}^{0} ds\Big\{ e^{2\eta s} \langle  [ \hat J^{5}_{\mu,p}, \hat \Delta_{k,k',s}(p_{1}, p_{2})  ] \rangle_{\beta, L}+ e^{\eta s}\big[ 
\langle  [ \hat \Delta^{5}_{\mu,k}(p,p_1), \hat J_{k',p_{2},s} ] \rangle_{\beta, L}\nonumber\\
&&\hskip2.8truecm+\ \langle  [ \hat \Delta^{5}_{\mu,k'}(p,p_2), \hat J_{k,p_{1},s} ] \rangle_{\beta, L}\big]\Big\}+\frac1{L^3}\langle\hat \Delta^{5}_{\mu,k,k'}(p,p_1,p_2)\rangle_{\beta,L}.\nonumber\end{eqnarray}
The next task is to Wick-rotate to imaginary times the various terms. The Wick rotation of the second term in the right side (the one expressed in terms of an integral over $s$)
can be performed as discussed in \cite{AMP, GMPhall, MPdrude}, see in particular \cite[Appendix B]{AMP} and \cite[Appendix B]{MPdrude}, and gives
\begin{eqnarray}&&\lim_{\beta,L\to\infty} \frac{i}{L^3}\int_{-\infty}^{0} ds\Big\{ e^{2\eta s} \langle  [ \hat J^{5}_{\mu,p}, \hat \Delta_{k,k',s}(p_{1}, p_{2})  ] \rangle_{\beta, L}+ e^{\eta s}\big[ 
\langle  [ \hat \Delta^{5}_{\mu,k}(p,p_1), \hat J_{k',p_{2},s} ] \rangle_{\beta, L}+\nonumber\\ 
&& \hskip2.4truecm+\ \langle  [ \hat \Delta^{5}_{\mu,k'}(p,p_2), \hat J_{k,p_{1},s} ] \rangle_{\beta, L}\big]\Big\}=\label{eq:G4wick} \\
&&\ =\ 
 \pmb{\langle} {\bf T}\, \hat J^{5}_{\mu,\pp}\;; \hat \Delta_{k,k'}(\pp_{1}, \pp_{2}) \pmb{\rangle}_{\infty} + \pmb{\langle} {\bf T}\,\hat \Delta^{5}_{\mu,k}(\pp, \pp_{1})\;; \hat J_{k',\pp_{2}} \pmb{\rangle}_{\infty} + \pmb{\langle} {\bf T}\, \hat \Delta^{5}_{\mu,k'}(\pp, \pp_{2})\;; \hat J_{k,\pp_{1}}  \pmb{\rangle}_{\infty},\nonumber\end{eqnarray}
where: $\pp_{i} = (\eta, p_{i})\equiv(p_{i,0}, p_i)$, $\pp = (-2\eta, -p_{1} - p_{2})\equiv(p_0,p)$, 
$$\hat J^5_{\mu,\pp}=\int_{0}^\beta dx_0 e^{-ip_0x_0}\hat J^5_{\mu,(x_0,p)} \quad \text{and}\quad \hat \Delta_{k,k'}(\pp_{1}, \pp_{2})=\int_{0}^\beta dx_0 e^{-i(p_{1,0}+p_{2,0})x_0} [\hat \Delta_{k,k'}(p_1,p_2)]_{x_0}, $$
with $\hat J^5_{\mu,(x_0,p)}=e^{x_0\mathcal H_L}\hat J^5_{\mu,p}e^{-x_0\mathcal H_L}$ and 
$[\hat \Delta_{k,k'}(p_1,p_2)]_{x_0}=e^{x_0\mathcal H_L}
\hat \Delta_{k,k'}(p_1,p_2)e^{-x_0\mathcal H_L}$ ($\hat J_{k,\pp_1}$ and $\hat \Delta^{5}_{\mu,k'}(\pp, \pp_{2})$ are defined analogously). Moreover, 
$\bf T$ is the imaginary-time-ordering operator, ordering the operators in decreasing imaginary-time order, and 
$\pmb{\langle} \cdot \pmb{\rangle}_{\infty} = \lim_{\beta, L \to \infty} (\beta L^{3})^{-1} \langle \cdot \rangle_{\beta,L}$. 
The semicolon symbol in the right-hand side of (\ref{eq:G4wick}) denotes truncation in the correlation functions. To introduce it, we used that in the left side 
we can freely subtract to every observable in the commutators the corresponding statistical average (of course, the subtraction leaves the commutator invariant). 

We are left with discussing the Wick rotation of the first term in the right side of \eqref{eq:secondcomm}. This will be done in the next subsection, where we prove that 
\begin{eqnarray} &&\hskip-.2truecm -\lim_{\beta,L\to\infty}\frac1{L^3}\int_{-\infty}^{0} ds_1 \int_{-\infty}^{s_1} ds_{2}\, e^{\eta s_1 + \eta s_{2}}\Big\{
\langle  [ [ \hat J^{5}_{\mu,p}, \hat J_{k,p_{1},s_1} ], \hat J_{k',p_{2},s_{2}} ] \rangle_{\beta, L}+\big[(k,p_1)\otto(k',p_2)\big]\Big\}=\nonumber\\
&&\qquad =\ \pmb{\langle} {\bf T}\, \hat J^{5}_{\mu,\pp}\;; \hat J_{k,\pp_{1}}\;; \hat J_{k',\pp_{2}} \pmb{\rangle}_{\infty},\label{proveC.2}\end{eqnarray}
with the same notations as in \eqref{eq:G4wick}. All in all, we get:
\begin{eqnarray}\label{eq:G5wick}
&&\hat\Pi^{5}_{\mu, k, k'}(\pp_{1}, \pp_{2})= \pmb{\langle} {\bf T}\, \hat J^{5}_{\mu,\pp}\;; \hat J_{k,\pp_{1}}\;; \hat J_{k',\pp_{2}} \pmb{\rangle}_{\infty}+
 \pmb{\langle} {\bf T}\, \hat J^{5}_{\mu,\pp}\;; \hat \Delta_{k,k'}(\pp_{1}, \pp_{2}) \pmb{\rangle}_{\infty}\\
 && + \pmb{\langle} {\bf T}\,\hat \Delta^{5}_{\mu,k}(\pp, \pp_{1})\;; \hat J_{k',\pp_{2}} \pmb{\rangle}_{\infty} + \pmb{\langle} {\bf T}\, \hat \Delta^{5}_{\mu,k'}(\pp, \pp_{2})\;; \hat J_{k,\pp_{1}}  \pmb{\rangle}_{\infty}+\,\lim_{\beta,L\to\infty}  \langle\hat \Delta^{5}_{\mu,k,k'}(p,p_1,p_2)\rangle_{\beta,L}. \nonumber
\end{eqnarray}

\subsection{Wick rotation for correlations of three observables}\label{sec:wick3}

In this section we prove \eqref{proveC.2}. Let 
$$T_{\beta,L}(\eta)=-\frac1{L^3}\int_{-\infty}^{0} ds_1 \int_{-\infty}^{s_1} ds_{2}\, e^{\eta s_1 + \eta s_{2}}\Big\{
\langle  [ [ \hat J^{5}_{\mu,p}, \hat J_{k,p_{1},s_1} ], \hat J_{k',p_{2},s_{2}} ] \rangle_{\beta, L}+\big[(k,p_1)\otto(k',p_2)\big]\Big\}$$
be the function of interest, thought of as a function of $\eta$. We add and subtract $T_{\beta,L}(\eta_\beta)$, where $\eta_\beta=\frac{2\pi}{\beta}\lceil \frac{\beta\eta}{2\pi}\rceil\in 
\frac{2\pi}{\beta}\mathbb N$. We will prove below that $T_{\beta,L}(\eta)-T_{\beta,L}(\eta_\beta)$ is bounded by (const.)$\beta^{-1}$, 
uniformly in $L$. As for $T_{\beta,L}(\eta_\beta)$, it is equal to a double integral over imaginary times of the appropriate Euclidean correlation function, as implied by the following 
proposition. 

\begin{prop}\label{prp:wick} Let $A, B, C$ be bounded fermionic operators, of the form:
\begin{equation}\label{eq:OX}
O = \sum_{X \subset \Lambda_{L}} O_{X}\;,
\end{equation}
with $O_{X}$ even in the fermionic operators, commuting with the total number operator. We assume that the sum in (\ref{eq:OX}) runs over subsets $X$ such that $|X|$ is bounded uniformly in $L$. Let $A(z) = e^{i \mathcal{H}_L z} A e^{-i\mathcal{H}_Lz}$, for $z\in \mathbb{C}$. Let $\eta_{i} \in (2\pi/\beta) \mathbb{N}$ (with the convention that $\mathbb N$ is the set of positive integers) and consider:
\begin{equation}\label{eq:IABC}
\mathcal{I}^{\beta,L}_{ABC} := \!\!-\int_{-\infty}^{0}\!\! ds_{1} \int_{-\infty}^{s_{1}} \!\!ds_{2}\Big(e^{\eta_{1} s_{1} + \eta_{2} s_{2}}
\langle[ [C, A(s_1)],B(s_{2})] \rangle_{\beta, L} + e^{\eta_{1} s_{2} + \eta_{2} s_{1}} \langle[ [ C, B(s_1)],A(s_{2})] \rangle_{\beta, L}\Big).\end{equation}
One has $\mathcal{I}^{\beta,L}_{ABC}=\mathcal{J}^{\beta,L}_{ABC}$, where 
\begin{equation}\label{eq:JABC}
\mathcal{J}^{\beta,L}_{ABC}:= \int_{0}^{\beta} ds_{1} \int_{0}^{\beta} ds_{2}\, e^{-is_{1} \eta_{1} - is_{2}\eta_{2}}\langle {\bf T}\,A(-is_{1}) B(-is_{2}) C \rangle_{\beta, L}\;,
\end{equation}
with $\bf T$ the time-ordering operator, which orders the operator in the decreasing imaginary-time order. 
\end{prop}

Before giving the proof of Proposition \ref{prp:wick}, let us explain how to adapt it to the case at hand. We let: $C=\hat J^{5}_{\mu,p}-\langle \hat J^{5}_{\mu,p}\rangle_{\beta,L}$, 
$A(s_1)=\hat J_{k,p_{1},s_1}-\langle \hat J_{k,p_{1}} \rangle_{\beta,L}$, and $B(s_2)=\hat J_{k',p_{2},s_2}-\langle \hat J_{k,p_{2}} \rangle_{\beta,L}$, so that, 
for $\eta_1=\eta_2=\eta_\beta$, $T_{\beta,L}(\eta_\beta)=\mathcal{I}^{\beta,L}_{ABC}$. Using the proposition, we get
\begin{eqnarray}T_{\beta,L}(\eta_\beta)&=&\frac1{L^3}\int_{0}^{\beta} dx_{1,0} \int_{0}^{\beta} dx_{2,0}\, e^{-i \eta_\beta(x_{1,0}+x_{2,0})}
\langle {\bf T}\, \hat J_{k,(x_{1,0},p_1)}; \hat J_{k',(x_{2,0},p_2)}; \hat J^5_{\mu,p}\rangle_{\beta, L}\nonumber\\
&\equiv&\frac1{\beta L^3}\langle {\bf T}\, \hat J_{k,\tilde\pp_1}; \hat J_{k',\tilde\pp_2}; \hat J^5_{\mu,\tilde\pp}\rangle_{\beta, L},\label{eq:C.19}
 \end{eqnarray}
with $\tilde\pp_{i} = (\eta_\beta, p_{i})$ and $\pp = (-2\eta_\beta, -p_{1} - p_{2})$.  If we now take $\beta,L\to\infty$ with $\eta_\beta\to \eta$, this expression tends to 
$\pmb{\langle} {\bf T}\,\hat J_{k,\pp_{1}}\;; \hat J_{k',\pp_{2}}\;;\hat J^{5}_{\mu,\pp} \pmb{\rangle}_{\infty}= \pmb{\langle} {\bf T}\, \hat J^{5}_{\mu,\pp}\;; \hat J_{k,\pp_{1}}\;; 
\hat J_{k',\pp_{2}} \pmb{\rangle}_{\infty}$, as desired (existence of the limit follows from the construction of the Euclidean correlation functions of Section \ref{sec:RG}). 

\begin{proof} By using the definition of $\bf T$, we rewrite \eqref{eq:JABC} as:
\begin{eqnarray}
\mathcal{J}^{\beta,L}_{ABC}&=& \int_{0}^{\beta} ds_{1} \int_{0}^{s_{1}} ds_{2}\, e^{-is_{1} \eta_{1} - is_{2}\eta_{2}}\langle A(-is_{1}) B(-is_{2}) C \rangle_{\beta, L} \nonumber\\
&+& \int_{0}^{\beta} ds_{1} \int_{s_{1}}^{\beta} ds_{2}\, e^{-is_{1} \eta_{1} - is_{2}\eta_{2}}\langle B(-is_{2}) A(-is_{1}) C \rangle_{\beta, L} \equiv \text{I} + \text{II}\;.
\end{eqnarray}
For notational convenience, we denote:
\begin{equation}
A_{\eta_{1}}(z) := e^{z \eta_{1}} A(z)\;,\quad B_{\eta_{2}}(z) := e^{z \eta_{2}} B(z)\;,\qquad z\in \mathbb{C}\;.
\end{equation}
Consider $\text{I}$. We apply Cauchy theorem to rewrite the integral over $s_{2}$ as follows:
\begin{eqnarray}\label{eq:cau2}
\int_{0}^{s_{1}} ds_{2}\, \langle A_{\eta_{1}}(-is_{1}) B_{\eta_{2}}(-is_{2}) C \rangle_{\beta, L} &=&- i\int_{-\infty}^{0} dt_{2}\, \langle A_{\eta_{1}}(-is_{1}) B_{\eta_{2}}(t_{2}) C \rangle_{\beta, L}\\
&+& \phantom{-}i\int_{-\infty}^{0} dt_{2}\, \langle A_{\eta_{1}}(-is_{1}) B_{\eta_{2}}(t_{2} - is_{1}) C \rangle_{\beta, L}\;,\nonumber
\end{eqnarray}
where we used the fact that $B_{\eta_{2}}(z) \to 0$ as $\text{Re}z\to -\infty$, thanks to the factor $e^{\eta_{2} \text{Re} z}$ and the fact that $\eta_2>0$; see Fig. \ref{fig:1}.

\begin{figure}\centering
\includegraphics[scale=0.9]{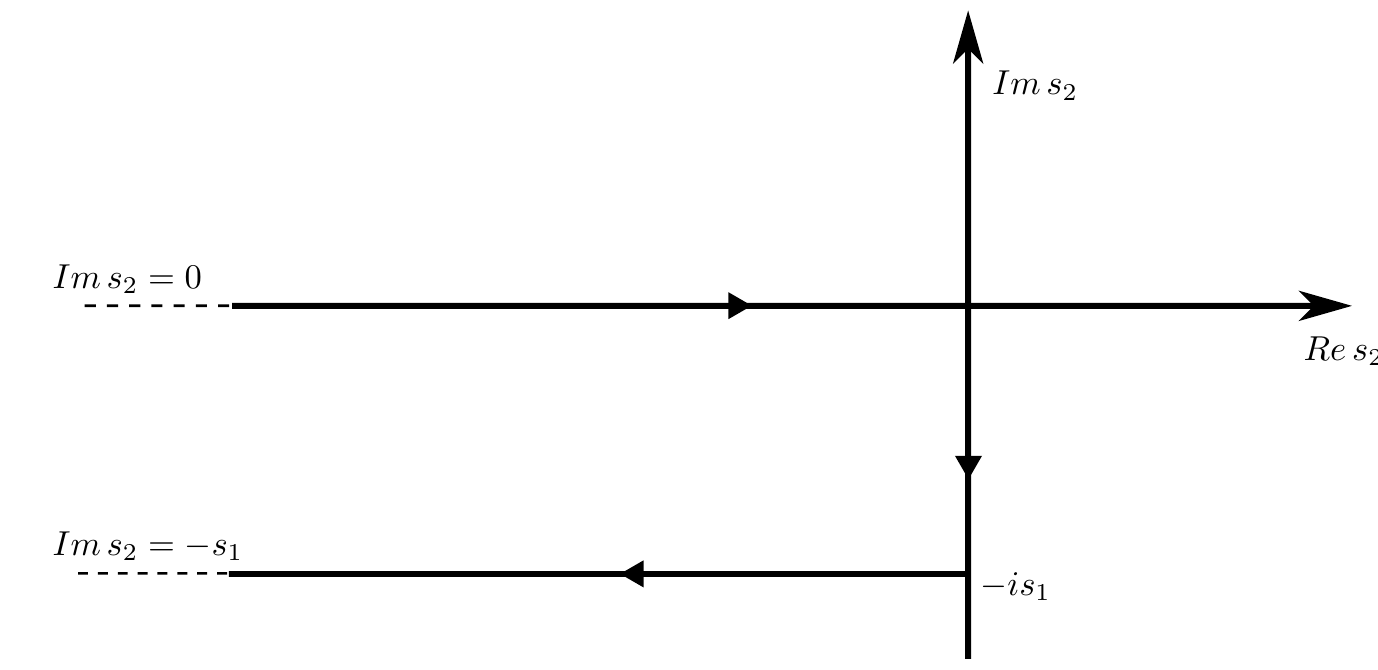}
 \caption{The integral over the complex contour is zero.}\label{fig:1}
\end{figure}

We now rewrite the integral over $s_{1}$ in a similar way, thus getting
\begin{eqnarray}\label{eq:cau}
\text{I} &=& - \int_{-\infty}^{0} dt_{1} \int_{-\infty}^{0} dt_{2}\, \Big( \langle A_{\eta_{1}}(t_{1}) B_{\eta_{2}}(t_{2}) C \rangle_{\beta, L} - \langle A_{\eta_{1}}(t_{1} - i\beta) B_{\eta_{2}}(t_{2}) C \rangle_{\beta, L}\\&&\qquad\qquad - \langle A_{\eta_{1}}(t_{1}) B_{\eta_{2}}(t_{1} + t_{2}) C \rangle_{\beta, L} + \langle A_{\eta_{1}}(t_{1} - i\beta) B_{\eta_{2}}(t_{1} + t_{2} - i\beta) C \rangle_{\beta, L}\Big)\;. \nonumber
\end{eqnarray}
Recalling that $e^{i\eta_{1}\beta}=1$, we have $\langle A_{\eta_{1}}(t_{1} - i \beta) B \rangle_{\beta, L} = \langle B A_{\eta_{1}}(t_{1}) \rangle_{\beta, L}$, 
and similarly for $B_{\eta_2}(t_1+t_2-i\beta)$, so that (\ref{eq:cau}) can 
be rewritten as:
\begin{equation}
\text{I} =
- \int_{-\infty}^{0} dt_{1} \int_{-\infty}^{0} dt_{2}\, \Big( \langle [ A_{\eta_{1}}(t_{1}), B_{\eta_{2}}(t_{2})C ] \rangle_{\beta,L} + \langle [ C, A_{\eta_{1}}(t_{1}) B_{\eta_{2}}(t_{1} + t_{2}) ] \rangle_{\beta,L}\Big)\;.
\end{equation}
By proceeding in the same way, we can rewrite $\text{II}$ analogously: 
\begin{equation}
\text{II} =-\int_{-\infty}^{0} dt_{1} \int_{-\infty}^{0} dt_{2}\, \Big( \langle [ C B_{\eta_{2}}(t_{2}), A_{\eta_{1}}(t_{1}) ]  \rangle_{\beta,L} + \langle [ B_{\eta_{2}}(t_{1} + t_{2}) A_{\eta_{1}}(t_{1}), C ] \rangle_{\beta,L} \Big)\;.
\end{equation}
If we now take the sum of $\text{I}$ and $\text{II}$, use the fact that $[A, BC] = B [A, C] + [A, B]C$ and recall the Jacobi identity $[[A, B], C]] + [[C, A], B] + [[ B, C], A]  = 0$, we find:
\begin{eqnarray}
&&\mathcal{J}^{\beta,L}_{ABC} = \int_{-\infty}^{0} dt_{1} \int_{-\infty}^{0} dt_{2}\, \Big( \langle [[ B_{\eta_{2}}(t_{2}),C ],A_{\eta_{1}}(t_{1})],  \rangle_{\beta,L} +
\langle [[ A_{\eta_{1}}(t_{1}) ,B_{\eta_{2}}(t_{1} + t_{2}) ],C] \rangle_{\beta,L} \Big)\nonumber\\
&&= \int_{-\infty}^{0} dt_{1} \int_{-\infty}^{0} dt_{2}\, \langle [[ B_{\eta_{2}}(t_{2}),C ],A_{\eta_{1}}(t_{1})],  \rangle_{\beta,L} +
\int_{-\infty}^{0} dt_{1} \int_{-\infty}^{t_1} dt_{2}\, 
\langle [[ A_{\eta_{1}}(t_{1}) ,B_{\eta_{2}}(t_{2}) ],C] \rangle_{\beta,L} .\nonumber
\end{eqnarray}
In the first term in the second line we split the integral over $t_2$ as $\int_{-\infty}^0dt_2=\int_{-\infty}^{t_1}dt_2+\int_{t_1}^0dt_2$. If we now combine together the two terms 
with the integral over $t_2$ from $-\infty$ to $t_1$ and use the Jacobi identity again, we get:
\begin{equation}
\mathcal{J}^{\beta,L}_{ABC} =  \int_{-\infty}^{0} dt_{1} \int_{t_1}^{0} dt_{2}\, \langle [[ B_{\eta_{2}}(t_{2}),C ],A_{\eta_{1}}(t_{1})],  \rangle_{\beta,L} -
\int_{-\infty}^{0} dt_{1} \int_{-\infty}^{t_1} dt_{2}\, 
\langle [[ C, A_{\eta_{1}}(t_{1}) ],B_{\eta_{2}}(t_{2}) ]\rangle_{\beta,L},\nonumber
\end{equation}
which is the same as the right side of \eqref{eq:IABC}. \end{proof}

In view of the discussion at the beginning of this subsection, as well as of \eqref{eq:C.19} and following lines, in order to conclude the proof of \eqref{proveC.2} we are left 
with proving that $T_{\beta,L}(\eta)-T_{\beta,L}(\eta_\beta)$ is bounded from above by (const.)$\beta^{-1}$, uniformly in $L$. Using the fact that $0\le \eta_\beta - \eta \leq (2\pi)/\beta$, 
we find that 
\begin{eqnarray} &&\big| T_{\beta,L}(\eta)-T_{\beta,L}(\eta_\beta)\big|\le\\
&&\qquad \le  \frac{2\pi}{\beta}
\frac1{L^3}\int_{-\infty}^{0} ds_1 \int_{-\infty}^{s_1} ds_{2}\,(|s_{1}|+|s_{2}|) e^{\eta(s_1 + s_{2})}\Big\{
\big\|  \big[ [ C, A(s_1)], B(s_2)\big]\big\| +(A\otto B)\Big\},\nonumber\end{eqnarray}
with $A,B,C$ as defined before \eqref{eq:C.19}. 
By the Lieb-Robinson bounds for multi-commutators \cite{BruPedra_multi}, see in particular \cite[item (ii) of Corollary 4.12]{BruPedra_multi}, 
one finds that there exist $C_{ABC}>0$ independent of $L$ such that 
the norm $\big\|  \big[ [ C, A(s_1)], B(s_2)\big]\big\|$ is bounded from above by $C_{ABC} L^{3}(1+ |s_1|+|s_{2}|)^{6}$, where $6$ should be understood as twice the spatial dimension. Therefore, 
\begin{equation}
\big| T_{\beta,L}(\eta)-T_{\beta,L}(\eta_\beta)\big|\le
\frac{4\pi C_{ABC}}{\beta}\int_{-\infty}^{0} ds_{1} \int_{-\infty}^{s_{1}} ds_{2}\, (1+|s_{1}|+|s_{2}|)^7 e^{\eta(s_1+s_2)}, 
\end{equation}
which vanishes in the limit as $\beta\to \infty$, uniformly in $L$, for any $\eta > 0$. This concludes the proof of \eqref{proveC.2}.

\bigskip

{\bf Acknowledgements.} This work has been supported by the European Research Council (ERC) under the European Union's Horizon 2020 research and innovation programme 
(ERC CoG UniCoSM, grant agreement n.724939 and ERC StG MaMBoQ, grant agreement n.802901). M.P. acknowledges financial support from the Swiss National Science 
Foundation, for the project ``Mathematical Aspects of Many-Body Quantum Systems''. A.G. and V.M. acknowledge financial support from MIUR, PRIN 2017 project MaQuMA, PRIN201719VMAST01.
We warmly thank the following colleagues for several enlightening discussions on different aspects of this work: 
J\"urg Fr\"ohlich, on the role of the chiral anomaly in several condensed matter systems;
Rafael Greenblatt, on the Wick rotation of the quadratic response; Nai Phuan Ong, on the experimental aspects of the 
chiral anomaly in Weyl semimetals.

\end{document}